\def\USEMOLPHYS{0} %
\if\USEMOLPHYS1
\documentclass[]{interact}
\else
\documentclass[pra,%
amsmath,amssymb,
floatfix,
nofootinbib,
reprint,
onecolumn,
]{revtex4-2}
\usepackage{booktabs}
\usepackage{libertine}
\usepackage[italic]{mathastext} %
\usepackage{graphicx}
\fi

\usepackage{microtype}

\usepackage{units}
\usepackage[caption=false]{subfig}%

\usepackage{xcolor}
\usepackage[hyperindex,colorlinks,bookmarksnumbered,ocgcolorlinks=false, plainpages=true,pdfstartview=FitH]{hyperref}
\definecolor{CBcyanlight}{RGB}{1,102,94}
\definecolor{CBblue}{RGB}{43,131,186}
\hypersetup{linkcolor=CBcyanlight,urlcolor=CBblue,citecolor=CBcyanlight}

\if\USEMOLPHYS1
\usepackage[numbers,sort&compress,merge]{natbib}
\bibpunct[, ]{[}{]}{,}{n}{,}{,}

\fi
\graphicspath{{pics/}}%

\usepackage{xspace}

\let\oldtheequation\theequation
\makeatletter
\def\tagform@#1{\maketag@@@{\ignorespaces#1\unskip\@@italiccorr}}
\renewcommand{\theequation}{(\oldtheequation)}
\makeatother 

\newcommand{\ie}{\mbox{i.\,e.}\xspace}
\newcommand{\eg}{\mbox{e.\,g.}\xspace}

\newcommand{\matrgreek}[1]{\ensuremath{\pmb{#1}}} %

\newcommand{\matr}[1]{\textbf{#1}}

\newcommand{\erw}[1]{\ensuremath {\langle{#1}\rangle}}

\newcommand{\braket}[2]{\ensuremath{ \langle #1 | \, #2  \rangle }}
\newcommand{\ketbra}[2]{\ensuremath{  | {#1} \rangle %
\langle {#2} |}}
\newcommand{\ket}[1]{\ensuremath{  | {#1} \rangle}}
\newcommand{\bra}[1]{\ensuremath{\langle {#1} | }}
\newcommand{\matrixe}[3]{\ensuremath{ \langle{#1} | \vphantom
        {#1 #3} {#2}
| {#3} \rangle }}
\newcommand{\Matrixe}[3]{\ensuremath{ \left \langle{#1} \left| \vphantom {#1 #3} {#2}  \right| {#3} \right \rangle }}
\newcommand{\Braket}[2]{\ensuremath{\left \langle #1 \vphantom{#1 #2}
    \left | \, #2 \vphantom{#1 #2} \right . \right \rangle }}

\newcommand{\partd}[2]{\ensuremath{ \frac{\partial {#1}}
{\partial {#2}} }}

\newcommand{\ii}{\ensuremath{\mathrm{i}}}

\definecolor{ocre}{RGB}{243,102,25}
\definecolor{mygray}{RGB}{243,243,244}
\definecolor{fzjred}{RGB}{175,90,80}

\definecolor{blau}{HTML}{1F78B4}
\definecolor{gruen}{HTML}{33A02C}
\definecolor{hellblau}{HTML}{A6CEE3}
\definecolor{hellgruen}{HTML}{B2DF8A}

\definecolor{nrot}{HTML}{d7191c}
\definecolor{norange}{RGB}{253,174,97}
\definecolor{ngruen}{HTML}{abdda4}
\definecolor{nblau}{HTML}{2b83ba}

\definecolor{nrot1}{RGB}{215,48,31}
\definecolor{nrot2}{RGB}{252,141,89}
\definecolor{nrot3}{RGB}{253,204,138}
\definecolor{nrot4}{RGB}{254,240,217}

\definecolor{CBred}{RGB}{215,25,28}
\definecolor{CBorange}{RGB}{253,174,97}
\definecolor{CByellow}{RGB}{255,255,191}
\definecolor{CBgreen}{RGB}{171,211,164}
\definecolor{CBlgreen}{RGB}{166,217,106}
\definecolor{CBdgreen}{RGB}{26,150,65}
\definecolor{CBblue}{RGB}{43,131,186}
\definecolor{CBblue2}{RGB}{146,197,222}
\definecolor{CBdblue}{RGB}{5,113,176}
\definecolor{CBgray60}{RGB}{102,102,102}
\definecolor{CBgray20}{RGB}{204,204,204}

\definecolor{idxgreen}{RGB}{0,158,115}
\definecolor{idxocre}{RGB}{213,94,0}
 
\newcommand{\nSPF}{\ensuremath{n_\text{SPF}}}

\begin{document}

\title{A tensor network view of multilayer multiconfiguration time-dependent Hartree methods}

\if\USEMOLPHYS1
\author{
\name{Henrik~R.~Larsson\textsuperscript{a}}%
\affil{\textsuperscript{a}Department of Chemistry and Biochemistry, University of California, Merced, CA 95343, USA}%
}

\else
\author{Henrik~R.~Larsson}
\affiliation{Department of Chemistry and Biochemistry, University of California, Merced, CA 95343, USA}
\email{larsson [at] ucmerced.e$\delta$u}
\fi

\if\USEMOLPHYS1
\maketitle
\fi

\begin{abstract}
The multilayer multiconfiguration time-dependent Hartree (ML-MCTDH) method and the density matrix renormalization group (DMRG)
are powerful workhorses applied mostly in different scientific fields.
Although both methods are based on tensor network states,
very different mathematical languages are used for describing them.
This severely limits knowledge transfer and sometimes leads to re-inventions of ideas well known in the other field.
Here, we review ML-MCTDH  and DMRG theory using both MCTDH expressions and tensor network diagrams.
We derive the ML-MCTDH equations of motions using diagrams and compare them with time-dependent and time-independent DMRG algorithms.
We further review two selected recent advancements. The first advancement is related to optimizing unoccupied single-particle functions in MCTDH, which corresponds to subspace enrichment in the DMRG. 
The second advancement is related to finding optimal tree structures and on highlighting similarities and differences of tensor networks used in MCTDH and DMRG theories.
We hope that this contribution will foster more fruitful cross-fertilization of ideas between ML-MCTDH and DMRG.
\end{abstract}
\if\USEMOLPHYS0
\maketitle
\fi

\section{Introduction}
\label{sec:intro}
The multiconfiguration time-dependent Hartree (MCTDH) method and
the density matrix renormalization group (DMRG) 
were developed nearly simultaneously in the fields of molecular quantum dynamics and condensed matter physics, respectively~\cite{Multiconfigurational1990meyer,Wave1992manthe,Density1992white,Densitymatrix1993white}.
Both methods have greatly influenced their respective field and continue to do so. This may be seen most clearly by noticing that MCTDH and DMRG 
are being used outside their initial field of research
and
are now established in other fields such as \textit{ab initio} molecular electronic  structure~\cite{Initio1999white,Quantum2001mitrushenkov,Highly2002chan,QCDMRG2003legeza} in the case of DMRG and 
strong-field electron dynamics in the case of MCTDH~\cite{MCTDHF2003zanghellini,Timedependent2004kato,Multiconfiguration2005nest,Correlated2005caillat,Multidimensional2009meyer,Timedependent2014hochstuhl,Colloquium2020lode}, among many other examples.
Initially, MCTDH and DMRG focused on very different areas and applications. MCTDH is based on the time-dependent Schrödinger equation (TDSE) whereas the DMRG is based on the time-independent Schrödinger equation (TISE).
Around 20 years after their foundations, both methods were extended.
The less flexible MCTDH wave function ansatz (relative to the DMRG), a Tucker decomposition~\cite{Tensor2009kolda}, was improved by the development of the multilayer MCTDH method (ML-MCTDH)~\cite{Quantum2003meyer,Multilayer2003wang,Multilayer2008manthe}, where a tree tensor network state (TTNS) or hierarchical Tucker format~\cite{Zerotemperature1978jullien,Classical2006shi,New2009hackbusch,Tensor2019hackbusch,Hierarchical2010grasedyck,Literature2013grasedyck,Tensor2015szalay} is used.
For the DMRG,
the focus on the TISE was broadened by extensions to real-time evolution~\cite{Efficient2004vidal,RealTime2004white,Timestep2005feiguin,Timeevolution2019paeckel}.
Furthermore, the wavefunction ansatz of the DMRG, a matrix product state (MPS)~\cite{Finitely1992fannes,Thermodynamic1995ostlund,Efficient2003vidal,Matrix2006verstraete,Densitymatrix2011schollwock},
which is a subset of TTNSs, was later extended to TTNSs~\cite{Concatenated2010hubener,Simulating2010murg,Heisenberg2013changlania,Unconstrained2014gerster,Tree2015murg}.\footnote{Note that we interpret ML-MCTDH and DMRG as particular algorithms to solve the Schrödinger equation using the same TTNS ansatz. With this interpretation, the  DMRG can be used for MPSs and TTNSs even though it has been introduced as a way for approximating ground states as MPSs.}
Inspired by the DMRG, other methods that extend TTNSs to general tensor network states (TNSs) have been developed~\cite{Renormalization2004verstraete,Entanglement2007vidal,Class2008vidal,Criticality2006verstraete,Approximating2009changlani,Completegraph2010marti,Low2012chan,Practical2014orus,Conversion2019haghshenas,Isometric2020zaletel}.
These different individual developments led to very different mathematical ``languages'' used to describe DMRG, ML-MCTDH, MPSs and TTNSs. 
In fact, describing DMRG using MPSs even is a more recent concept that differs dramatically from the original DMRG language~\cite{Densitymatrix2005schollwock,Densitymatrix2011schollwock}.
This language barrier is becoming more severe as more and more DMRG approaches are used in systems typically treated with (ML-)MCTDH and vice versa~\cite{Least2006ueda,Analytic2009dorando,TimeDependent2011haegeman,Linear2014nakatani,TimeDependent2011haegeman,Thouless2013wouters,Analytic2014kinder,Numerically2009wang,Computing2019larsson,Multiset2019kloss,
Unifying2016haegeman,Timedependent2019xie,LargeScale2019baiardi,Optimization2019baiardi,Matrix2018kurashige,Timedependent2021ren,Comparison2021mainali,Matrix2022larsson,Stochastic2022xu,Stateresolved2022larsson,Flexible2023glaser,Exploring2023kohler}.
It is very difficult to analyze similarities and differences of methods developed by these two communities, even when these methods actually are very similar or virtually identical.

Here, with a focus on ML-MCTDH, we will highlight connections between ML-MCTDH and DMRG 
by ``translating'' MCTDH expressions into the diagrammatic language that is extensively used in TNS community but so far has not come to full use in the ML-MCTDH community.
We hope that these translations will show more clearly the connections between ML-MCTDH, DMRG and TNSs.\footnote{
Some incomplete translations have already been given, \eg, in Refs.~\cite{Dynamical2007koch,Quantum2008lubich,Computing2019larsson,Tensor2019hackbusch,Time2021lindoya,Symmetries2021weike}.}
We will highlight the power of the diagrammatic language by deriving the ML-MCTDH equations of motions %
through diagrams.
We will discuss an alternative way to solve the ML-MCTDH equations of motions, which is given by time-dependent and time-independent DMRG approaches.
Furthermore, 
we will show connections between DMRG and ML-MCTDH advancements  exemplified by two advanced topics, (1) subspace enrichment (optimal unoccupied single particle functions, SPFs), and (2) optimal TTNS structures, which also includes a discussion of why MPSs are preferred in molecular electronic structure, compared to TTNSs in vibrational quantum dynamics.

Our outline is as follows: Sections \ref{sec:mlmctdh} and \ref{sec:diagrams} briefly review the ML-MCTDH  and tensor network languages, respectively. 
Sections \ref{sec:diagram_comparison}  and  \ref{sec:ml_mctdh_diagrams}  ``translate'' standard \makebox{(ML-)}MCTDH expressions and the ML-MCTDH equations of motions, respectively.
\autoref{sec:mctdh_deriv} fully derives the ML-MCTDH equations of motions solely using diagrams. 
\autoref{sec:canonicalization} reviews TTNS canonicalization/orthogonalization procedures and DMRG-typical sweeps. 
A time-dependent DMRG approach that can be viewed as counterpart to the ML-MCTDH equations is reviewed in \autoref{sec:tddmrg} and used for describing the time-independent DMRG in \autoref{sec:tise_dmrg}.
Sections \ref{sec:opt_unocc_spf} and \ref{sec:tree_opt} deal with two advanced topics  on unoccupied SPFs and on tree structure optimizations. 
We conclude in \autoref{sec:conclusions}.

\section{The multilayer multiconfiguration time-dependent Hartree ansatz}
\label{sec:mlmctdh}
To solve Schrödinger's equation, the standard full configuration interaction (FCI) approach is a Ritz-Galerkin ansatz~\cite{Numerical1987gottlieb,Tensor2015szalay},
where the $F$-dimensional wavefunction  
is represented by 
a direct product of ``primitive'' bases $\{\ket{\chi^{(\kappa)}_{j_\kappa}}\}_{j_\kappa=1}^{N_\kappa}$ of finite size $N_\kappa$ in each dimension $\kappa\in[1,F]$:
\begin{equation}
  \ket{\Psi_\text{FCI}} = \sum_{j_1=1}^{N_1}  \sum_{j_2=1}^{N_2} \dots \sum_{j_F=1}^{N_F} %
  C_{j_1,j_2,\dots,j_F} \bigotimes_{\kappa=1}^{F} \ket{\chi^{(\kappa)}_{j_\kappa}}.
  \label{eq:fci}
\end{equation}
Here, we use Dirac's notation~\cite{New1939dirac,Quantum2019cohen-tannoudji} 
and  assume an orthonormal primitive basis. 
Further, we will only discuss the case of distinguishable particles (or an appropriate mapping of indistinguishable particles, such as the Jordan-Wigner transform) and thus do not assume any (anti-)symmetry in the 
real- or complex-valued coefficient tensor $\matr C$, whose entries are then to be determined.

In most molecular quantum dynamics applications, the primitive basis is a spectral or pseudospectral grid-like basis given by the discrete variable representation (DVR)~\cite{Calculation1965harris,Calculation1968dickinson,Discrete1982lill,Discretevariable2000light,Introduction2008tannor,PhaseSpace2018tannor}, which, among others, is also known as Lagrange-mesh method~\cite{Lagrangemesh2015baye} in some Physics literature
and as cardinal basis in some applied mathematics literature~\cite{Chebyshev2001boyd,Introduction2008tannor}.
Note that nonlinear bases, in particular Gaussians, are possible as well~\cite{Multimode2008burghardt,Gaussianbased2013romer}.

The FCI ansatz in \autoref{eq:fci} is exact in the space of the primitive basis (given no further approximations on the Hamiltonian) but the number of entries in the coefficient tensor $\matr C$ scales as  $\widetilde N^F$, where $\widetilde N$ is the geometric mean of the basis sizes $N_\kappa$.
To avoid this prohibitive scaling,
the \mbox{(ML-)}MCTDH ansatz approximates \autoref{eq:fci}
by expanding $\ket{\Psi}$ in another orthonormal basis, the so-called \emph{single-particle functions (SPFs)} $\{\ket{\phi_{j_\kappa}^{1;\kappa}}\}$.%
\footnote{To avoid defining ``single-particle states,''
here we will use the term ``function'' for both the state $\ket{\phi_{j_\kappa}^{1;\kappa}}$ and its representation, \eg, in position space,  $\braket{x}{\phi_{j_\kappa}^{1;\kappa}}=\phi_{j_\kappa}^{1;\kappa}(x)$. Many MCTDH derivations typically start with defining SPFs as actual functions and then use Dirac's notation for matrix elements and projectors. Here, we prefer to use Dirac notation throughout.
}
The difference between the primitive basis and the SPF basis is that the latter is variationally optimized and thus fewer basis functions are required. The \mbox{(ML-)}MCTDH ansatz then takes the form of
\begin{equation}
\ket{\Psi} = \sum_{j_1=1}^{n_1} \sum_{j_2=1}^{n_2} \cdots \sum_{j_d=1}^{n_d} A^{1}_{j_1,j_2,\dots,j_d} \bigotimes_{\kappa=1}^{d} \ket{\phi_{j_\kappa}^{1;\kappa}}. \label{eq:mlmctdh}
\end{equation}
Note that the SPFs may be multidimensional (thus $d \le F$); see below.
The superscript $1$ in $ A^{1}_{j_1,j_2,\dots,j_d}$ and in $\ket{\phi_{j_\kappa}^{1;\kappa}}$  denotes the \emph{layer} in ML-MCTDH.
For time evolution,
both the $d$-dimensional tensor $\matr A^{1}$ and the  SPFs 
are optimized according to the Dirac-Frenkel-McLachlan time-dependent variational principle (TDVP)~\cite{Note1930dirac,Wave1934frenkel,Variational1964mclachlan,Geometry1981kramer,Multiconfiguration2000beck}.

Some remarks on the notation that is used here: For consistency with the main body of ML-MCTDH literature, here we closely but not exclusively follow the notation of Manthe~\cite{Multilayer2008manthe,Wavepacket2017manthe}.
It only marginally differs from that of Meyer and Vendrell~\cite{Multilayer2011vendrell},
but there are some differences to the notation used by Wang and Thoss (\eg, the counting of layers differs)~\cite{Multilayer2003wang,Multilayer2015wang}.
Notably, all of these notations are based on the original MCTDH formulations~\cite{Multiconfigurational1990meyer,Wave1992manthe}.
A different notation exists in the mathematics community~\cite{Dynamical2013lubich,Projectorsplitting2014lubich,Time2015lubich,Implementation2017kloss,Time2018lubich}.
A non-exhaustive comparison of used symbols in the MCTDH, mathematics, and tensor network communities is given in \autoref{tab:notation}.

The ansatz in \autoref{eq:mlmctdh} is very flexible and there are three main realizations of it:
(1) In the simplest MCTDH ansatz, the SPFs are one-dimensional and $d=F$.
The SPFs are then represented by the same time-independent, primitive basis used in the FCI ansatz, \autoref{eq:fci}. MCTDH then has the same exponential scaling as the FCI ansatz, but the base is reduced as the SPFs are optimized and thus fewer of them are needed, compared to the primitive basis.

(2) For MCTDH with \emph{mode combination}~\cite{Relaxation1998worth}, $d < F$ and the SPFs are expressed in terms of a multidimensional ($d_\kappa$-dimensional) basis.
Typically, this basis is a direct-product basis, but it also can be a non-direct-product basis~\cite{Effect2001sukiasyana,Rotational2016fuchsel,Dynamical2017larsson}.
The multidimensional degrees of freedom that describe the multidimensional SPFs are then denoted as ``\emph{logical}.''
Mode combination is useful for describing strongly correlated degrees of freedom. %
For systems with $F\succsim 4$, mode combination often leads to a better efficiency than the normal MCTDH ansatz.
In the context of molecular electronic structure, 
an antisymmetrized MCTDH ansatz with mode combination is equivalent to the time-dependent complete active space self-consistent field (TD-CASSCF) algorithm ~\cite{Selfconsistent1939hartree,MC1973hinze,Extended1966das,Complete1980roos,Molecular2013helgaker}. 
This is also called multiconfiguration time-dependent Hartree-Fock, MCTDHF~\cite{MCTDHF2003zanghellini,Timedependent2004kato,Multiconfiguration2005nest,Correlated2005caillat,Multidimensional2009meyer,Timedependent2014hochstuhl,Colloquium2020lode}.%
\footnote{
Note that a method dubbed MCTDHF by Yeager and J\o{}rgensen actually is a response CASSCF approach and thus not explicitly time-dependent~\cite{Multiconfigurational1979yeager}.
However, 
related publications already utilized the time-dependent variational principle~\cite{Static1974moccia,Time1980dalgaard,Remarks1983mcweeny}.}
In TD-CASSCF, the three-dimensional SPFs describing one electron
would be called molecular orbitals and the primitive basis functions would be called atomic orbitals.

(3) For ML-MCTDH, $d < F$ holds as well, but the $d_\kappa$-dimensional SPFs in   \autoref{eq:mlmctdh}
are expanded recursively using the very same MCTDH ansatz, \ie,
\begin{equation}
  \ket{\phi_i^{1; \kappa}} = \sum_{j_1=1}^{n_{\kappa,1}}  \sum_{j_2=1}^{n_{\kappa,2}}\dots  \sum_{j_{d_\kappa}=1}^{n_{\kappa,{d_\kappa}}} A^{2;\kappa}_{i; j_1,j_2,\dots,j_{d_\kappa}} \bigotimes_{\lambda=1}^{d_\kappa} \ket{\phi_{j_\lambda}^{2;\kappa, \lambda}}.
\end{equation}
$\ket{\phi_{j_\lambda}^{2;\kappa, \lambda}}$ are then either described by the primitive basis (possibly with mode combination) or they are again expanded using the MCTDH ansatz. This is repeated recursively until the last layer $L$ is reached.

\begin{figure}
\centering
\includegraphics{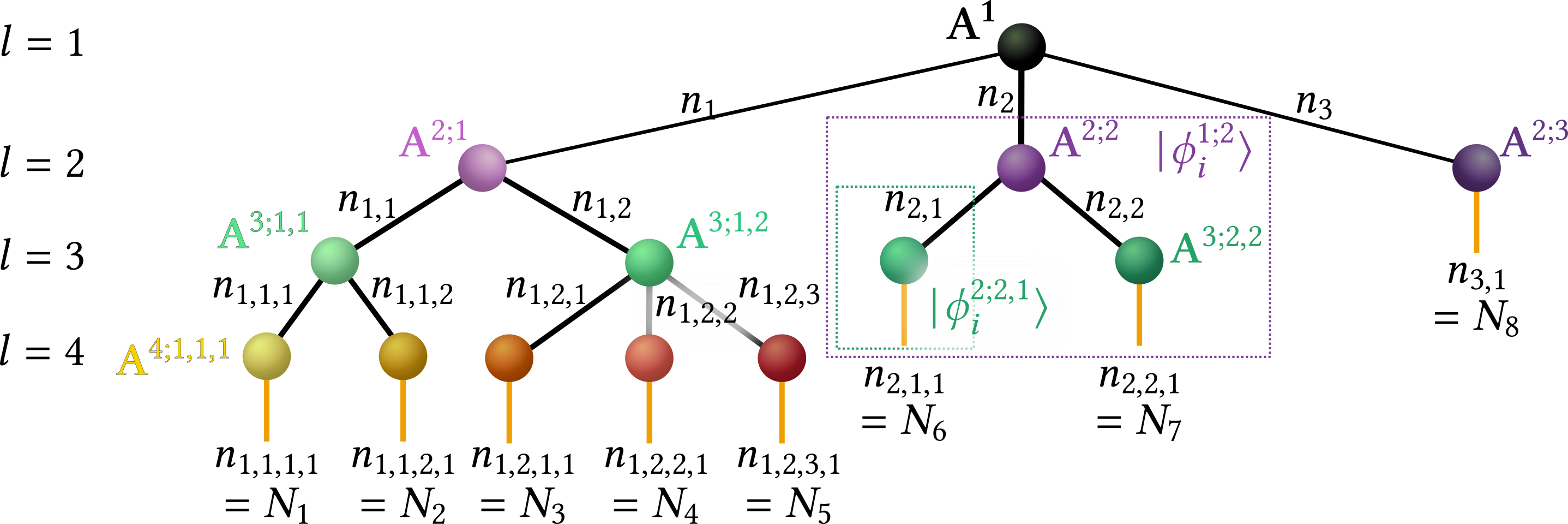}
\caption{Example ML-MCTDH tree/TTNS displaying the notation.
Bonds (vertices) denoting physical indices 
corresponding to primitive basis functions are shown in orange.
Their size is given by $N_i$ but also implicitly through $n_{\kappa_1,\kappa_2,\dots,\kappa_l}$. 
Bonds denoting ''virtual`` 
indices
for single-particle functions (SPFs)
are shown in black.
The tree has $L=4$ layers. 
Their basis sizes are $n_{\kappa_1,\kappa_2,\dots,\kappa_l}$,
their dimension is $d_{\kappa_1,\kappa_2,\dots,\kappa_{l-1}}$, and the total number of SPFs at one node is  $n_{\kappa_1,\kappa_2,\dots,\kappa_{l-1}}$.
Note that the SPF 
$\ket{\phi_i^{l; \kappa_1,\kappa_2,\dots,\kappa_{l}}}\equiv \ket{\phi_i^{z}}$ is described by the tensor  $\matr A^{l+1; \kappa_1,\kappa_2,\dots,\kappa_{l}}\equiv \matr A^{z'+1}$ and those from lower nodes connecting to $\matr A^{z'+1}$.
Some of the SPFs and SPF coefficient tensors are shown as well. 
The tensors that describe two particular SPFs are marked by dotted rectangles.
}
  \label{fig:mltree_notation}
\end{figure}

The recursive structure of the ML-MCTDH ansatz can be represented using tree diagrams~\cite{Multilayer2008manthe} as shown in \autoref{fig:mltree_notation}, which also gives an example of some ML-MCTDH notation.
In the ML-MCTDH context, these diagrams are almost exclusively used to depict the tree structure of a particular ML-MCTDH wavefunction. Only recently have they been used to represent
some other MCTDH-related quantities~\cite{Computing2019larsson,Symmetries2021weike,Time2021lindoya,Time2021lindoy}.
As we will show below in \autoref{sec:diagrams}, there is a direct connection of these ML-MCTDH diagrams to the tensor network diagrams used in DMRG literature%
~\cite{Densitymatrix2011schollwock,Practical2014orus,Computing2019larsson,Timeevolution2019paeckel}.
Among others, we will further show how these diagrams can be used for deriving the ML-MCTDH equations of motions.

\begin{table}
  \caption{Some notation used in this work and in MCTDH literature. A non-exhaustive list of alternative symbols used in MCTDH, mathematics, and DMRG literature is given, also.
  See \autoref{fig:mltree_notation} for a concrete example of the usage of some of the symbols.
  \label{tab:notation}
  }
 \begin{tabular}{lll}
 \toprule
   description & symbol & alternative symbols\\ \midrule
physical dimension               & $F$          & $D$, $f$, $L$\\
physical (``primitive``) basis   state                & $\ket{\chi^{(\kappa)}_{j_\kappa}}$ & $\ket{\sigma}$\\
physical basis dimension         & $N_\kappa$          & $N^{[\kappa]}, n_\kappa$, $k$, $\sigma$ \\
number of layers & $L$ \\
layer        & $l$ \\
SPF basis state &$\ket{\phi_{i}^{l;\kappa_1,\kappa_2,\dots,\kappa_{l}}}$ & $\ket{\nu_{i_q}^{(\kappa,q)}}$, $\ket{\xi_{a_\gamma}^{(\kappa,q,\gamma)}}$\\
horizontal position in layer $l$ & $\kappa$           & $n$ \\
node/SPF position & $\kappa_1, \kappa_2, \dots,\kappa_L$ & $\kappa$ \\
layer and node/SPF position & $z\equiv l; \kappa_1, \kappa_2, \dots,\kappa_l$ &%
 $\lambda$\\ %
&%
& $z\equiv l; \kappa_1, \kappa_2, \dots,\kappa_{l-1}$\\
coefficient tensor position & 
 $z'\equiv l; \kappa_1,\kappa_2,\dots,\kappa_{l-1}$ & \\
& $z'+1\equiv l+1; \kappa_1,\kappa_2,\dots,\kappa_{l}$ &$z'\equiv l+1; \kappa_1,\kappa_2,\dots,\kappa_{l}$ \\
node/tensor/site/SPF dimension     & $d_{\kappa_1,\kappa_2,\dots,\kappa_{l-1}}$ & $d^{[l;\kappa]}$, $p_{\kappa}$ , $F(\kappa)$, $Q(\kappa)$, $M(\kappa,q)$   \\
bond dim./rank/SPF basis size   & $n_{\kappa_1,\kappa_2,\dots \kappa_{l}}$  &$n^{[l;\kappa]}$ ,  $m$, $M$, $D$, $\chi$, $r$\\
generic SPF basis size & $\nSPF$ &  $m$, $M$, $D$, $\chi$, $r$\\
total size of basis representing SPF  & $\bar n_{\kappa_1,\kappa_2,\dots,\kappa_{l-1}}$ \\
SPF index                   & $i,j,k,x,j_1,j_2,\dots$                & $m,n$\\
SPF 
composite index &  $J\equiv j_1,j_2,\dots,j_{d_{\kappa_1,\kappa_2,\dots \kappa_{l}}}$\\
SHF composite index  
 & $J^{\kappa_\lambda} \equiv j_1,j_2,\dots,j_{\kappa_\lambda-1},$\\
 &\hphantom{$J^{\kappa_\lambda} \equiv$}$j_{\kappa_\lambda+1},\dots,j_{d_{\kappa_1,\kappa_2,\dots \kappa_{l}}}$\\
SPF tensor/site             & $\matr A^{l;\kappa_1,\kappa_2,\dots \kappa_{l-1}} \equiv \matr A^{z'}$   &$\matr A^{[l;\kappa]}$, $\matr B^{\kappa, j_\kappa}$, $\matr C^{\kappa, q, i_q}$, $\chi^{[l;\kappa]}$, $\Lambda^{[l;\kappa]}$\\
configuration & $\ket{\Phi_{J}^{z'}}$ & $\ket{\Phi_{J}^{z}}$  \\ 
single-hole function (SHF) & $\ket{\Psi^z_i}$\\
SHF configuration & $\ket{\Xi_{J^{\kappa_l}}^{z}}$ & $\ket{\Phi_{J^{\kappa_l}}^{z}}$  \\ 
SPF projector & $\hat P^z$ & $\hat P^z_{\kappa_l}$\\
SHF projector & $\mathcal{\hat P}^z$ \\ 
single-particle density matrix & $\matrgreek \rho^z$\\
mean-field operator & $\erw{\hat H^z}_{ij}$\\
gauge operator & $\hat g^z$ & $\hat h^z$\\
one-dimensional Hamiltonian term &  $\hat h^{(\kappa)}_s$\\
\bottomrule
 \end{tabular}
\end{table}

In the commonly used notation,
the $d_{\kappa_1,\kappa_2,\dots,\kappa_{l-1}}$-dimensional SPFs of layer $l$ are denoted by $\ket{\phi_{i}^{l;\kappa_1, \kappa_2,\dots,\kappa_l}}$. 
The additional symbols, $\kappa_1, \kappa_2,\dots,\kappa_l$ specify the exact location of the SPFs in the tree.
The indices $\kappa_1, \kappa_2,\dots,\kappa_l$  are determined by traversing the tree using depth-first search starting at $l=1$.
During the traversal, the horizontal position of the path through the tree taken at layer $\lambda$ is given by 
$\kappa_{\lambda}$.

To avoid dealing with the cluttered notation used for indexing and specifying the locations of tensors and functions in the tree, MCTDH developers often simplify the notation by defining the composite index $J\equiv j_1,j_2,\dots,j_{d_{\kappa_1,\kappa_2,\dots, \kappa_{l}}}$, which specifies the entries of the tensors $A^{l+1; \kappa_1, \kappa_2,\dots,\kappa_{l}}$.
Depending on the context, $J$ can also specify a raveled index with range $[1,\bar n_{\kappa_1,\kappa_2,\dots,\kappa_{l}}]$ with
\begin{equation}
  \bar{n}_{\kappa_1,\kappa_2,\dots,\kappa_{l}} = \prod_{\lambda=1}^{d_{\kappa_1,\kappa_2,\dots,\kappa_{l}}} n_{\kappa_1,\kappa_2,\dots,\kappa_l,\kappa_\lambda}.%
  \label{eq:nbar}
\end{equation}
We further simplify the notation\footnote{Note that in Ref.~\cite{Computing2019larsson} we use a different notation where $\kappa$ means the horizontal position in layer $l$. Then giving $l$ and $\kappa$ suffices to specify the location of an SPF in the tree.} %
by defining  the compound indices 
$z\equiv l; \kappa_1,\kappa_2,\dots,\kappa_{l}$,
$z'\equiv l; \kappa_1,\kappa_2,\dots,\kappa_{l-1}$,
and $z'+1\equiv l+1; \kappa_1,\kappa_2,\dots,\kappa_{l}$. $z$ specifies the layer and location of an SPF, and $z'$ that of an SPF tensor.\footnote{Note that Vendrell and Meyer use the same symbol but a different definition~\cite{Multilayer2011vendrell}, where $z$ does not contain $\kappa_l$, which is added to the symbols separately.}
This simplified notation
leads to a more compact description of the SPFs as 
\begin{equation}
  \ket{\phi_i^{z}} = \sum_{J=1} A_{i; J}^{z'+1} \ket{\Phi_J^{z'+1}},
  \label{eq:spf_simplified}
\end{equation}
where
\begin{equation}
\ket{\Phi_J^{z'+1}} \equiv %
\ket{\Phi_{ j_1,j_2,\dots,j_{d_{\kappa_1,\kappa_2,\dots \kappa_{l}}} }^{l+1, \kappa_1,\kappa_2,\dots,\kappa_l}} %
= \bigotimes_{\kappa_{l+1}=1}^{d_{\kappa_1,\kappa_2,\dots,\kappa_l}} \ket{\phi_{j_{\kappa_{l+1}}}^{l+1; \kappa_1,\kappa_2,\dots, \kappa_l,\kappa_{l+1}}}\label{eq:spf_conf}
\end{equation}
are the \emph{configurations}.

Armed with this simplified notation, we introduce 
the counterpart of the SPFs,  the \emph{single-hole functions (SHFs)} $\ket{\Psi_{i}^{z}}$, which 
are defined by projecting 
the corresponding SPFs $\bra{\phi_{i}^{z}}$ 
onto 
the ML-MCTDH state $\ket{\Psi}$, \ie, 
\begin{equation}
   \ket{\Psi_{i}^{z}} = \braket{\phi_{i}^{z}}{\Psi}.
\end{equation}
Using the orthonormality conditions of the SPFs, we 
can describe $\ket{\Psi}$ as sum of product of SHFs and SPFs,
\begin{equation}
  \ket{\Psi}  = \sum_{i=1}^{n_{\kappa_1,\kappa_2,\dots,\kappa_l}}  \braket{\phi_{i}^{z}}{\Psi} \otimes \ket{\phi_{i}^{z}} %
  \equiv  \sum_{i=1}^{n_{\kappa_1,\kappa_2,\dots,\kappa_l}} \ket{\Psi_{i}^{z}}\otimes \ket{\phi_{i}^{z}}.
  \label{eq:shf_expansion}
\end{equation}
The SHFs are useful for defining effective operators, as shown below in \autoref{sec:rdms_mean_fields}.

In traditional DMRG language~\cite{Densitymatrix1993white,Densitymatrix2005schollwock,Densitymatrix2011schollwock}, \autoref{eq:shf_expansion}, corresponds to using single-site DMRG and expanding the state in terms of a \emph{system block}, $\{\ket{A_i}\}$, and an \emph{environment block}, $\{\ket{B_i}\}$:
\begin{equation}
  \ket{\Psi_\text{DMRG}} = \sum_{i} \ket{A_i} \otimes \ket{B_i},\label{eq:dmrg_state_expansion}
\end{equation}
where the system block is given in MCTDH notation by the SHFs and the environment block is given by the SPFs.
\section{Tensor network diagrams}
\label{sec:diagrams}

\begin{figure}
\centering
\includegraphics{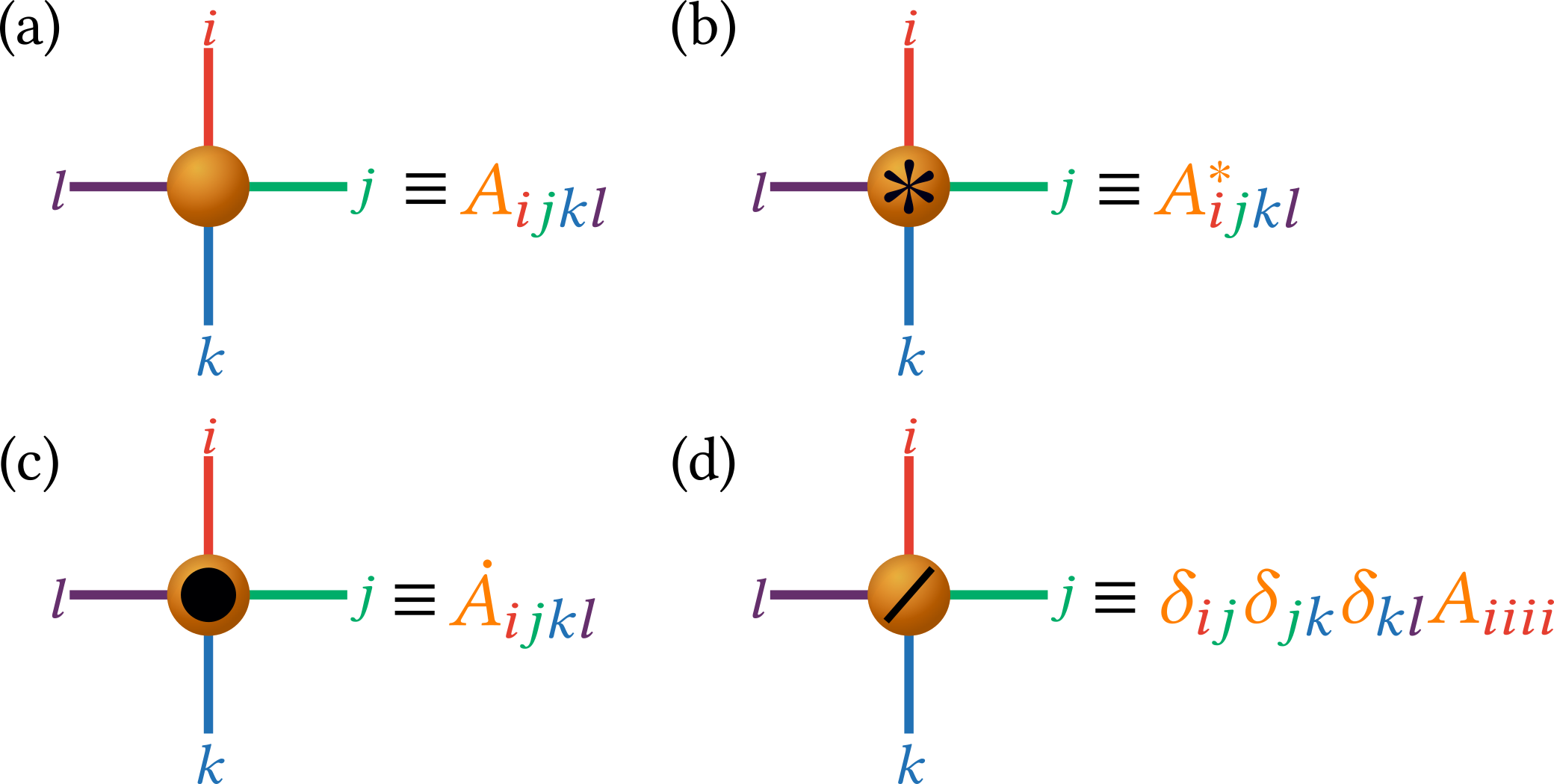}
\caption{Example of mapping a tensor $A_{ijkl}$ to a node with four vertices/bonds in a tensor network diagram. Panels (b)-(d) show  additional notations introduced here for specifying complex conjugation (b), time-derivative (c), and diagonal tensors (d).}
  \label{fig:definitions1}
\end{figure}

The complete tensor contraction pattern to retrieve the full ML-MCTDH state/TTNS represented by the primitive basis from the tensors/SPF representations $\matr A^{z'}$ actually is given by the diagram shown in \autoref{fig:mltree_notation}.
This is possible by introducing a means of translating these types of tensor network diagrams to mathematical equations. 
Since the 19th century, scholars have described tensor algebra using
graphical notations~\cite{Application1885kempe,Extract1878clifford,Application1878sylvester}; see Ref.~\cite{Boosting2021kim} for an overview.
Tensor network diagrams were first used by Penrose in the 1950s~\cite{Applications1971penrose} and 
can be highly useful not only to visualize but also to derive mathematical expressions~\cite{Densitymatrix2011schollwock,Low2012chan,Practical2014orus,Tensor2016cichocki,Handwaving2017bridgeman,Timeevolution2019paeckel}.
The notion of tensor networks leads to TNSs.
In a tensor network diagram, each tensor is represented as a node and each of the tensor's indices are represented as edges or ``bonds;'' see \autoref{fig:definitions1} for a representative example.
Some additional definitions are often introduced.
In the following, an asterisk on top of the node denotes complex conjugation. A dot denotes the time-derivative of a tensor and a diagonal line denotes diagonal tensors, as shown in \autoref{fig:definitions1}.
As the TTNS  is the wavefunction ansatz of the ML-MCTDH method, 
in the following, the terms TTNS and ML-MCTDH state are used interchangeably.

\begin{figure}
\centering
\includegraphics{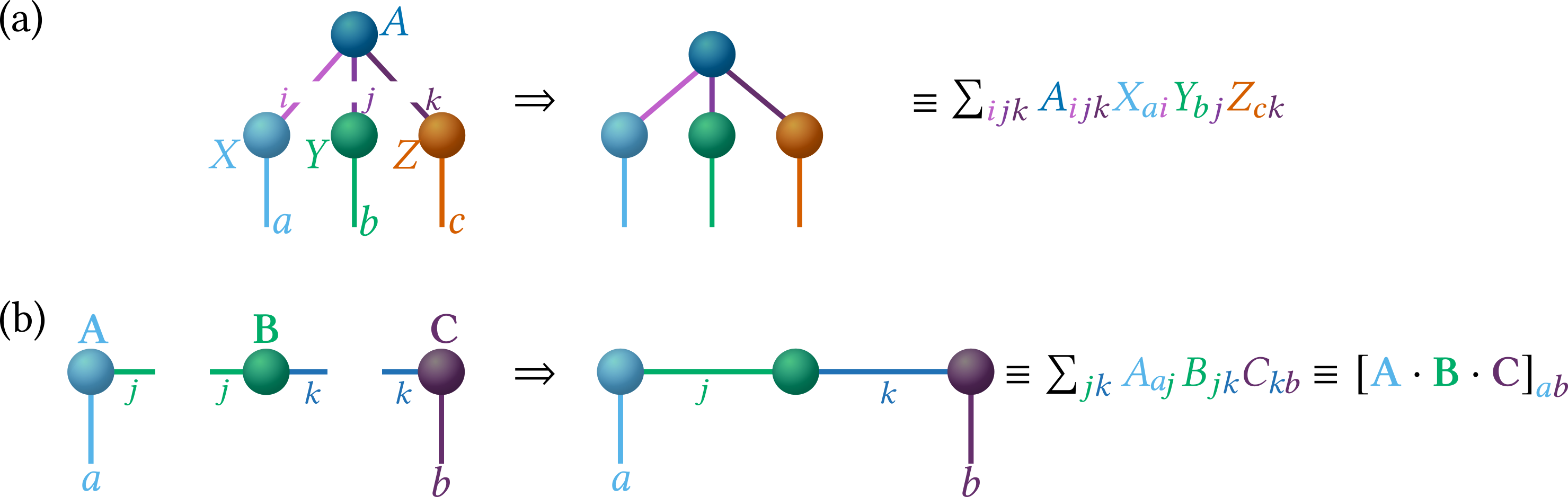}
\caption{Two exemplary tensor contractions. Shown are the individual tensors, the actual tensor network diagrams, where bonds/vertices corresponding to the indices that are contracted over are connected, as well as the mathematical equations. 
(a) shows the contraction of an MCTDH state/Tucker tensor decomposition. (b) shows the $a,b$ths entry of the product of the matrices $\matr A$, $\matr B$, and $\matr C$.}
  \label{fig:notation_mat_prods}
\end{figure}

Similar to Einstein's summation convention~\cite{Grundlage1916einstein},
contractions of tensors (summations over common indices) are defined by connecting the common bonds (edges) of those nodes whose corresponding indices should be contracted. 
\autoref{fig:notation_mat_prods} depicts two examples.

\begin{figure}
\centering
\includegraphics{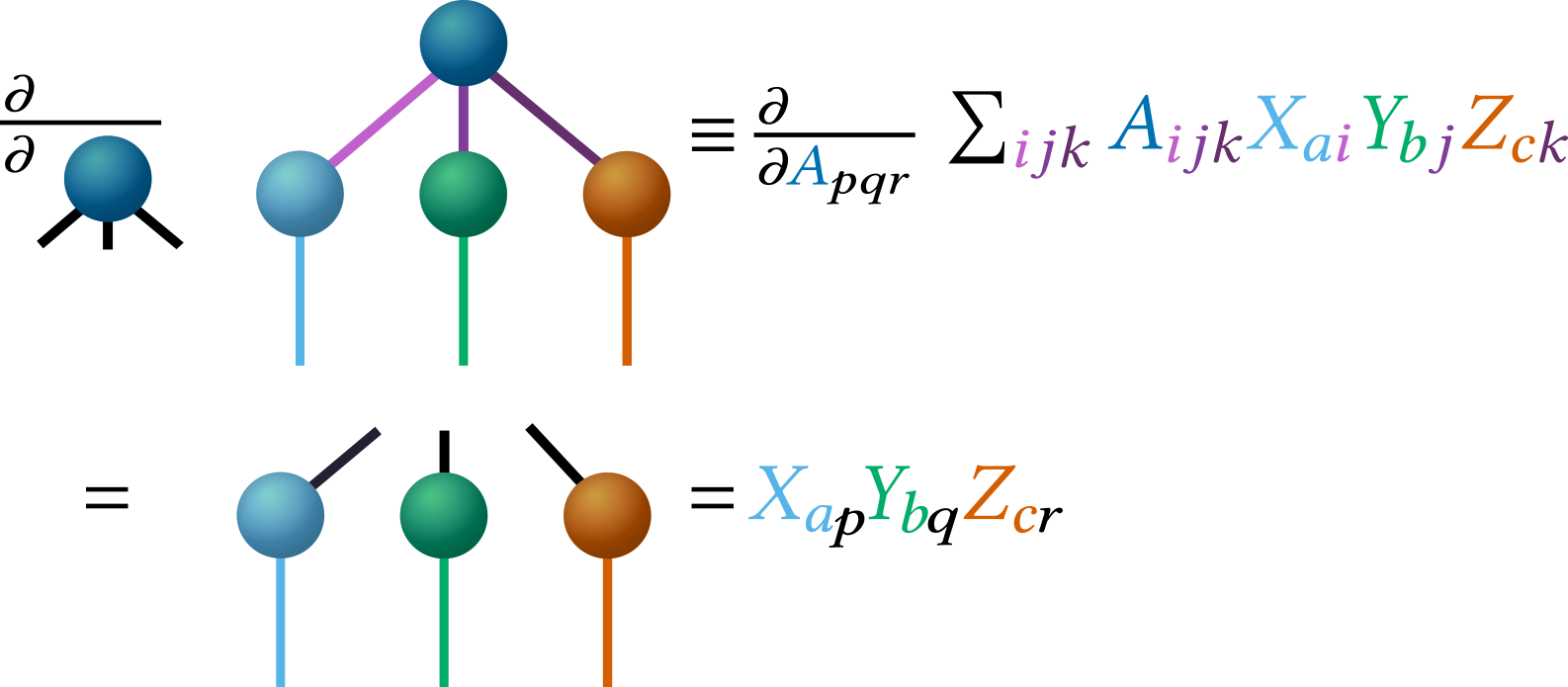}
\caption{Example of how to use diagrams to derive mathematical expressions. Shown is the derivative of the tensor network from \autoref{fig:notation_mat_prods} (a) with respect to $A_{pqr}$, i.e., 
$\left.\partd{}{A_{ijk}}\right|_{i=p,j=q,k=r}$.}
  \label{fig:math}
\end{figure}

Tensor network diagrams are not only helpful to visualize tensor contractions, but they also can be used to derive equations and to denote specific properties. \autoref{fig:math} shows an example of calculating the derivative of a tensor network through a diagram. 

\section{Tensor network diagrams in comparison to MCTDH language}
\label{sec:diagram_comparison}
\begin{figure}
\centering
\includegraphics{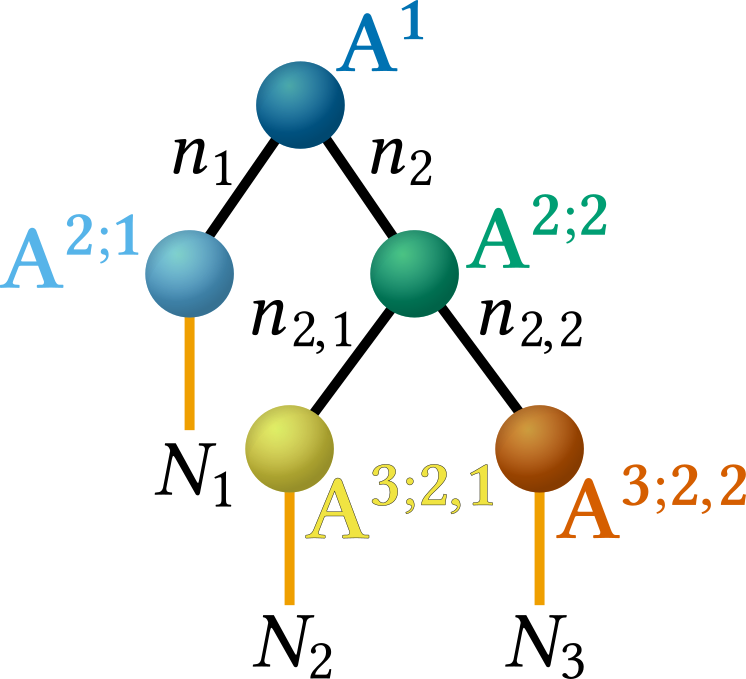}
\caption{Example ML tree/TTNS  that is used here for deriving ML-MCTDH expressions. The TTNS has three layers and describes a three-dimensional state.}
  \label{fig:mltree}
\end{figure}

Aiming at comparing the MCTDH language with tensor diagrams, here we give a translation of commonly used MCTDH expressions.
We will use the ML-MCTDH state/TTNS shown in \autoref{fig:mltree} as example. Note that, given the flexibility of TTNSs in terms of their structure, TTNS equations are best represented with the help of example trees. Generalizations to other tree structures are straightforward. This is different to MPSs  or projected entangled pair states, since those TNSs have simpler structures (one- or two-dimensional grids, respectively).

In the diagram of a TTNS, indices that are free ``dangling'' bonds, \ie, that are not connected to another node, are called physical. 
In tensor network diagrams, all other bonds of a TTNS are called virtual.
Physical and virtual bonds cannot always directly be distinguished 
in more general diagrammatic expressions 
and need to be inferred from the initial TTNS. %
To simplify this, in the following diagrams,  we will always color physical bonds orange and virtual bonds black. Here, each tensor has its own specific color, as given in \autoref{fig:mltree}. 
In tensor network diagrams, the number of SPFs, $n_{\kappa_1,\kappa_2,\dots,\kappa_l}$, corresponds to the dimension of a bond and hence is also called bond dimension $D$ (sometimes also rank $r$ in mathematics or number of kept states, $M$, in traditional DMRG language; compare with \autoref{tab:notation}).
In the following we will use $\nSPF$ when referring to the bond dimension/number of SPFs in general and $n_{\kappa_1,\kappa_2,\dots,\kappa_l}$  when referring to a specific node. 
The dimensions of the physical bonds are given as that of the primitive basis.

Unlike a complete TTNS, functions such as SPFs or SHFs do have free virtual bonds. 
Any function must contain one or more free physical bonds and, typically, one single free virtual bond that denotes the function number (\eg, $i$ in $\ket{\phi_i^z}$).\footnote{Several virtual bonds are possible but rare. See below in \autoref{eq:shf_doubly_indexed} for an example given by the doubly-indexed SHF.}
The physical bonds then denote the function values.\footnote{A physical bond can also be interpreted as the vector subspace that the state is element of. Then a TTNS diagram actually represents a state $\ket{\Psi}$ and not a particular representation.}
Orthonormality is often depicted either by using arrows instead of bonds or by using special symbols for the nodes such as triangles. %
Here, we follow the convention from ML-MCTDH and some mathematics literature~\cite{Dynamical2013lubich,Tensor2019hackbusch,Tensor2015szalay} %
and depict orthonormality by the direction of the virtual bonds.
If the virtual bond is pointing upward (toward layer $1$, away from the physical bonds), then the function/tensor is orthonormal. 
Note that there always is only one single virtual bond pointing upward.
If the virtual bond is pointing downward (toward the physical bonds/the last layer), then the function/tensor is nonorthonormal.\footnote{In special cases such as natural SHFs, virtual downward pointing bonds can be orthogonal but not orthonormal, see \autoref{sec:ml_mctdh_diagrams}.}

\begin{figure}
\centering
\includegraphics{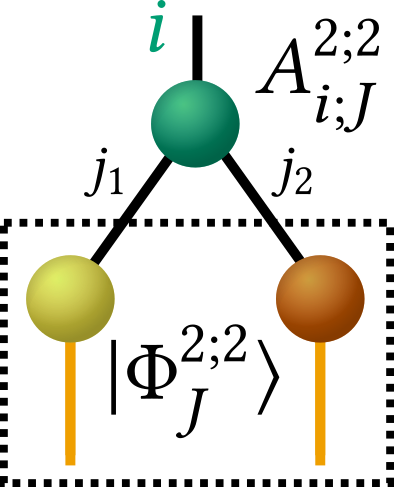}
\caption{Example configuration. Shown are $\ket{\Phi_J^{2;2}}$ from \autoref{fig:mltree} in the dashed rectangle. A contraction of $\ket{\Phi_J^{2;2}}$  with $A^{2;2}_{i;J}$, leads to the SPF $\ket{\phi_i^{1;2}}$.}
  \label{fig:config_def}
\end{figure}

\begin{figure}
\centering
\includegraphics{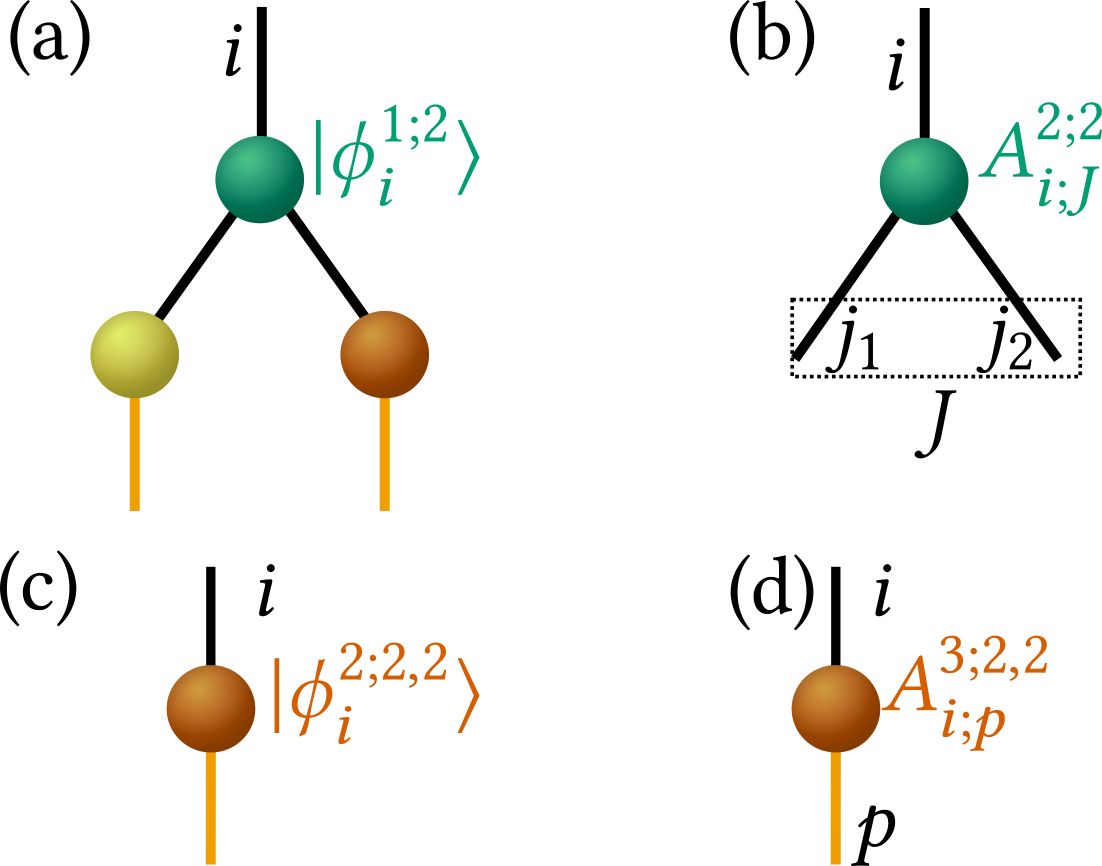}
\caption{Diagrams of SPFs in comparison to tensors. Four examples of SPFs $\ket{\phi_i^z}$ and tensors $\matr A^{z'+1}$ taken from \autoref{fig:mltree} are shown. (a) and (c) are SPFs as they contain physical bonds. (b) and (d) are tensors as the diagrams only contain a single node. The node in (c) is identical to that in (d) and can be interpreted either as SPF or tensor, depending on the context.}
  \label{fig:spf_def}
\end{figure}

An example of a configuration  (group of SPFs/nodes) is shown in 
\autoref{fig:config_def} together with a contraction of the configurations with a coefficient tensor, which leads to an  SPF. 
Examples of SPFs in comparison to single coefficient tensors are shown in \autoref{fig:spf_def}. 
A coefficient tensor $\matr A^{z'}$ is just a single node and its connecting bonds.
The values of $A^{z'+1}_{i;J}$ can be obtained from the SPFs $\{\ket{\phi_i^z}\}_i$ by projecting all SPFs connecting to $z$ onto $\{\ket{\phi_i^z}\}_i$, compare with \autoref{eq:spf_simplified} and \ref{eq:spf_conf}.
Note that for the last layer in the tree, a node can be interpreted either as tensor or as SPF,
see \autoref{fig:spf_def}(c) and (d). 
Which interpretation is used depends on the context.
$\matr A^1$ is called root tensor and the corresponding node is called root node.

\begin{figure}
\centering
\includegraphics{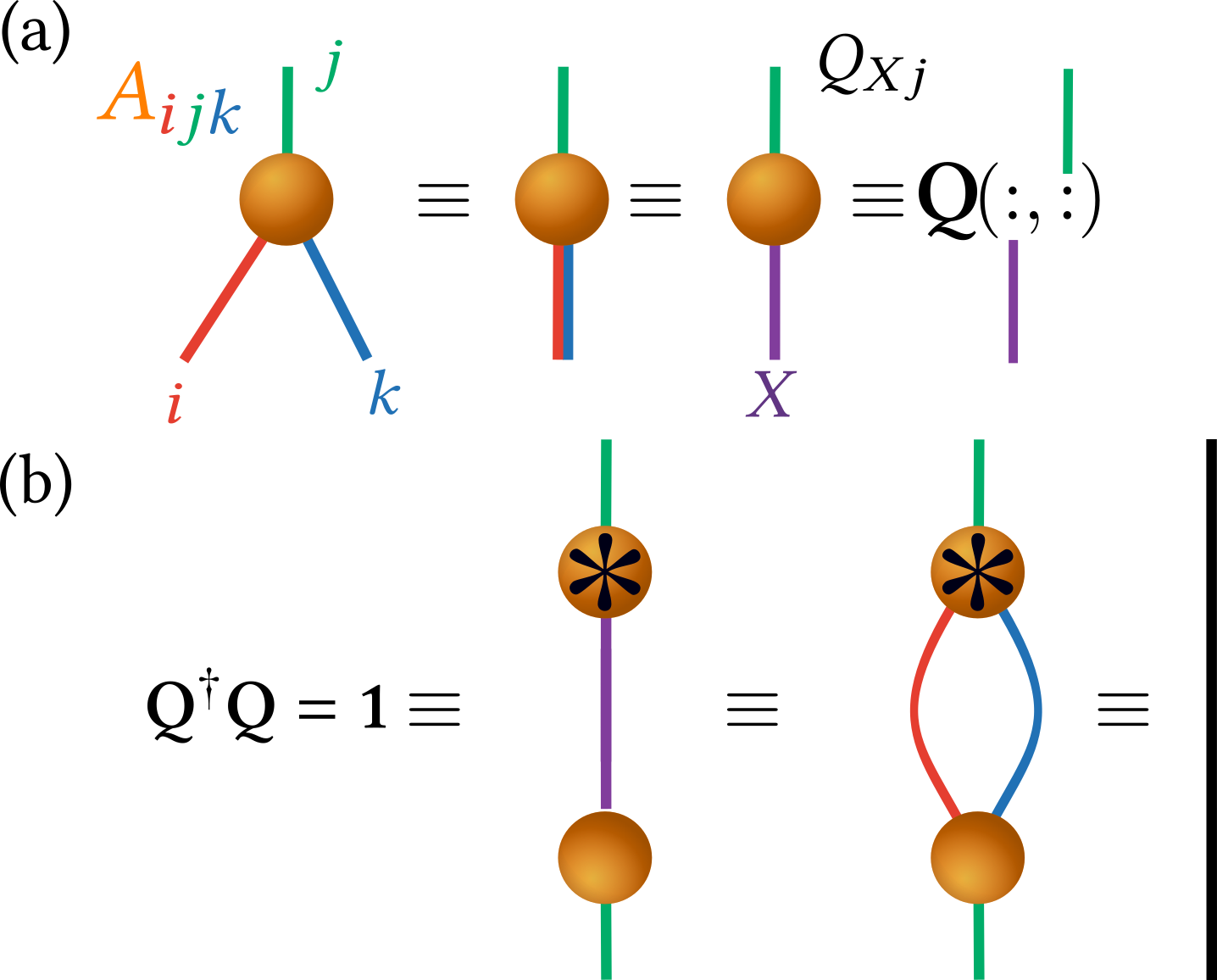}
\caption{Example of matricization and orthonormal matrices as diagrams. (a) Matricization of the tensor $A_{ijk}$ into $Q_{Xj}$, where $X$ is a composite index for $i$ and $k$. (b) Diagrammatic notation of the orthonormality of $\matr Q$ and, equivalently, $\matr A$. The straight line denotes a unit matrix.} 
  \label{fig:orth_mat}
\end{figure}

It is often useful to reshape (and permute) a tensor into a matrix, which is called matricization (or flattening or matrix unfolding)~\cite{Tensor2019hackbusch}.
For example,  a tensor $A_{ijk}$ of size $n_1 \times n_2 \times n_3$ may be permuted and reshaped into a matrix $\matr Q$ by using the mapping $A_{ijk}\equiv Q_{Xj}$ where $X$ is a  composite index $X=i \cdot n_3 + k$. This is helpful for realizing orthonormality relationships, \eg, $\matr Q^\dagger \matr Q = \matr 1$, or more explicitly, $\sum_{X} Q^\ast_{Xj} Q_{X\tilde j} =\delta_{j\tilde j}$ and, equivalently, $\sum_{i k} A^\ast_{ijk} A_{i\tilde jk} = \delta_{j\tilde j}.$
\autoref{fig:orth_mat} gives an example of matricization and orthonormality relationships in diagrams.

\begin{figure}
\centering
\includegraphics{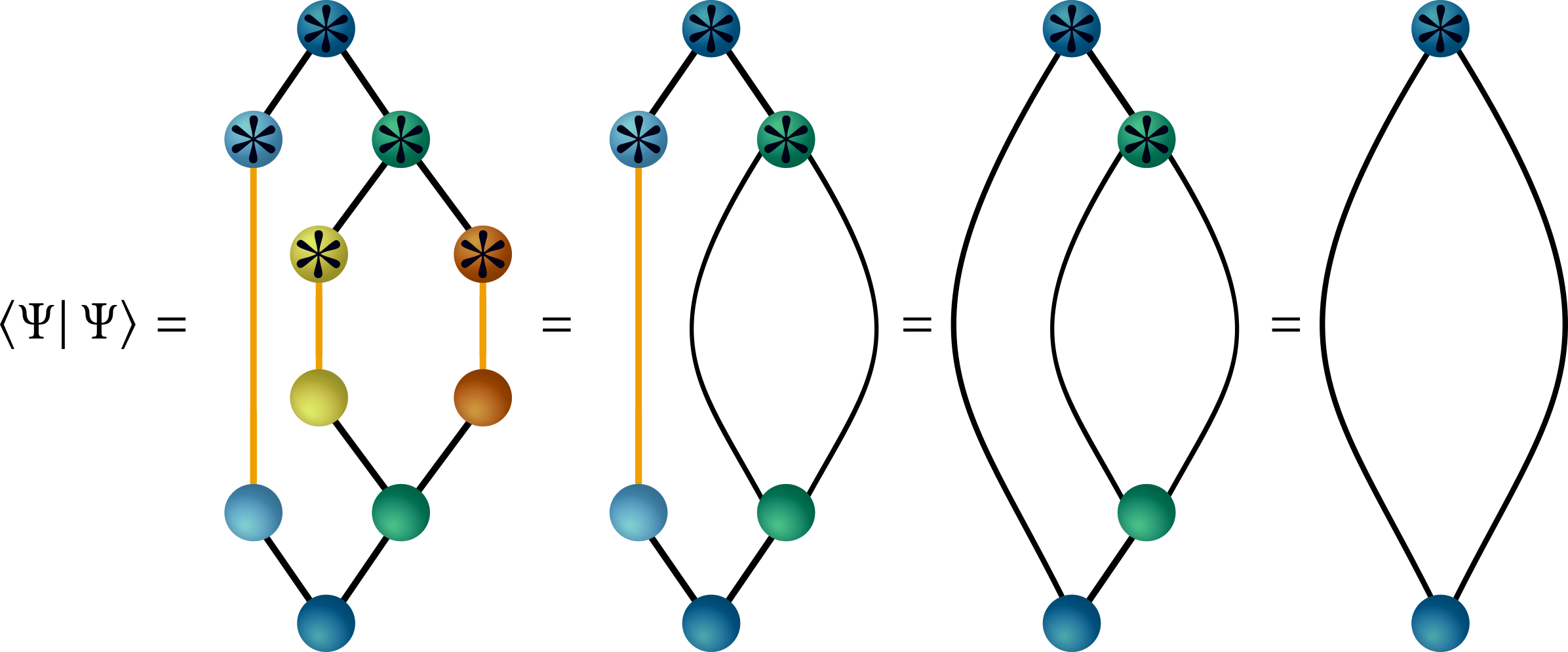}
\caption{Diagram of the overlap of a TTNS with itself (squared norm). The TTNS from \autoref{fig:mltree} is used as example.
The orthonormality relationships of the SPFs greatly simplify the diagram. Compare with \autoref{fig:orth_mat} (b).}
  \label{fig:ttns_bracket}
\end{figure}

In ML-MCTDH, the orthonormality of the SPFs means that each of the tensors $\matr A^{z'}$  used to construct the SPFs in a diagram can be matricized into orthonormal matrices $\matr Q^{z'}\equiv \matr Q^{l+1; \kappa_1,\kappa_2,\dots,\kappa_{l}}$ of size $\bar{n}_{\kappa_1,\kappa_2,\dots,\kappa_{l}} \times  n^{\kappa_1,\kappa_2,\dots,\kappa_{l}}
$, where $\bar{n}_{\kappa_1,\kappa_2,\dots,\kappa_{l}}$ is the total size of all the bonds pointing downward, as  defined in \autoref{eq:nbar}.
In this case, the matricization is performed by using the composite index $J$ (corresponding to the bonds pointing downward, toward the physical bonds) in $A^{z'}_{i;J}$ as row index and the SPF index (the bond pointing upward to the root node) $i$ as column index; thus $A^{z'}_{i;J} \equiv Q^{z'}_{J,i}$.
This leads to extremely useful simplifications of the diagrams. For example, the squared norm of an ML-MCTDH state, $\braket{\Psi}{\Psi}$, simply is given by that of the root node coefficient tensor, $\| \matr A^1 \|_2^2$. This is shown diagrammatically in  \autoref{fig:ttns_bracket}.

\begin{figure}
\centering
\includegraphics{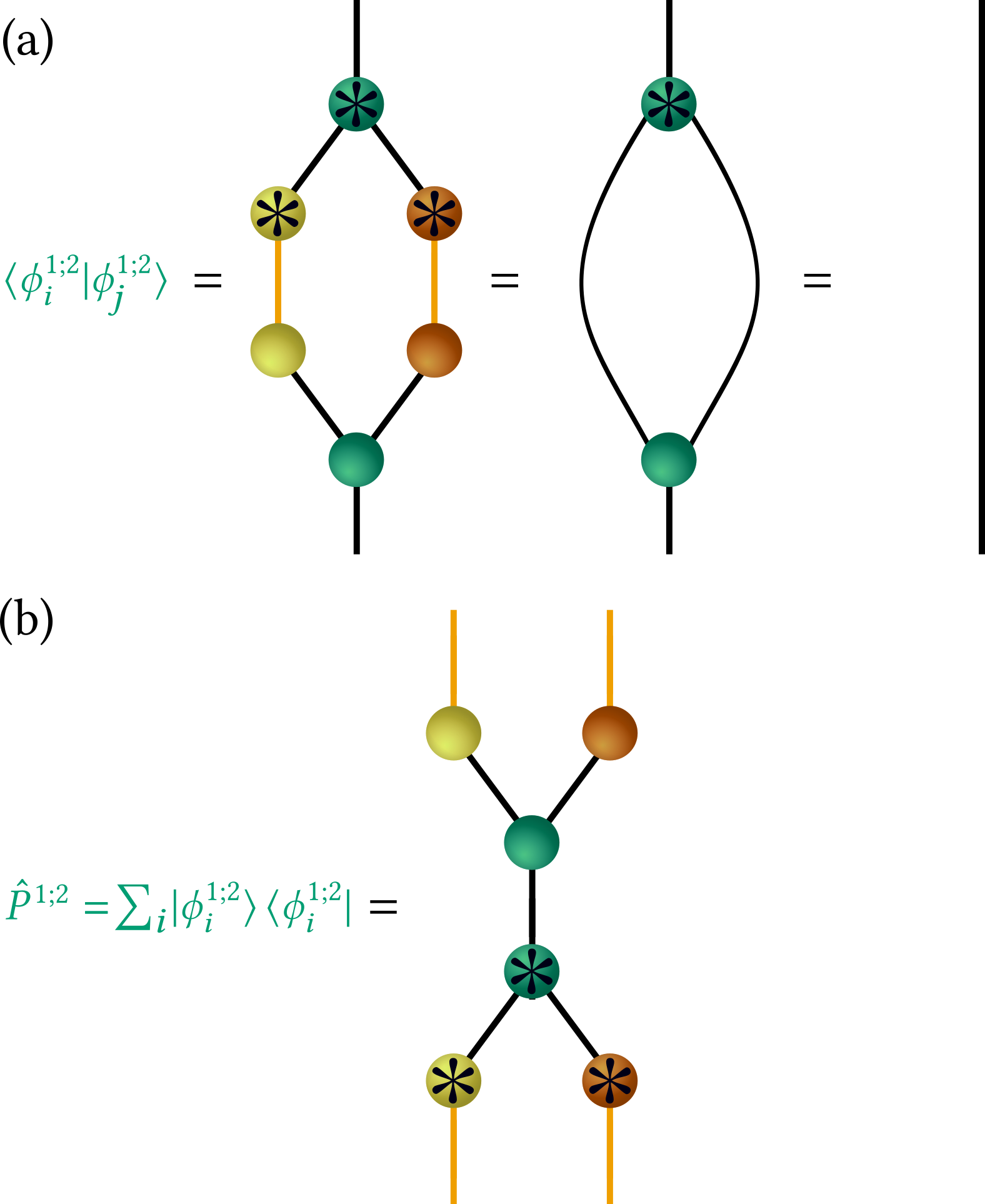}
\caption{SPF overlap (a) and projection (b). $\ket{\phi^{1;2}_i}$ from \autoref{fig:mltree} is used as example. For the overlap of orthonormal SPFs,  compare with \autoref{fig:orth_mat} (b). }
  \label{fig:spf_properties}
\end{figure}

The computation of the overlap, $\braket{\phi^z_{i}}{\phi^z_j}$, and the projector onto an SPF space,
\begin{equation}
\hat P^z = \sum_{i=1}^{n_{\kappa_1,\kappa_2,\dots,\kappa_{l}}} \ketbra{\phi_i^z}{\phi_i^z}
\label{eq:spf_projector}
\end{equation}
are shown in \autoref{fig:spf_properties}. Due to the orthonormality,   $\braket{\phi^z_{i}}{\phi^z_j}$ reduces to $\delta_{ij}$.
Note that even though formally the SPFs are orthonormal, in practice the finite accuracy of the  ML-MCTDH differential equation solver leads to small nonorthonormality errors of the SPFs. In MCTDH implementations, this error is neglected in the tensor contractions, but the SPF projector is modified to take the spurious nonorthonormalities into account via~\cite{Efficient1997beck}
\begin{equation}
\hat P^z_{\text{nonorth}} = \sum_{ij} \ket{\phi_i^z} [\matr S^{z}]^{-1}_{ij} \bra{\phi_j^z},
\end{equation}
where $\matr S^{z}$ is the overlap matrix of $\{\ket{\phi_i^z}\}$.

\begin{figure}
\centering
\includegraphics{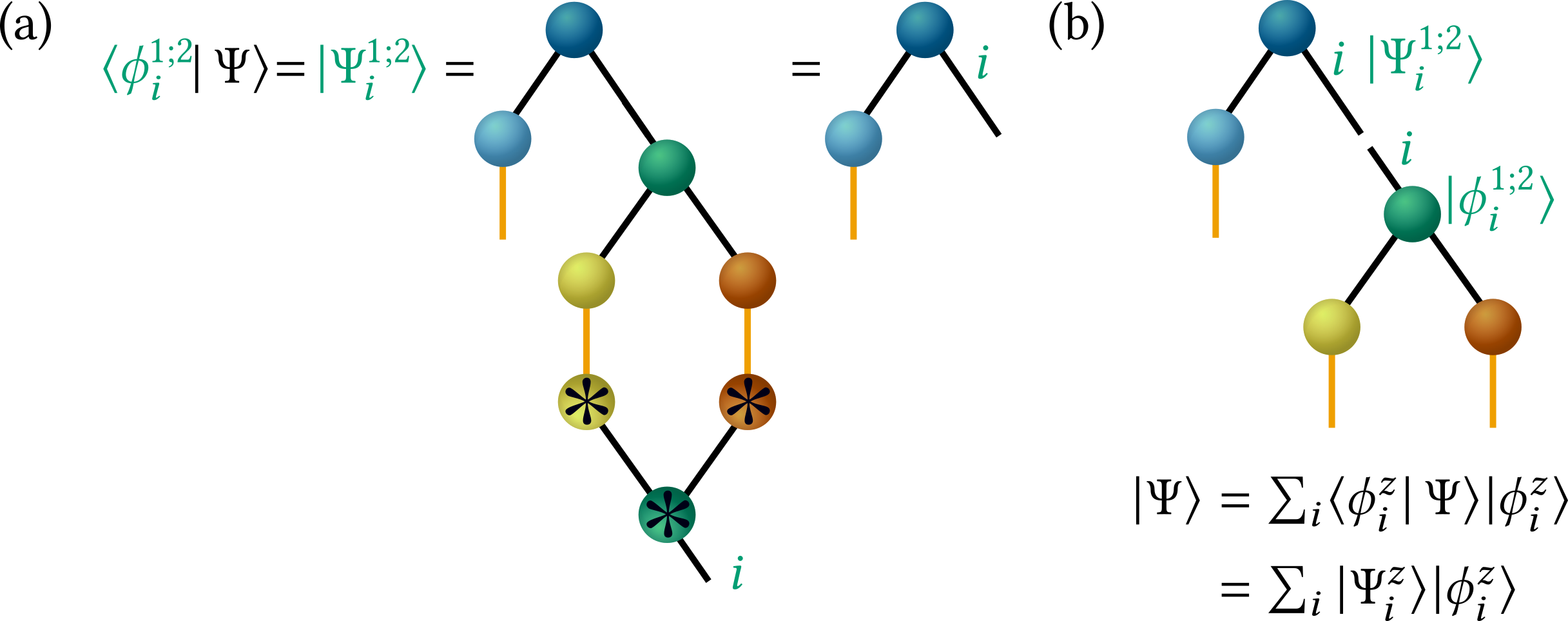}
\caption{Example of a single-hole function (SHF), its construction and usage.
(a) Construction of SHF $\ket{\Psi^{1;2}_i}$ from \autoref{fig:mltree}
through projection of $\ket{\phi^{1;2}_i}$ onto $\ket{\Psi}$.
Orthonormality relations simplify the SHF, compare with \autoref{fig:orth_mat} (b).
(b) Simplified $\ket{\Psi^{1;2}_i}$ together with its counterpart, the SPF $\ket{\phi^{1;2}_i}$. Contracted over $i$, they form the total TTNS $\ket{\Psi}$.
}
  \label{fig:shf_def}
\end{figure}

\begin{figure}
\centering
\includegraphics{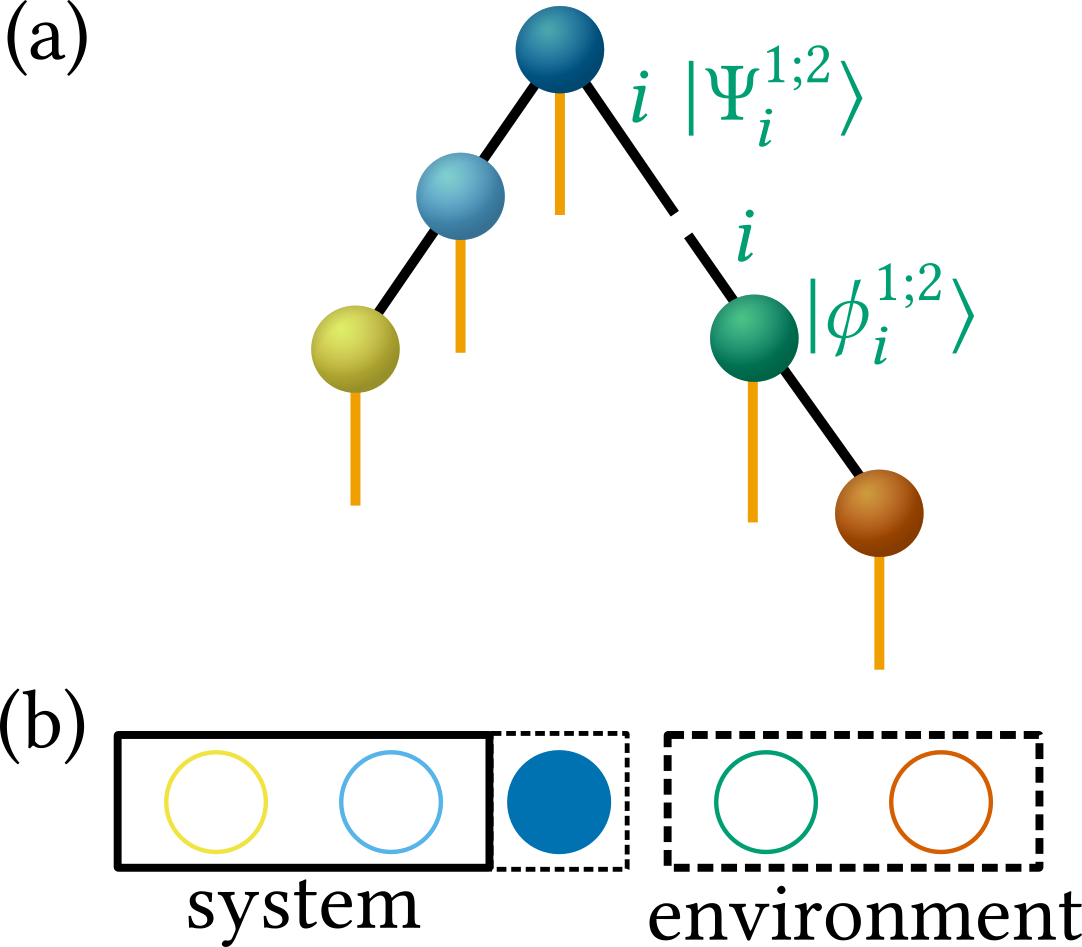}
\caption{Comparison of ML-MCTDH diagram and traditional DMRG language. 
(a) The decomposition of an ML-MCTDH state/MPS in terms of a single-hole function (SHF) and a single-particle function (SPF); compare with \autoref{fig:shf_def}. 
(b) Single-site DMRG setting with a system and an environment block. 
The system (environment) block is spanned by the SHF (SPF) space;
compare with \autoref{eq:dmrg_state_expansion}.
}
  \label{fig:dmrg_state_expansion}
\end{figure}

An example of an SHF $\ket{\Psi^{z}_i}$ is shown in \autoref{fig:shf_def}.
In contrast to the SPF, for an SHF the function index $i$ corresponds to a downward (and not upward) pointing virtual bond and thus an SHF is nonorthonormal. They can be made orthogonal, however;  see \autoref{sec:rdms_mean_fields}.
The relationship between the SHF-SPF decomposition in ML-MCTDH and the system-environment decomposition in traditional DMRG, \autoref{eq:dmrg_state_expansion}, is shown in \autoref{fig:dmrg_state_expansion} (in this case we make an exception and show an MPS instead of the TTNS from \autoref{fig:mltree}).

\begin{figure}
\centering
\includegraphics{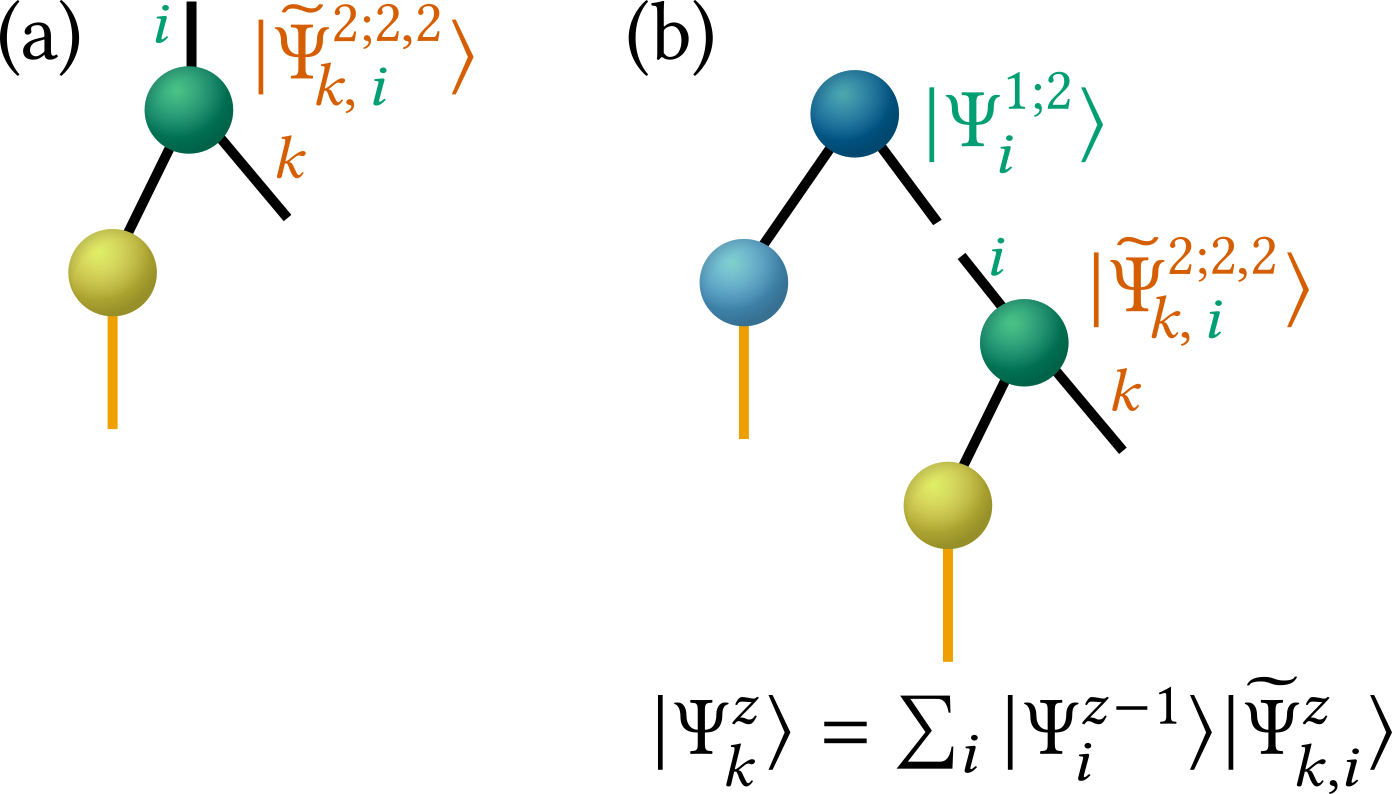}
\caption{Example of a doubly-indexed single-hole function (SHF) and its usage.
(a) $\ket{\widetilde \Psi^{2;2,2}_{k,i}}$ from \autoref{fig:mltree} as example of a doubly-indexed SHF. 
(b) $\ket{\Psi^{1;2}_i}$ (see \autoref{fig:shf_def}) together with $\ket{\widetilde \Psi^{2;2,2}_{k,i}}$. Contracted over $i$, they form the SHF $\ket{\Psi^{2; 2,2}_k}$.
}
  \label{fig:dshf_def}
\end{figure}

The SHF definition can be generalized to doubly-indexed SHFs defined as~\cite{Multilayer2008manthe} 
\begin{equation}
\begin{split}
  \ket{\widetilde\Psi^{l; \kappa_1, \kappa_2,\dots \kappa_l}_{\textcolor{idxocre}{k},\textcolor{idxgreen}{i}}} =&  \sum_{j_1} \sum_{j_2} \dots \sum_{j_{\kappa_l-1}} \sum_{j_{\kappa_l+1}} \dots \sum_{j_{d_{\kappa_1,\kappa_2,\dots,\kappa_{l-1}}}} A^{l;\kappa_1,\kappa_2,\dots,\kappa_{l-1}}_{\textcolor{idxgreen}i;j_1,j_2,\dots,j_{\kappa_l-1},\textcolor{idxocre}k, j_{\kappa_l+1},\dots,j_{{d_{\kappa_1,\kappa_2,\dots,\kappa_{l-1}}}}}\\ %
  &   \bigotimes_{\tau = 1}^{\kappa_{l-1}}
  \ket{\phi_{j_\tau}^{l;\kappa_1,\kappa_2,\dots,\kappa_{l-1}, \tau}} %
     \bigotimes_{\gamma = \kappa_{l+1}}^{d_{\kappa_1,\kappa_2,\dots,\kappa_{l-1}}}   \ket{\phi_{j_\gamma}^{l;\kappa_1,\kappa_2,\dots,\kappa_{l-1}, \gamma}},
\label{eq:shf_doubly_indexed}
\end{split}
\end{equation}
where we have highlighted the two ``hole indices'' $i$ and $k$ by colors. 
An example is shown in \autoref{fig:dshf_def}.
Doubly-indexed SHFs allow for a recursive definition of the SHFs: 
\begin{equation}
  \ket{\Psi^z_k} = \sum_i \ket{\Psi^{z-1}_i} \ket{\widetilde \Psi^z_{k,i}}.
\end{equation}
In a diagram, the doubly-indexed SHF simply is represented by two virtual bonds, one pointing downward and one pointing upward; see  \autoref{fig:dshf_def} (b).

\section{ML-MCTDH equations and connection to diagrams}
\label{sec:ml_mctdh_diagrams}
The ML-MCTDH equations are based on the ML-MCTDH ansatz, \autoref{eq:mlmctdh}, %
and the TDVP, %
\begin{equation}
 \Matrixe{\delta \Psi}{\hat H - \ii \hbar \partd{}{t}}{\Psi}=0,\label{eq:dirac_principle}
\end{equation}
where $\delta \Psi$ denotes first-order variations of $\Psi$ with respect to its parameters. %
In the following, we will set $\hbar=1$.
For the ML-MCTDH wavefunction ansatz, \autoref{eq:dirac_principle} is solved subject to two constraints:
Firstly, the orthonormality conditions of the SPFs, and secondly, their gauge invariance, 
\begin{equation}
  \braket{\phi^z_i}{\dot\phi^z_j} = -\ii  \matrixe{\phi^z_i}{\hat g^{z}}{\phi^z_j},\label{eq:gauge_invariance}
\end{equation}
where $\hat g^{z}$ is some arbitrary, hermitian constraint operator and $\ket{\dot \phi_j^z}$  denotes the time derivative of $\ket{\phi_j^z}$.
The gauge invariance is discussed more in \autoref{sec:canonicalization}. In short, unitary transformations of the SPFs/SHFs alter neither orthonormality nor the state. This leads to an arbitrariness of the SPFs that is manifested by the gauge invariance. This arbitrariness is  fixed by \autoref{eq:gauge_invariance}.  
Various choices of $\hat g^z$ exist~\cite{Comment1994manthe,Multiconfiguration2000beck,Symmetries2021weike}
and here, for simplicity, we will use one of the most common choices and simply set it to zero~\cite{Multiconfigurational1990meyer}, \ie, 
\begin{equation}
  \braket{\phi^z_i}{\dot\phi^z_j} \stackrel!= 0.\label{eq:gauge_invariance_zero}
\end{equation}
Generalizing the following equations to non-zero $\hat g^z$ is straightforward and some aspects of using a different gauge will be discussed in \autoref{sec:tddmrg}. 

\subsection{Density matrices and mean fields}
\label{sec:rdms_mean_fields}

\begin{figure}
\centering
\includegraphics{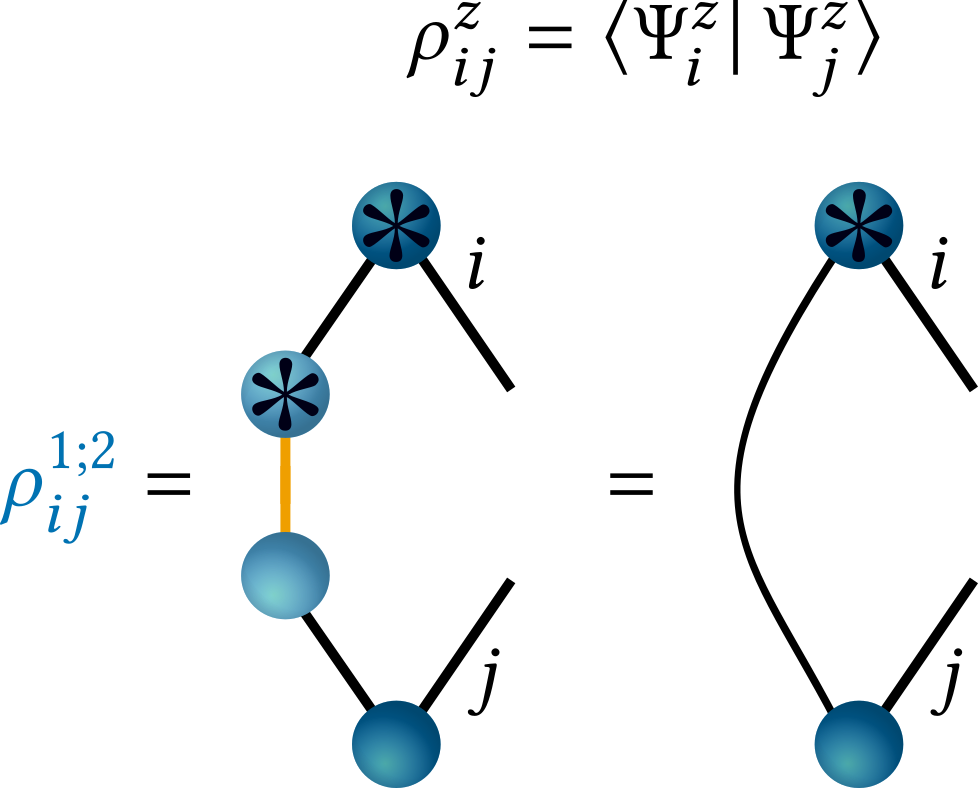}
\caption{Diagram of a single-particle density matrix for the TTNS from \autoref{fig:mltree}. Compare with \autoref{eq:rdm_def}.}
  \label{fig:rdm_def}
\end{figure}

Before we proceed with defining the equations of motions, we will define two additional quantities and  provide diagrammatic expressions for them.
The ``single-particle'' density matrix %
is defined as
\begin{equation}
  \rho^z_{ij} = \braket{\Psi^z_i}{\Psi^z_j},
  \label{eq:rdm_def}
\end{equation}
and it is equal to the transpose of the one-particle reduced-density matrix in electronic structure theory~\cite{Molecular2013helgaker}.
In addition, it is related to the Gram matrix $\matr G$  with entries $G_{ij}  = \braket{\partial_{\lambda_i} \Psi}{\partial_{\lambda_j} \Psi}$
of $\ket{\Psi}$ with parameters $\lambda_i$, which is
used in similar theories~\cite{Tensor2019hackbusch,Matrix1989kay,TimeDependent2011haegeman,Linear2014nakatani,Analytic2014kinder,Tangentspace2019vanderstraeten}.
SPF orthonormality simplifies the computation of $\rho^z$ and an example is shown in \autoref{fig:rdm_def}.
Diagonalizing it yields natural SPFs and SHFs. 
The eigenvalues of $\rho^z$ are called natural occupations.
The SHFs with nonvanishing natural occupations are then orthogonal, but not orthonormal. 

\begin{figure}
\centering
\includegraphics{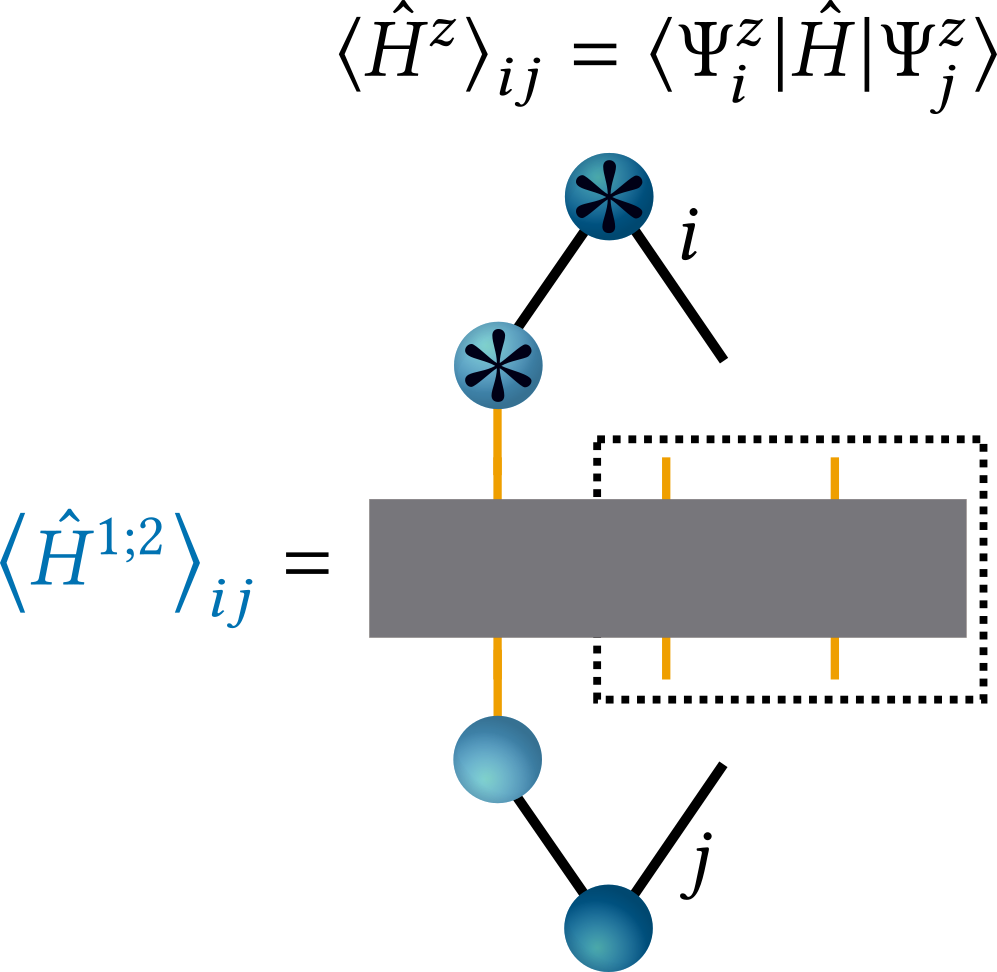}
\caption{Diagram of a mean-field matrix for the TTNS from \autoref{fig:mltree}. 
The dashed rectangle corresponds to the ``operator part'' of the mean-field matrix and the indices specify the ``matrix part.''
Compare with \autoref{eq:mf_def}.}
  \label{fig:mfdef}
\end{figure}

The mean-field operators are defined as 
\begin{equation}
  \erw{\hat H^z}_{ij} = \matrixe{\Psi^z_i}{\hat H}{\Psi^z_j} = \braket{\Psi}{\phi_i^z}\hat H \braket{\phi_j^z}{\Psi},\label{eq:mf_def}
\end{equation}
where we have inserted the definitions for the SHFs in the second equation to highlight that \autoref{eq:mf_def} still defines an operator and not a scalar.
Hence, $\erw{\hat H^z}_{ij}$ is a matrix of operators.
An example diagram that highlights the ``operator part'' and the ``matrix part'' (indices $i$ and $j$) of \autoref{eq:mf_def} is depicted in \autoref{fig:mfdef}.

\subsection{Equations of motions}

Using the definitions of the mean-field matrices and density matrices, we now give concise expressions of the ML-MCTDH equations of motions.
\autoref{eq:dirac_principle} together with \autoref{eq:gauge_invariance_zero} leads to the following equations of motions for the SPFs~\cite{Wave1992manthe}:%
\begin{equation}
  \ii \ket{\dot \phi_i^{z}} = (1-\hat P^z) \sum_{j,k} 
   (\rho^{z})^{-1}_{ij} \erw{\hat H}^{z}_{jk} \ket{\phi_k^z}.
   \label{eq:spf_eom}
\end{equation}
Note that in practice the single-particle density matrix  $\matrgreek \rho^z$ can be singular and typically needs to be regularized~\cite{Wave1992manthe}, but alternatives exist~\cite{Multiconfigurational1990meyer,Multiconfigurational2015manthe,Regularizing2018wang}; see also \autoref{sec:opt_unocc_spf}.
We will make \autoref{eq:spf_eom} more explicit by projecting the corresponding configurations onto  \autoref{eq:spf_eom} so that the equations of motions for the SPF coefficient tensor are obtained as 
\begin{equation}
  \ii \dot{A}_{i; J}^{z'+1} = \matrixe{\Phi^{z'+1}_J}{(1 - \hat P^{z})\sum_{jk} (\rho^{z})^{-1}_{ij} \erw{\hat H}^{z}_{jk}}{\phi_k^z}.
  \label{eq:spftensor_eom}
\end{equation}
For the root tensor in the first layer %
we obtain standard  Ritz-Galerkin-like~\cite{Numerical1987gottlieb,Tensor2015szalay} 
equations of motions for the TDSE expressed in a finite (but time-dependent) basis:
\begin{equation}
  \ii \dot{A}_{I}^{1} =\sum_J \matrixe{\Phi^{1}_I}{\hat H}{\Phi^1_J} A_J^1.
  \label{eq:roottensor_eom}
\end{equation}

\subsection{Some common forms of the Hamiltonian}

The most common form of $\hat H$ and other operators used by MCTDH and many other methods %
is a sum-of-product form (SoP), where the Hamiltonian is decomposed as a sum of $S$ direct products of one-dimensional operators:
\begin{equation}
\hat H_\text{SoP} = \sum_{s=1}^{S} c_s \bigotimes_{\kappa=1}^F \hat{\tilde h}_s^{(\kappa)} = \sum_{s=1}^{S} \bigotimes_{\kappa=1}^F \hat h_s^{(\kappa)}.\label{eq:sop}
\end{equation}
In the second part of \autoref{eq:sop} we have absorbed the complex-valued coefficients $c_s$ into one of the operators $\hat{\tilde h}^{(\kappa)}_s$, for clarity.
This Hamiltonian has the advantage that its application on a state has a relatively simple but favorably-scaling implementation~\cite{New1990manthe,General1993bramley}  based on one single loop over well-optimized matrix-matrix multiplications~\cite{Efficient2016larsson}.

The counterpart to the SoP Hamiltonian in the DMRG community is the matrix product operator (MPO)~\cite{Matrix2010pirvu,Densitymatrix2011schollwock,Tensor2015szalay,Matrix2016chan}, which is defined as
\begin{equation}
  \hat H_\text{MPO} = \sum_{\beta_1}\sum_{\beta_2}\dots\sum_{\beta_{F-1}}%
  \hat W_{\beta_1}^1 \otimes \bigotimes_{\kappa=2}^{F-1} \hat W_{\beta_{\kappa-1} \beta_{\kappa}}^{\kappa}  \otimes    \hat W_{\beta_{F-1}}^{F}.\label{eq:mpo}
\end{equation}
An MPO is a generalization of an MPS to an operator. Likewise, a tree tensor network operator can be defined as generalization of a TTNS to an operator~\cite{Tensor2015szalay}. In the MCTDH community, for diagonal potential operators this is known as multilayer potfit format~\cite{Multilayer2014otto}, whose advantages and disadvantages so far have not been fully explored~\cite{Accuracy2018otto}, however.

\begin{figure}
\centering
\includegraphics{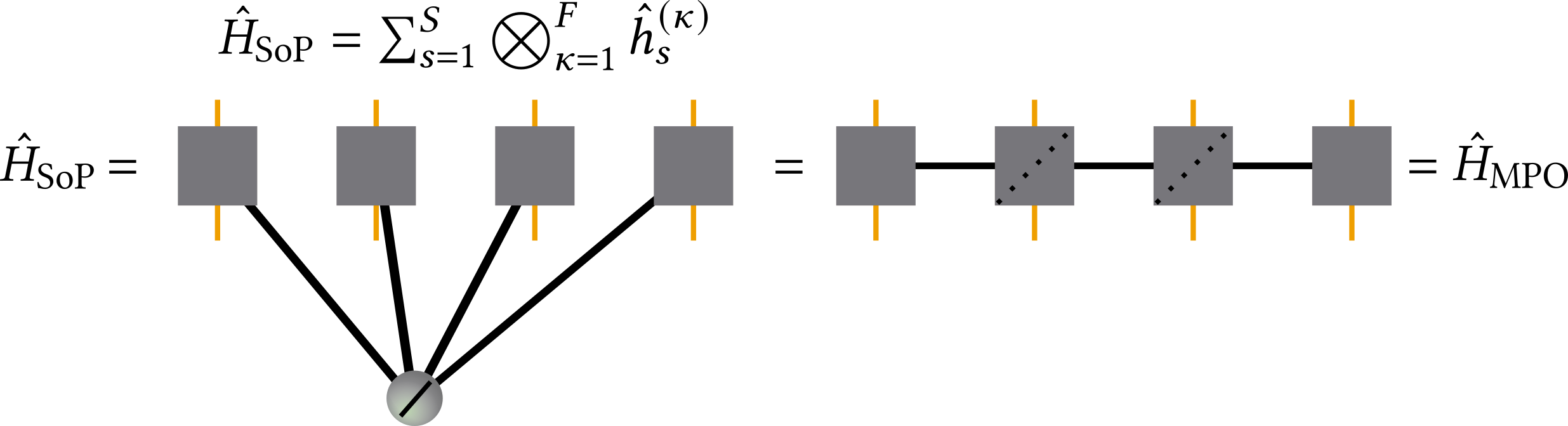}
\caption{Diagram of a sum-of-product (SoP) operator and mapping to a particular matrix-product operator (MPO) with diagonal tensors. 
Here, the dotted, diagonal line in the MPO tensors indicate tensors diagonal only with respect to the virtual bond dimension (black lines).
The orange lines denote the physical space.
Compare with Eqns.~\ref{eq:sop}-\ref{eq:sop_mpo_mapping}.
}
  \label{fig:sop}
\end{figure}

The SoP Hamiltonian can be represented as a  MPO that is diagonal in bond dimension~\cite{SpinProjected2017li,Minimal2020larsson}: %
Defining $s$ as raveled compound index $\beta_1,\beta_2,\dots,\beta_{F-1}$, \autoref{eq:sop} can be mapped onto \autoref{eq:mpo} by the relation
\begin{equation}
\hat W_{\beta_{\kappa-1} \beta_{\kappa}}^{\kappa\text{, SoP}} = \delta_{\beta_{\kappa-1} \beta_{\kappa}} \hat h^{(\kappa)}_s.\label{eq:sop_mpo_mapping}
\end{equation}
The SoP Hamiltonian and its mapping to an MPO is shown diagrammatically in \autoref{fig:sop}.

\begin{figure}
\centering
\includegraphics{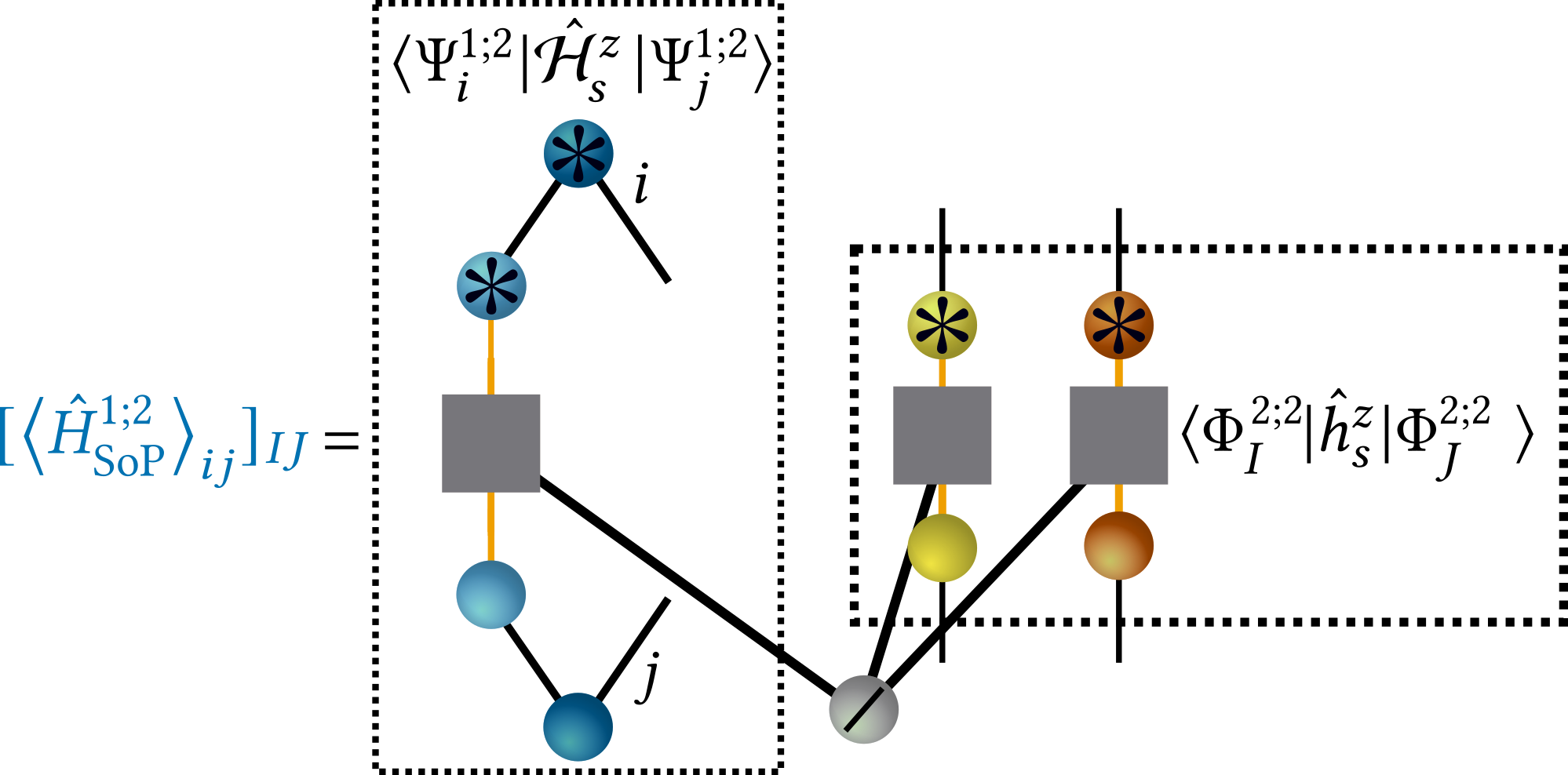}
\caption{Diagram of a matrix representation of a mean-field sum-of-product operator for the TTNS from \autoref{fig:mltree}.
The partitioning of this diagram  into one single-hole function term (mean field ``matrix part'') and one configuration term  (mean field ``operator part'')
is visualized by dotted rectangles.
Compare with \autoref{eq:sop_speciality_mf} and \autoref{fig:mfdef}.}
  \label{fig:sop_specialities}
\end{figure}

For the ML-MCTDH equations, \autoref{eq:spftensor_eom}, %
it is useful to rewrite $\hat H$ %
in terms of an operator 
$\hat h^{z}$ that acts solely on the same space as that spanned by the configurations $\{\ket{\Phi^{z'+1}_J}\}_J$, and in terms of an operator $\mathcal{\hat H}^z$ that acts on the remaining space~\cite{Multilayer2008manthe,Multilayer2011vendrell}:
\begin{equation}
\hat H = \sum_{s=1} \mathcal{\hat H}^z_s \hat h^{z}_s =  \sum_{s=1} \mathcal{\hat H}^z_s  \prod_{\kappa_{l+1}=1}^{d_{\kappa_1, \kappa_2,\dots,\kappa_{l}}} \hat h_s^{l+1; \kappa_1, \kappa_2,\dots,\kappa_{l+1}}.\label{eq:H_partition}
\end{equation}
$\hat h_s^z =  \hat h_s^{l; \kappa_1, \kappa_2,\dots,\kappa_{l}}$ is defined recursively in the same way as the SPFs,
making this 
partitioning similar to the SHF-SPF expansion of  a TTNS, \autoref{eq:shf_expansion}.
A virtually identical partitioning exists for the DMRG and 
this partitioning is trivially fulfilled both for SoPs and MPOs.
Then
the ``matrix'' representation of the mean-field operator, \autoref{eq:mf_def}, which is a given by a four-dimensional tensor, can be represented as
\begin{equation}
  [\erw{\hat H^z}_{ij}]_{IJ} = \sum_{s=1} \matrixe{\Phi_I^{z'+1}}{\hat h^{z}_s}{\Phi_J^{z'+1}} \matrixe{\Psi_i^z}{\mathcal{\hat H}^z_{s}}{\Psi_j^z}.
  \label{eq:sop_speciality_mf}
\end{equation}
An example diagram is shown in \autoref{fig:sop_specialities}.
Note that both the mean-field operators and the density matrices can be evaluated efficiently in a recursive way starting at the first layer for the density part and starting at the last layer for the operator part~\cite{Multilayer2008manthe}. This is related to the site-per-site contraction of overlaps in the  DMRG~\cite{Densitymatrix2011schollwock,Practical2014orus}.

\section{Derivation of ML-MCTDH using diagrams}
\label{sec:mctdh_deriv}
Here we will use tensor network diagrams to derive the ML-MCTDH equations of motion \ref{eq:spftensor_eom} and \ref{eq:roottensor_eom}.
For solving the TDVP, %
we need the first-order variations $\delta \Psi$. They are related to the derivatives of the TTNS with respect to all of its individual tensors components, which we specify here as $\lambda_x$:
\begin{equation}
  \delta \Psi =\sum_{x} \partd{\Psi}{\lambda_x} \delta \lambda_x.
\end{equation}
Inserting $\delta \Psi$ into the TDVP, %
\autoref{eq:dirac_principle}, leads to as many terms as there are parameters $\lambda_x$ %
in the TTNS for $\delta \Psi$, and we get\footnote{
    Note that we treat  $\lambda_x^\ast$ as independent of $\lambda_x$.}
\begin{equation}
\sum_{x} \delta \lambda_x^\ast \Matrixe{\partd{\Psi}{\lambda_x}}{\hat H - \ii \partd{}{t}}{\Psi} = 0.
\label{eq:var_principle_mat}
\end{equation}
Since the variations $\delta \lambda_x$ are arbitrary and independent of each other,
each of the terms in \autoref{eq:var_principle_mat} must vanish independently of each other and independently of the variations~\cite{Wave1934frenkel}.
Hence, we get the following set of equations
\begin{align}
 \Matrixe{\partd{\Psi}{\lambda_x}}{\hat H - \ii \partd{}{t}}{\Psi} &= 0  \quad \text{for all } x,\\
 \Leftrightarrow \ii \Braket{\partd{\Psi}{\lambda_x}}{\dot \Psi} &= \Matrixe{\partd{\Psi}{\lambda_x}}{\hat H}{\Psi}.\label{eq:var_principle_lambda}
\end{align}
Here, it will be useful to use the tensor structure to rewrite \autoref{eq:var_principle_lambda} in terms of the individual tensors of the TTNS,
\begin{equation}
\ii \Braket{\partd{\Psi}{\matr A^{z'}}}{\dot \Psi} = \Matrixe{\partd{\Psi}{\matr A^{z'}}}{\hat H}{\Psi}  \quad \text{for all possible } z'.
\label{eq:dirac_principle_tensors}
\end{equation}
Together with the gauge conditions  \autoref{eq:gauge_invariance_zero},
each of these equations will lead to the equations of motions for a particular tensor.

\begin{figure}
\centering
\includegraphics{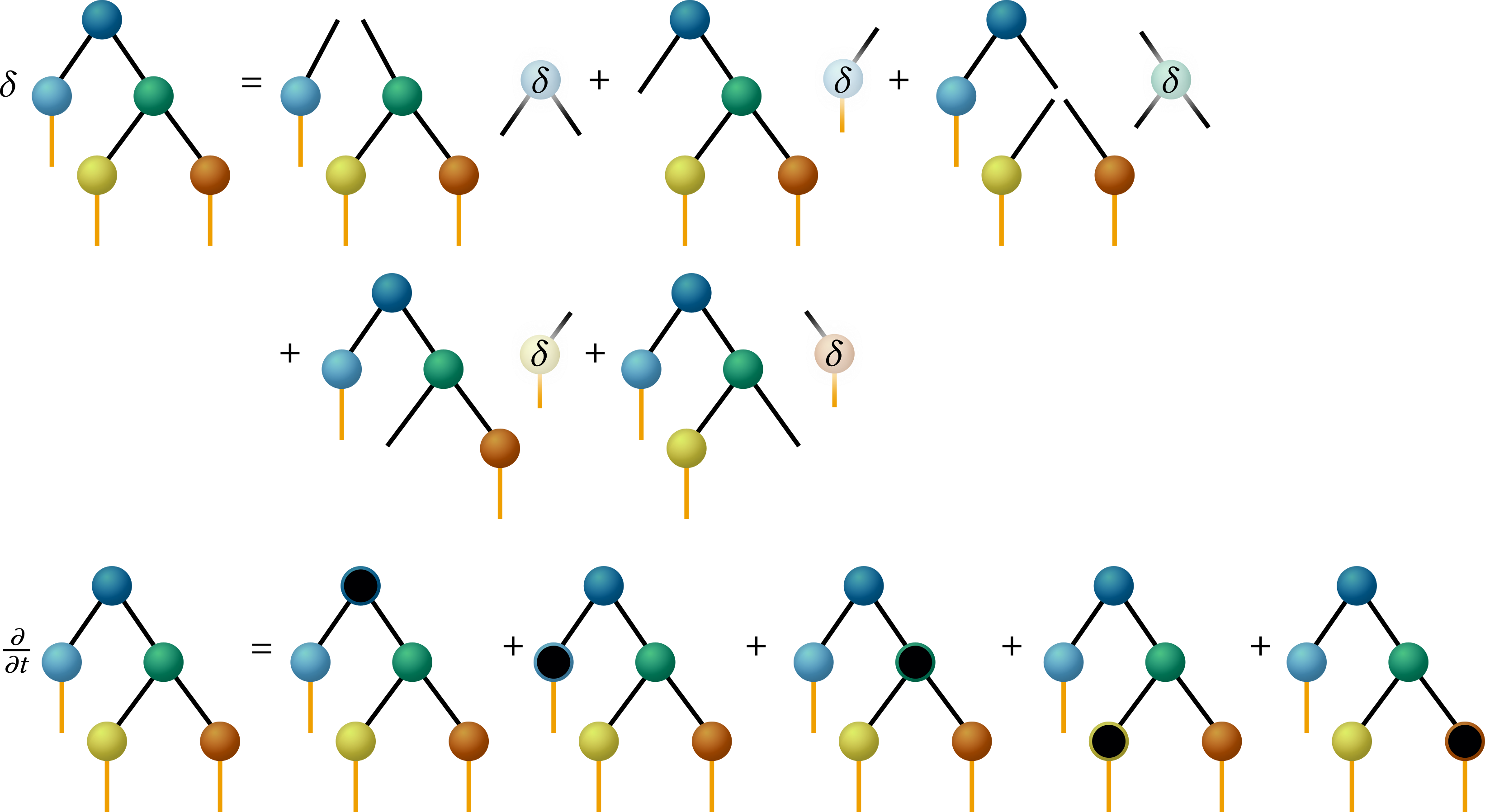}
\caption{Derivatives of a TTNS.
First-order variation of $\Psi$, which contain derivatives with respect to the individual tensors $\matr A^{z'}$ (upper panel), and time derivative (lower panel) 
for the TTNS from \autoref{fig:mltree}.
Tensors denoting variations are marked with a $\delta$.
}
  \label{fig:ml_deriv_derivs}
\end{figure}

For the derivations of the equations of motions for each tensor,
we again use the TTNS from \autoref{fig:mltree} as a simple example but note that this can be easily generalized to any TTNS, including TTNSs with more layers.
The first-order variation and time-derivative of $\Psi$ are shown in \autoref{fig:ml_deriv_derivs}.

\begin{figure}
\centering
\includegraphics{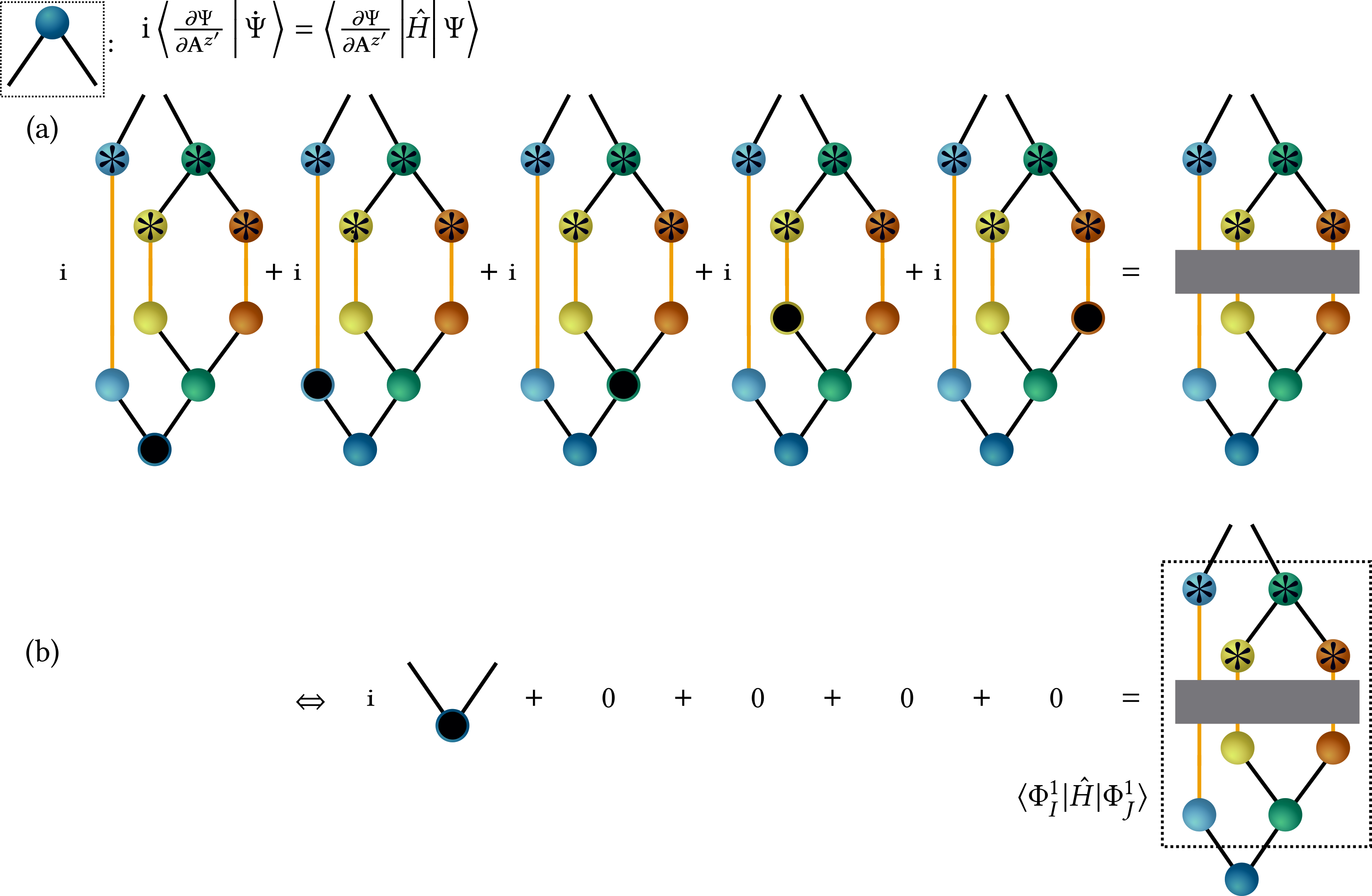}
\caption{Derivations of the equations of motions for the root tensor $\matr A^1$ (shown in the upper left rectangle) for the TTNS from \autoref{fig:mltree}. 
The Hamiltonian is denoted by the gray rectangle. 
The individual terms come from \autoref{fig:ml_deriv_derivs}.
After making use of the  SPF orthogonality relationships, \autoref{fig:spf_properties}, and the SPF gauge, \autoref{eq:gauge_invariance_zero}, one arrives at panel (b), which corresponds to  \autoref{eq:roottensor_eom}.
The dotted rectangle in the lower right of panel (b) 
highlights $\matrixe{\Phi^{1}_I}{\hat H}{\Phi^1_J}$.
}
  \label{fig:ml_deriv_root}
\end{figure}

We start the derivation with the first term of $\delta \Psi$ in \autoref{fig:ml_deriv_derivs}, which is for the root node, $\matr A^{1}$. 
Inserting this into \autoref{eq:dirac_principle_tensors} leads to the expression in \autoref{fig:ml_deriv_root}(a). Because we demand that the projection of an SPF onto its time-derivative is zero, \autoref{eq:gauge_invariance_zero}, 
most of the terms arising from $\dot \Psi$ vanish, 
and, using the SPF orthogonality relationships, \autoref{fig:spf_properties},
straightforwardly we obtain \autoref{eq:roottensor_eom}, which is shown diagrammatically in \autoref{fig:ml_deriv_root}(b).

\begin{figure}
\centering
\includegraphics{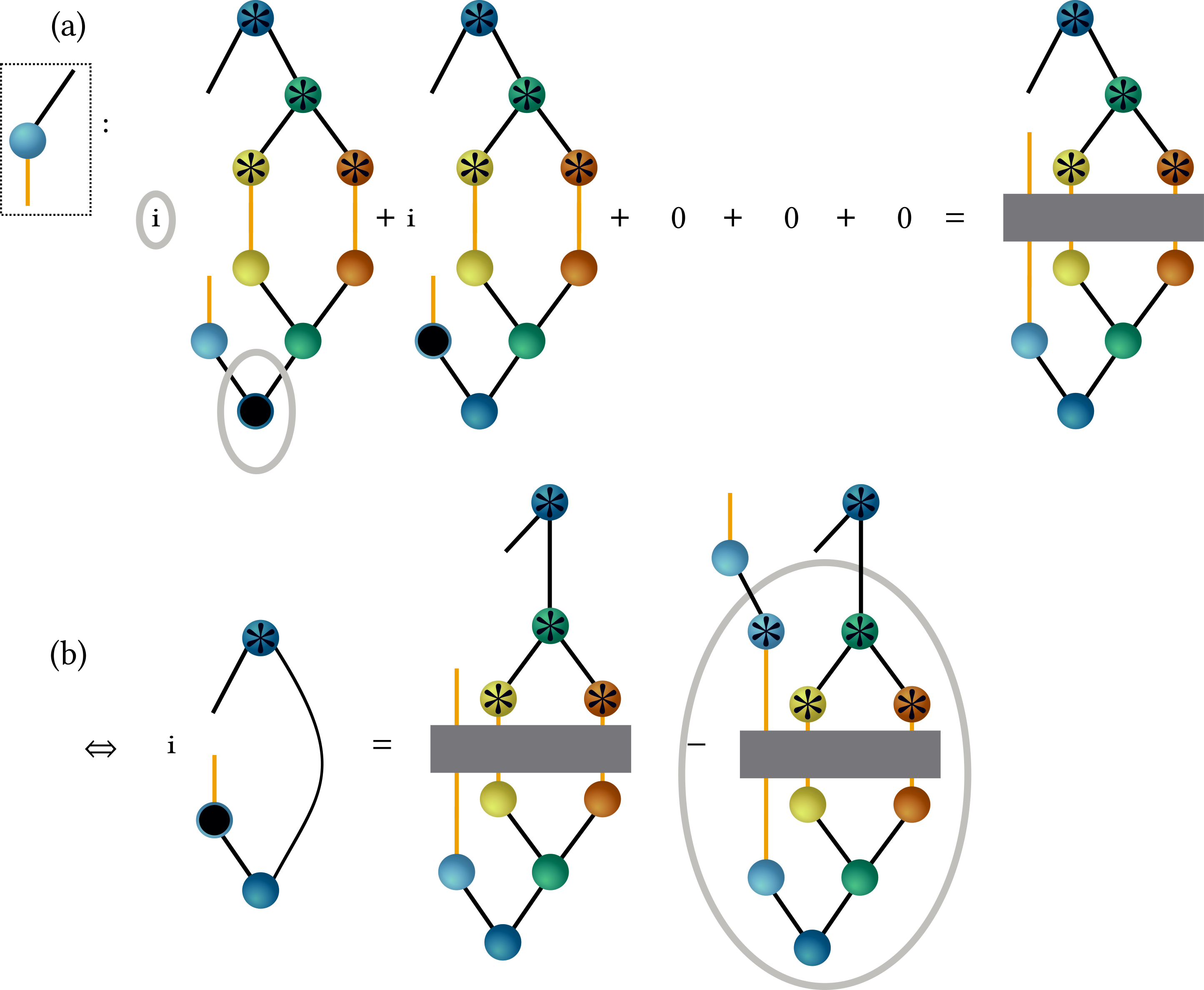}
\caption{Derivations of the equations of motions for ${\matr A}^{2;1}$ (shown in the upper left rectangle) from the TTNS shown in  \autoref{fig:mltree}. 
Terms that are trivially zero due to the gauge conditions have been set to zero immediately (compare with \autoref{fig:ml_deriv_root}). 
The  SPF orthogonality relationships, \autoref{fig:spf_properties}, 
and the expression from \autoref{fig:ml_deriv_root} have  been used in panel (b) to simplify the expression in (a). This is marked by gray ellipses.
}
  \label{fig:ml_deriv_mctdh}
\end{figure}

Next we turn to the equations for $\dot{\matr A}^{2;1}$ from the TTNS in \autoref{fig:mltree}. The derivation is shown in \autoref{fig:ml_deriv_mctdh}. Inserting the just-derived \autoref{eq:roottensor_eom}, we obtain a specific form of \autoref{eq:spftensor_eom} for this node.
Compared to the more general \autoref{eq:spftensor_eom}, there are two variations: Firstly, since a next layer $l+1$ does not exist for $\matr A^{2;1}$, here the configurations $\ket{\Phi^{z+1}_I}$ from \autoref{eq:roottensor_eom} just correspond to the primitive basis representation. Secondly, in \autoref{fig:ml_deriv_mctdh} we moved $\matrgreek \rho^z$ to the left-hand-side in \autoref{eq:spftensor_eom}, which is closer to the original MCTDH derivation and hints at other ways to solve for this equation, namely by using the pseudo-inverse of $\matr A^{2;1}$~\cite{Multiconfigurational1990meyer,Regularizing2018wang} (see also Refs.~\cite{Time1988meyer,Matrix1989kay} and  \autoref{sec:spf_regularization}).
Note that the just-derived equations for $\matr A^{1}$ and $\matr A^{2;1}$ are sufficient to fully program the plain MCTDH equations of motions without the multilayer form.

\begin{figure}
\centering
\includegraphics{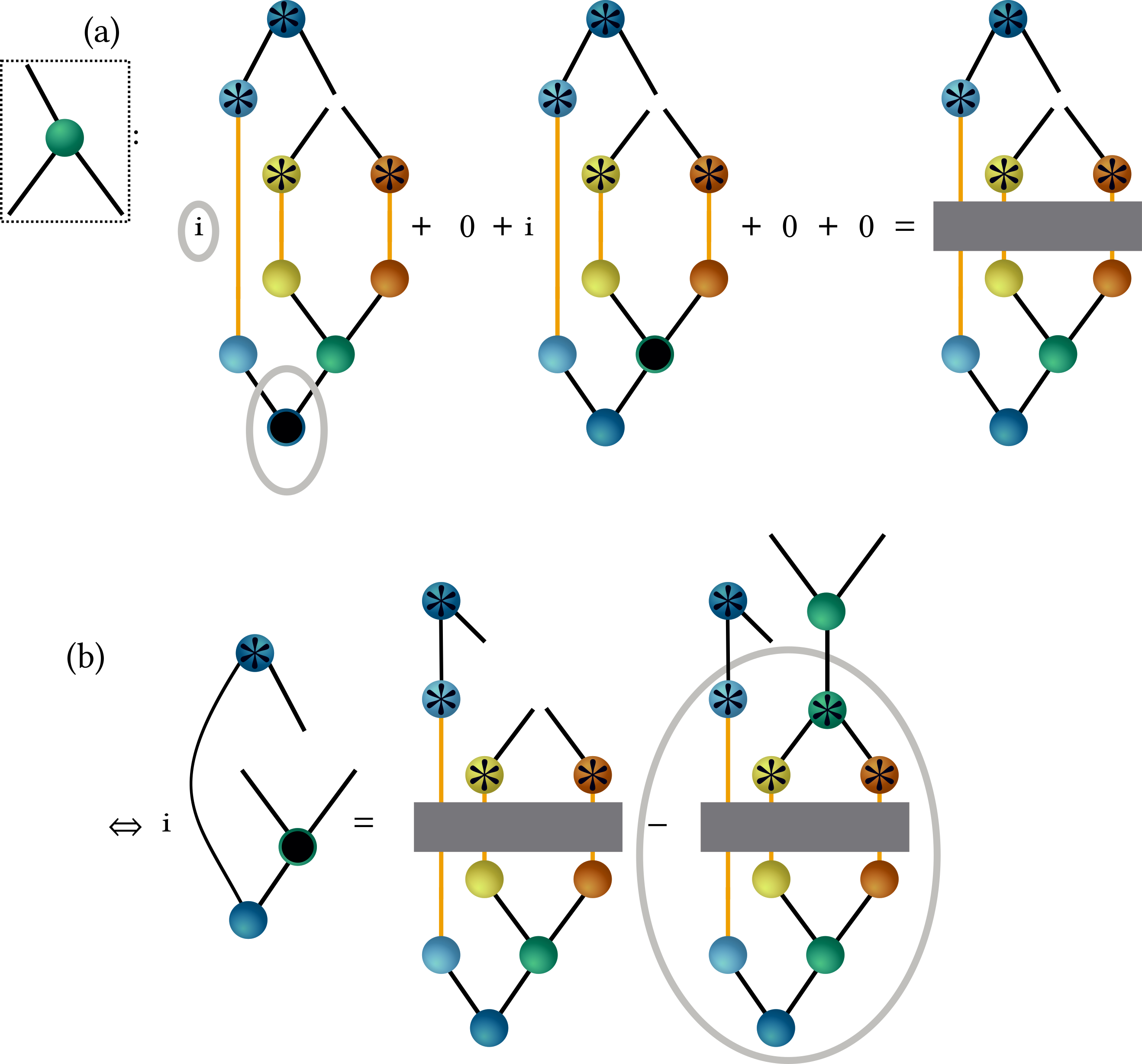}
\caption{Same as \autoref{fig:ml_deriv_mctdh} but for  ${\matr A}^{2;2}$.}
  \label{fig:ml_deriv_mc2}
\end{figure}

The derivations for the equations for   $\dot{\matr A}^{2;2}$ is very similar to that of $\dot{\matr A}^{2;1}$ and shown in 
 \autoref{fig:ml_deriv_mc2}. The main difference is that here we actually have configurations $\ket{\Phi^{z+1}_I}$ appearing in the diagrams.  This is because  ${\matr A}^{2;2}$ is not connected to any primitive basis, which only happens in ML-MCTDH and not in plain MCTDH for non-root-tensors.

\begin{figure}
\centering
\includegraphics{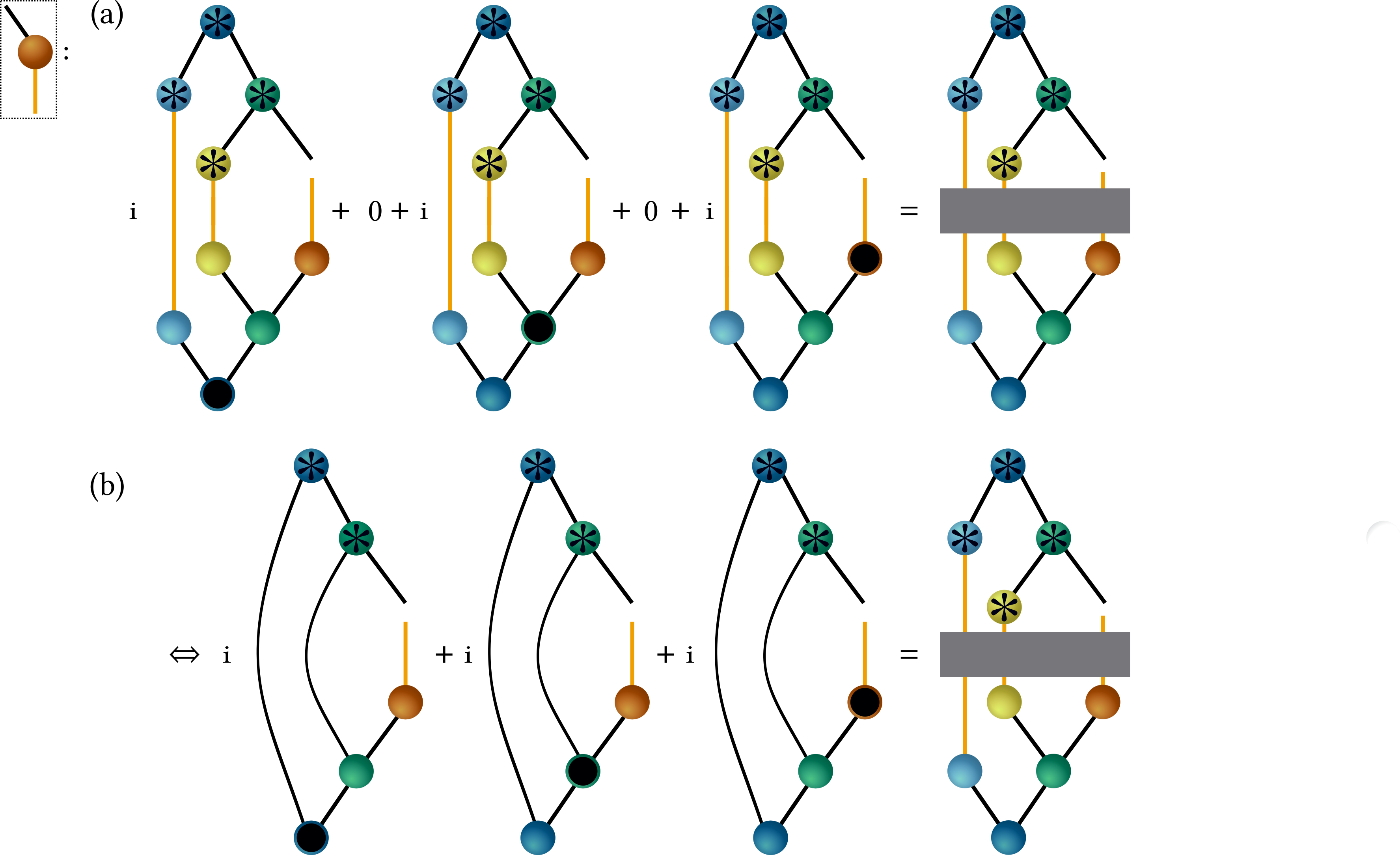}
\caption{First part of the derivation of the equations of motions for  $\matr A^{3; 2,2}$ (shown in the upper left rectangle)
 from the TTNS shown in  \autoref{fig:mltree}. 
Terms that are trivially zero due to the gauge conditions have been set to zero immediately (compare with \autoref{fig:ml_deriv_root}). 
SPF orthonormality simplifies the expression in (a), as shown in (b) (compare with \autoref{fig:ttns_bracket}).
}
  \label{fig:ml_deriv_ml1}
\end{figure}

\begin{figure}
\centering
\includegraphics{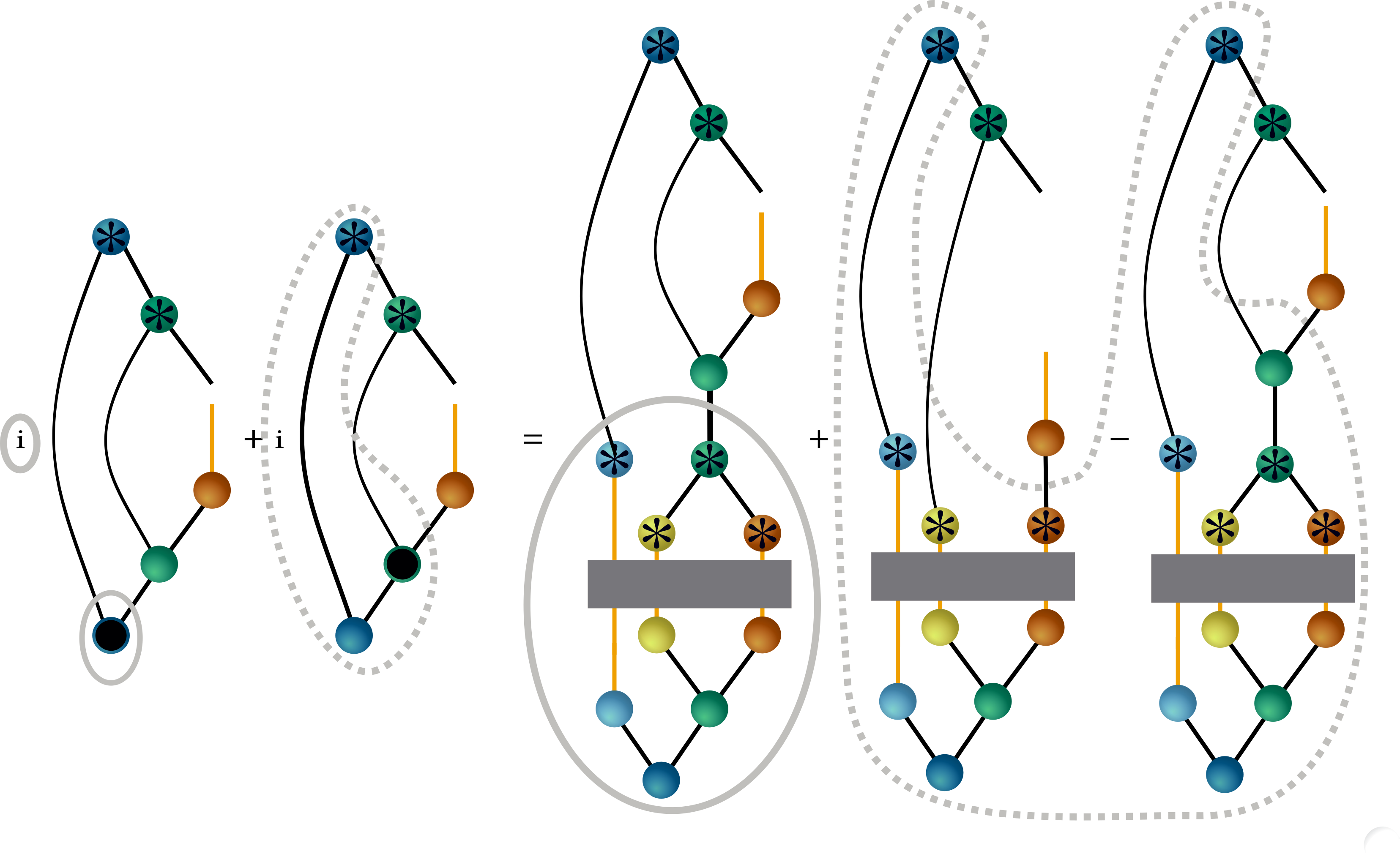}
\caption{Expression for two diagrams on the left-hand site of \autoref{fig:ml_deriv_ml1}(b). 
The gray-lined (gray-dashed) ellipses show the expression from 
 \autoref{fig:ml_deriv_root} (\autoref{fig:ml_deriv_mc2}).
 Note that the first and last term on the right-hand site cancel each other.
}
  \label{fig:ml_deriv_ml1b}
\end{figure}

\begin{figure}
\centering
\includegraphics{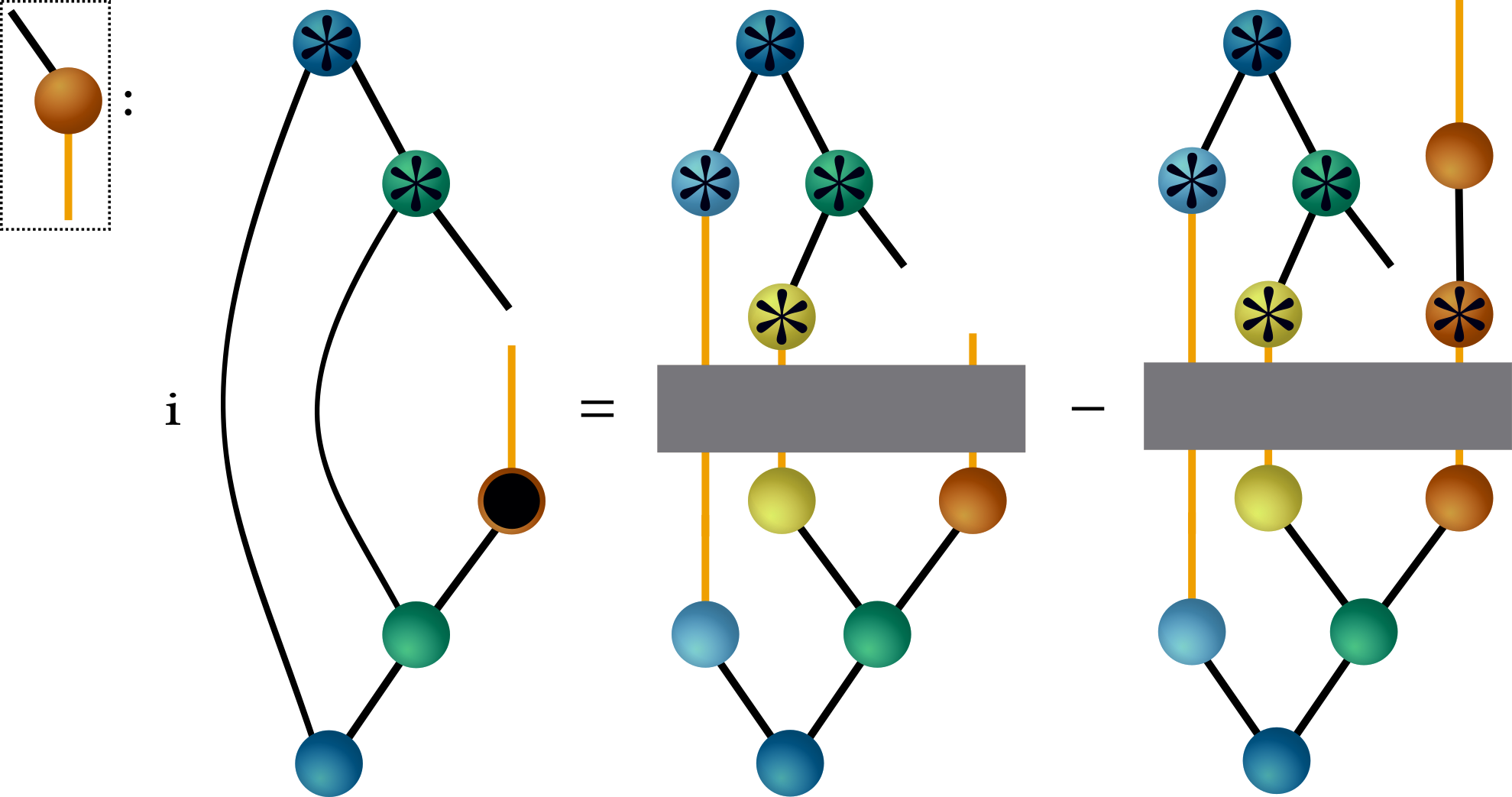}
\caption{Final expression of the equations of motions for $\matr A^{3; 2,2}$ (shown in the upper left rectangle) 
from the TTNS shown in  \autoref{fig:mltree}. 
To arrive at this, 
the expression from  \autoref{fig:ml_deriv_ml1b} is inserted in \autoref{fig:ml_deriv_ml1}.
}
  \label{fig:ml_deriv_ml2}
\end{figure}

We now turn to the derivations for the tensors in the third layer.
Since $\matr A^{3; 2,1}$ is not any different from $\matr A^{3; 2, 2}$, we will focus on only one of them. Inserting $\matr A^{3; 2,2}$ into \autoref{eq:dirac_principle_tensors} leads to the diagrams shown in \autoref{fig:ml_deriv_ml1}. Compared to the derivation for the first and second layer, here the derivation is  slightly more involved as we have to take the equations of the previous two layers into account. 
We first simplify two terms appearing on the left-hand side of  \autoref{fig:ml_deriv_ml1}(b). For that we combine the results from \autoref{fig:ml_deriv_root} and \autoref{fig:ml_deriv_mc2}, which gives the expression shown in \autoref{fig:ml_deriv_ml1b}.
 Note that two terms on the right-hand site  of \autoref{fig:ml_deriv_ml1b}
 cancel each other.
Inserting this back into  \autoref{fig:ml_deriv_ml1}(b) leads to the final equations of motions as shown in \autoref{fig:ml_deriv_ml2}. This corresponds to \autoref{eq:spftensor_eom} for layer $3$.

We note that next to the ``standard'' approach that we use here~\cite{Multiconfigurational1990meyer,Wave1992manthe,Multilayer2003wang,Multilayer2008manthe},
an alternative approach to derive the ML-MCTDH equations of motions exist~\cite{Dynamical2007koch,Quantum2008lubich,Time2015lubich,Time2015lubicha,Discretized2016kieri,Tangent2018bonfanti,Tangentspace2019vanderstraeten}. 
There, one first derives the projector $\hat P_T$ that projects onto the tangent space, that is, the space spanned by the first-order variations.
Then one solves
\begin{equation}
\ii \ket{\dot \Psi} = {\hat P_T}  \hat H \ket{\Psi}.
\label{eq:tdse_p}
\end{equation}
$\hat P_T$ can be written as
$\hat P_T \propto  \sum_{ij} \ket{\partial_{\lambda_i} \Psi}G^{-1}_{ij} \bra{\partial_{\lambda_j} \Psi}$, where $\matr G^{-1}$ is the inverse of the Gram matrix. 
With this formulation of $\hat P_T$
one obtains the standard ML-MCTDH equations of motions~\cite{Dynamical2010koch,Discretized2016kieri,Tangent2018bonfanti}.
Using a different expression of $\hat P_T$,
this approach will become important for the TDVP-DMRG scheme discussed in \autoref{sec:tddmrg}.

\section{Canonicalization: Change of root node and sweeps}
\label{sec:canonicalization}
\begin{figure}
\centering
\includegraphics{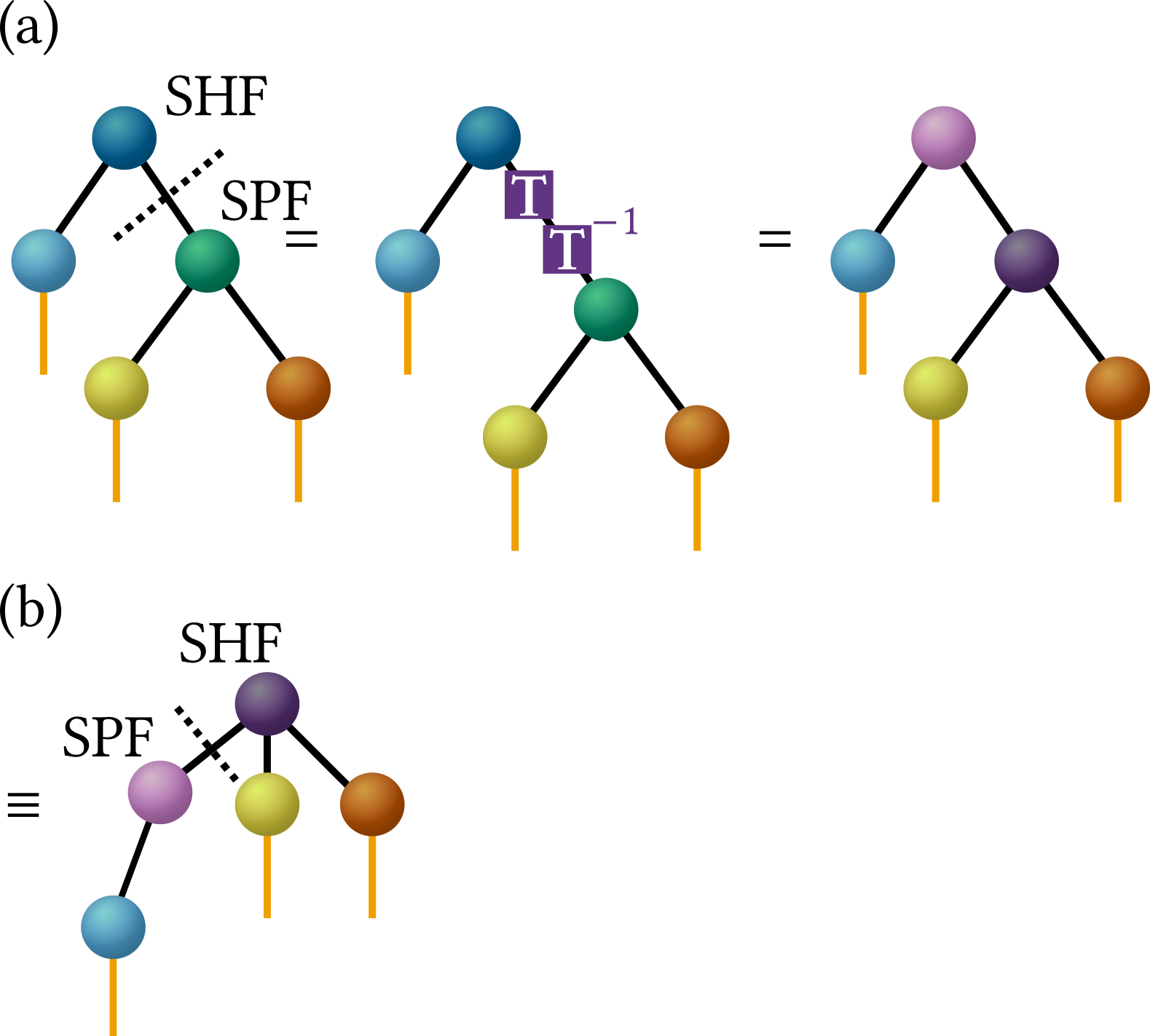}
\caption{Gauge-invariance in a ML-MCTDH state/TTNS.
(a) Inserting a matrix $\matr T$ and its inverse along a virtual bond does not change the state. The transformation matrix $\matr T$ and its inverse are denoted as purple rectangles. 
They are absorbed into the purple tensors;
compare with Eqs.~\eqref{eq:t_transformed_psi}-\eqref{eq:invariance_2}.
(b) If $\matr T$ is chosen to orthogonalize and normalize the SHFs $\ket{\Psi_i^{1;2}}$ then the tree changes its structure and the SHFs turn into SPFs and vice versa; see text for details.
The dotted lines mark the splitting of the SHFs and SPFs.
}
  \label{fig:ttns_gauge}
\end{figure}

Here, we will discuss how to change the root node of the ML-MCTDH state/TTNS \emph{without} changing the actual state that is represented, which is at the heart of DMRG algorithms.  
This ``canonicalization'' exploits that the TTNS is invariant under transformation by an invertible matrix $\matr T$. 
This is seen by inserting $\matr 1 = \matr T \matr T^{-1}$ into the SHF expansion, \autoref{eq:shf_expansion}:
\begin{align}
  \ket{\Psi}  &= \sum_{i} \ket{\Psi_{i}^{z}}\otimes \ket{\phi_{i}^{z}} %
  = \sum_{i k} \delta_{ik}\ket{\Psi_{i}^{z}}\otimes \ket{\phi_{k}^{z}} %
  = \sum_{i j k}  T_{ij} T^{-1}_{jk} \ket{\Psi_{i}^{z}}\otimes \ket{\phi_{k}^{z}} \\
  &=  \sum_{j} \ket{\widetilde \Psi_{j}^{z}}\otimes \ket{\tilde\phi_{j}^{z}},\label{eq:t_transformed_psi}
\end{align}
where we introduced
the $\matr T$-transformed 
SHFs $\ket{\widetilde \Psi^z_j}$ and
SPFs $\ket{\tilde \phi^z_j}$, 
\begin{align}
  \ket{\widetilde \Psi^z_j} &= \sum_{i} T_{ij} \ket{\Psi^z_i},
  \label{eq:invariance_1}
  \\
  \ket{\tilde \phi^z_j} &= \sum_{k} T^{-1}_{jk} \ket{\phi^z_k}.
  \label{eq:invariance_2}
\end{align}
This is depicted diagrammatically in \autoref{fig:ttns_gauge}(a).

$\matr T$ can be chosen to orthogonalize and normalize the SHFs. 
This then leads to \emph{nonorthogonal} $\matr T$-transformed SPFs.
Since SPFs should be orthonormal,
however, now we can re-interpret 
\autoref{eq:t_transformed_psi}
and just exchange the meanings of the $\matr T$-transformed SHFs and SPFs by formally turning the now orthonormal SHFs into SPFs and vice versa. This leads to a change of the tree structure. 
Additionally, 
if the transformation is done for the first layer, the root node changes. This is shown in \autoref{fig:ttns_gauge}(b).

\begin{figure}
\centering
\includegraphics{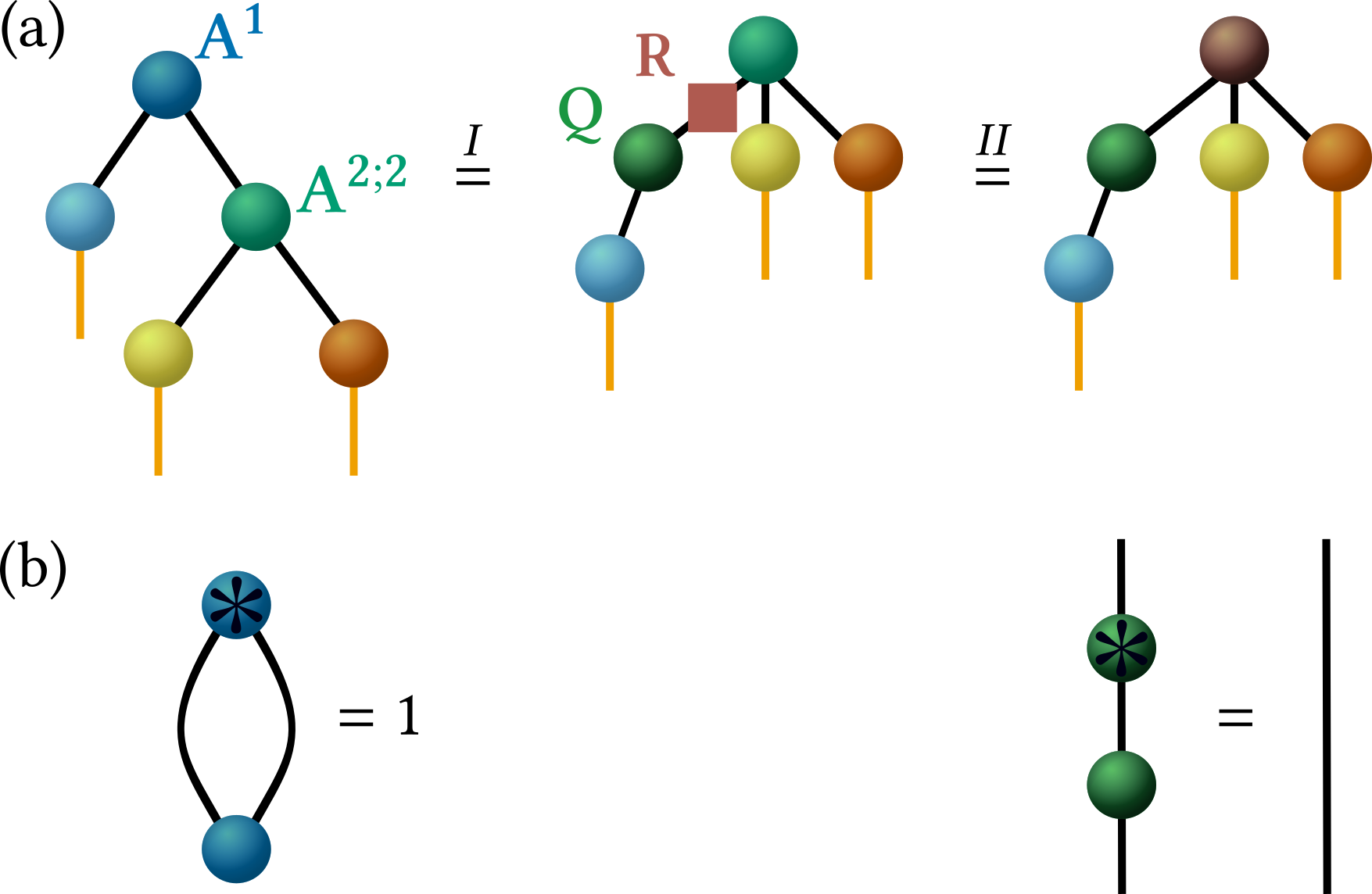}
\caption{Change of root node (canonicalization) from tensor $\matr A^1$ to $\matr A^{2;2}$. 
(a) Steps performed in the canonicalization: (I) Reshaping and
QR decomposition and (II) absorption of $\matr R$ into $\matr A^{2;2}$. This leads to a different tree structure. 
(b)  Orthogonality conditions (isometries) of $\matr A^1$ for the
two tree structures (canonical forms) where $\matr A^1$ is the root node (left) and where $\matr A^{2;2}$ is the root node (right).
The straight line denotes a unit matrix.
}
  \label{fig:ttns_qr}
\end{figure}

In practice, due to orthonormality of all SPFs in the TTNS, to orthogonalize and normalize the SHFs $\ket{\Psi_{j_{\kappa}}^{ 1; \kappa } }$ one only needs to orthogonalize the tensor $\matr A^1$ with entries $A^{1}_{j_1,j_2,\dots,j_{\kappa},\dots,j_d }$.
This is done by
first transposing it to 
$\matr A^{\tilde z'}$ with entries
$A^{\tilde z'}_{j_{\kappa}; \tilde J}$, where
$\tilde z'$ and $\tilde J$ correspond to the labels for  the tree structure with changed root node. 
Then $A^{\tilde z'}_{j_{\kappa}; \tilde J}$ is matricized (compare with \autoref{fig:orth_mat})
and orthogonalized. 
The matrix orthogonalization and normalization can be done, %
\eg, using a QR matrix decomposition, which corresponds to Gram-Schmidt orthogonalization~\cite{Matrix2013golub,Tensor2019hackbusch}.
The $\matr R$ in the QR matrix then corresponds to $\matr T^{-1}$ in \autoref{eq:invariance_2} and is absorbed in the tensor connecting to  $\matr A^{\tilde z'}$. %
This procedure, also sometimes dubbed isometrization~\cite{Unconstrained2014gerster},
together with  the different orthonormality conditions (isometries) of the initial root tensor is shown diagrammatically in \autoref{fig:ttns_qr}.

\begin{figure}
\centering
\includegraphics{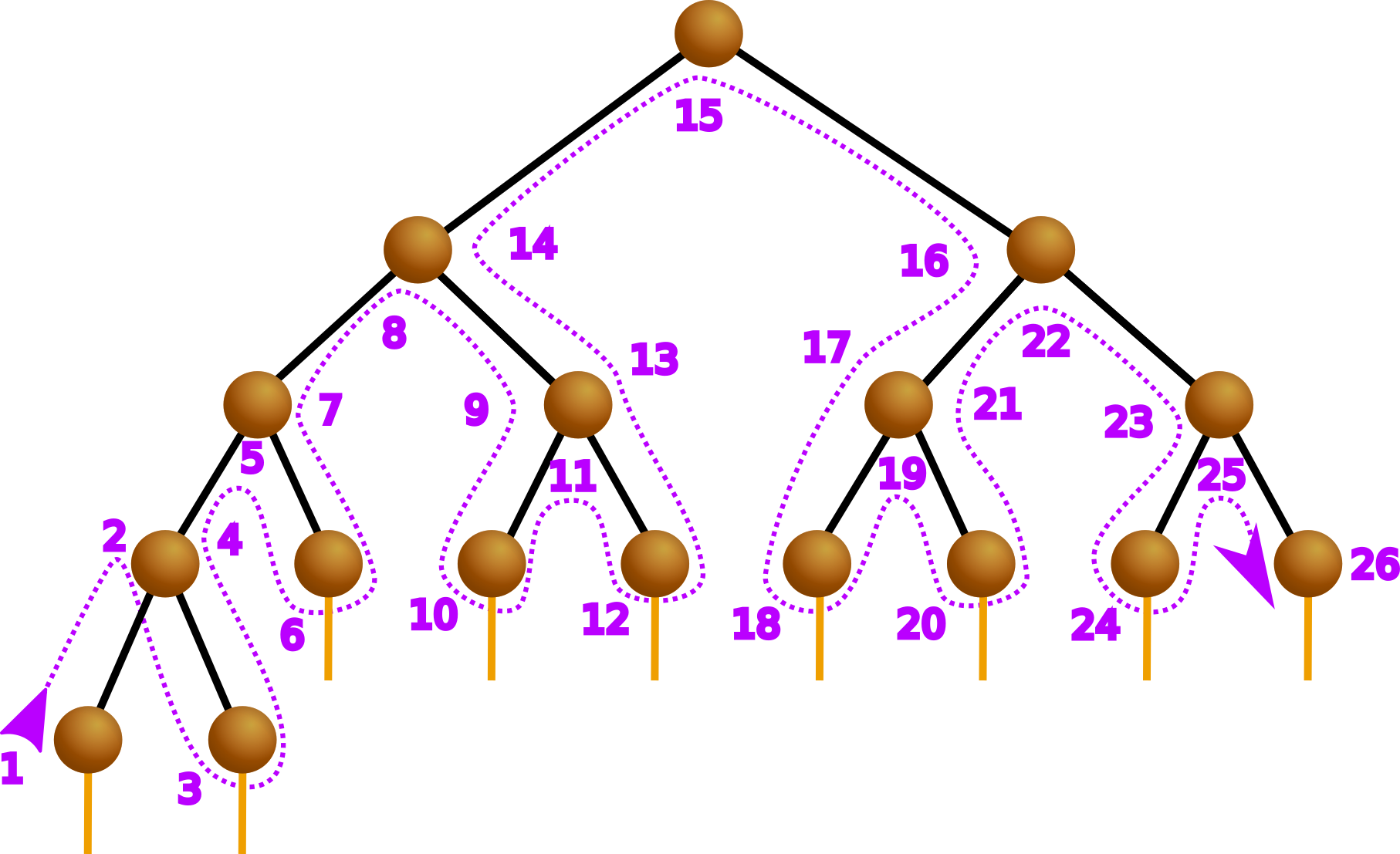}
\caption{Example of a sweep through a tree.
Here, the sweep is defined to start at a node in the last layer (lower left). 
At each ``stop'' of the sweep, which is denoted by numbers, the corresponding tensor is orthogonalized, which changes the canonical form of the tree. 
This defines one forward sweep. Reversing the direction leads to a backward sweep that ends at the node that initializes the forward sweep.
}
  \label{fig:sweep}
\end{figure}

Note that this change of canonicalization can be done with any SPF-SHF pair in the tree and not only that of layer 1. Further, the canonicalization allows us to systematically traverse the tree in a way that every tensor at some point will become at least once a root node. This is an important ingredient in any DMRG algorithm and called \emph{sweep}. It can be implemented using graph algorithms such as depth-first search; see \autoref{fig:sweep} for an example. 
Other ways to traverse a TTNS in one sweep are possible as well~\cite{Unconstrained2014gerster}.

As technical remark,
instead of using a QR decomposition to orthogonalize the SHFs one can also use other orthogonalization procedures such as singular-value-decomposition (SVD)~\cite{Matrix2013golub,Tensor2019hackbusch}
(using $\matr U$ in the SVD $\matr A = \matr U \matr S \matr V^\dagger$ as orthogonal matrix and $\matr S \matr V^\dagger$ as $\matr T^{-1}$). This is equivalent to using the eigenvectors of the reduced density matrix. Hence, this leads to natural SPFs, which often result in  improved numerics as then the Hamiltonian represented by the basis of the configurations often is diagonally dominant. 
SPFs that diagonalize the separable part of the Hamiltonian are another commonly used choice to improve numerics~\cite{Computation2008doriol}.
All of these choices realize a particular gauge.

\section{Time-dependent DMRG approximation}
\label{sec:tddmrg}

One major problem of the ML-MCTDH equations of motions is that they can become highly nonlinear. Hence, sophisticated integrators need to be used for efficient propagation~\cite{Multiconfiguration2000beck}. 
To overcome this for standard MCTDH,
based on insights that SPFs typically evolve slower than the actual wavefunction, an efficient integrator targeting the structure of the MCTDH equations has been developed and dubbed constant mean-field (CMF) integrator~\cite{Efficient1997beck,Integration2006manthe}.
Therein, the equations of motions 
for the SPFs, \autoref{eq:spf_eom},
and those for the root tensor, \autoref{eq:roottensor_eom}, 
are decoupled by assuming that the configurations, density matrices and mean fields are constant in time during the propagation. 
This has the advantage that the root tensor can be propagated using standard exponential propagators~\cite{Introduction2008tannor} such as the short iterative Lanczos (SIL) propagator~\cite{Unitary1986park}. 
The SIL propagation 
leads to 
efficient error estimates~\cite{Efficient1997beck,Integration2006manthe}.
The equations for the SPFs are then solved using a general purpose propagator.  %
The CMF integration scheme can be an order of magnitude faster than propagating all MCTDH tensors simultaneously. The error is of  order $\Delta t^3$ with respect to the adaptive time step $\Delta t$. %

\begin{figure}
\centering
\includegraphics{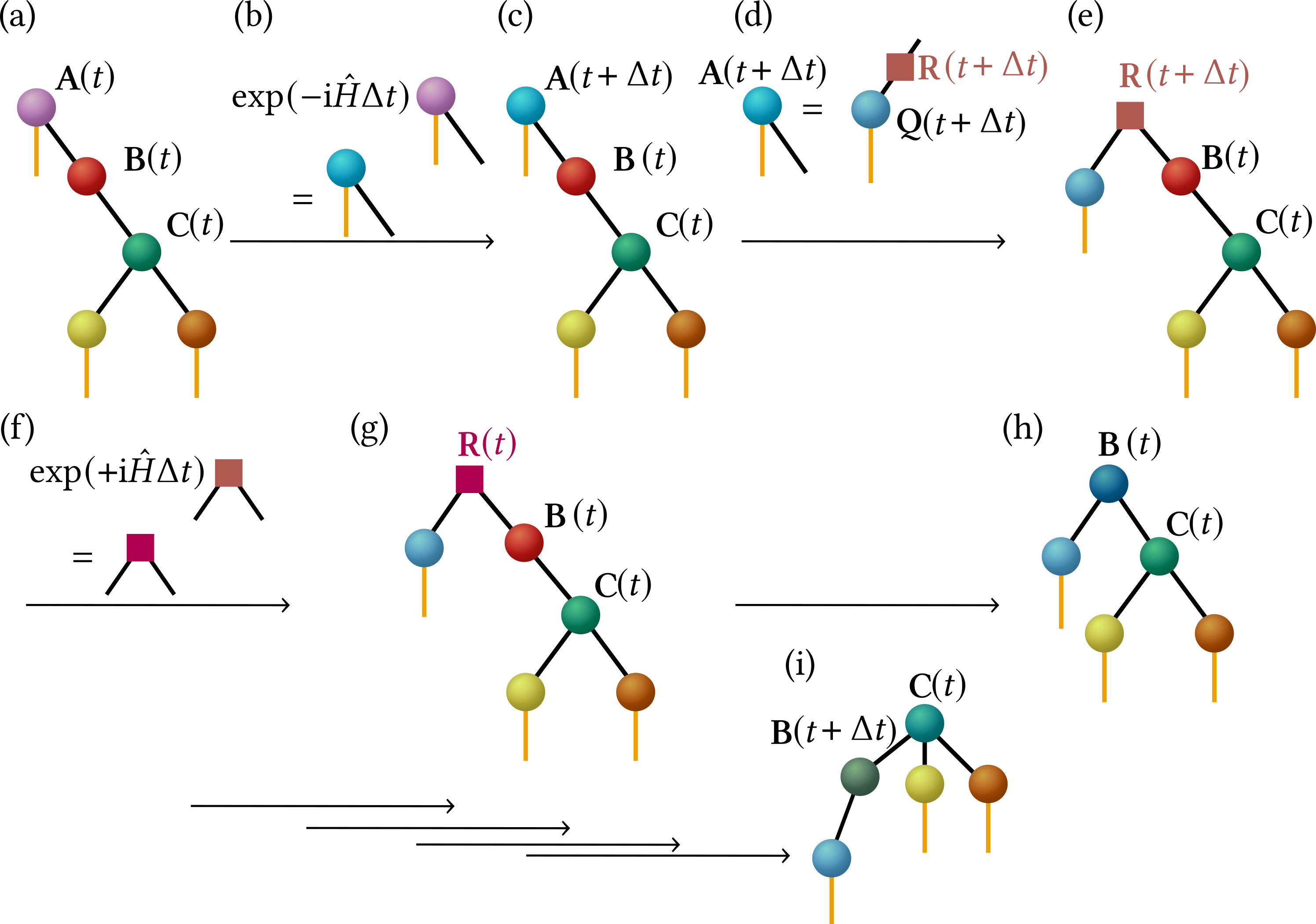}
\caption{TDVP-DMRG scheme for the TTNS from \autoref{fig:mltree}. Panels (a)-(h) show different steps in the algorithm for propagating the first tensor $\matr A$ and for changing the root node to the neighboring tensor $\matr B$.
Panel (i) shows the TTNS for propagated tensors $\matr A$ and $\matr B$.
As the tree structure changes during the sweep, we here use generic symbols $\matr A$, $\matr B$, $\matr C$, $\matr Q$, $\matr R$ for labeling specific tensors. 
Here, $\matr C(t)$ is the only tensor with $3$ virtual bonds. Hence, during the sweep, $\matr C(t)$ will appear twice as root node, but it will only be propagated once to time $t+\Delta t$. $\matr C(t+\Delta t)$ then will be skipped the next time it is root node during the sweep. This is the main difference when applying TDVP-DMRG to TTNSs instead of MPSs.
See the text for details.}
  \label{fig:tddmrg}
\end{figure}

As CMF is designed for normal MCTDH, its direct usage for ML-MCTDH is limited to the first two layers in a TTNS.
While in principle the CMF idea can be extended to any layer in ML-MCTDH, there it is not as powerful as only the root tensor can be propagated using an exponential integrator. 
Hence, the standard CMF integrator may not offer significant improvements for TTNSs with many layers.
An integrator that we call here TDVP-DMRG puts the CMF integrator to its extreme and allows for using exponential integrators \emph{for all} tensors by making use of sweeps~\cite{Time2015lubich,Unifying2016haegeman}.
The idea of TDVP-DMRG 
was motivated by finding an alternative way to solve the MCTDH equations of motions~\cite{Projectorsplitting2014lubich,Time2015lubich}, but it 
has been introduced in several fields almost simultaneously. Thus, it is also known as projector splitting integrator (PSI)-MCTDH~\cite{Implementation2017kloss,Tangent2018bonfanti,Time2021lindoya,Time2021lindoy}, Lubich integrator~\cite{Regularizing2018meyer,Symmetries2021weike} 
(named after its inventor~\cite{Projectorsplitting2014lubich,Time2015lubich,Time2018lubich})  as well as ``tensor-train KSL''  %
for the case of MPSs~\cite{Time2015lubicha,TensorTrain2022lyu}.
TDVP-DMRG has first been used for MPSs (and later for standard MCTDH~\cite{Implementation2017kloss,Time2018lubich}).
The first application of TDVP-DMRG to TTNSs we are aware of is Ref.~\cite{Tensor2019schroder}.
For a mathematical analysis, see Ref.~\cite{Time2021ceruti}.
We have recently used it for the challenging case of large-amplitude dynamics of  the fluxional Zundel ion~\cite{Stateresolved2022larsson}.
Here, before sketching the derivation, 
we first explain the basic ideas of TDVP-DMRG using somewhat hand-waving arguments. 
Since the TTNS changes its root node during the TDVP-DMRG sweep, the ML-MCTDH labeling scheme, which is tied to a particular rood node, is difficult to use. Instead, we will use generic symbols such as $\matr A$ and $\matr B$ to denote tensors.
A diagrammatic illustration for our example tree (\autoref{fig:mltree}) is shown in \autoref{fig:tddmrg}. 
We start the sweep in \autoref{fig:tddmrg}(a) with a tensor in one of the last layers,
which we dub here    $\matr A(t)$. 
The first propagation step in TDVP-DMRG, \autoref{fig:tddmrg}(b), follows the CMF scheme and propagates  $\matr A(t)$ to time $t+\Delta t$ using \autoref{eq:roottensor_eom} with  the approximation of time-independent   
configurations $\ket{\Phi_I^1}$. %

Instead of following the CMF scheme and solving the complicated \autoref{eq:spftensor_eom} for propagating the tensor $\matr B(t)$ adjacent to $\matr A(t+\Delta t)$ in \autoref{fig:tddmrg}(c), %
we orthogonalize $\matr A(t+\Delta t)$ using, \eg, a QR decomposition. This  gives $\matr A(t+\Delta t)=\matr Q(t+\Delta t)\matr R(t+\Delta t)$, as shown in \autoref{fig:tddmrg}(d). 
Then $\matr A(t+\Delta t)$ is exchanged by 
$\matr Q(t+\Delta t)$ and $\matr R(t+\Delta t)$, which is a new, \emph{additional} node in the tree, see \autoref{fig:tddmrg}(e). 
For changing the root node to $\matr B(t)$, we need to absorb $\matr R(t+\Delta t)$ into $\matr B(t)$. 
However, $\matr R(t+\Delta t)$ is at a different time step than $\matr B(t)$.
In order to have matching times for $\matr R$ and $\matr B$, 
we interpret  $\matr R(t+\Delta t)$ as actual root node and \emph{back-propagate}  $\matr R(t+\Delta t)$ to  $\matr R(t)$, which is shown in  \autoref{fig:tddmrg}(f) and (g).
This back-propagation again uses the root tensor MCTDH equations, \autoref{eq:roottensor_eom}, 
but note that the configurations used in  \autoref{eq:roottensor_eom} 
for propagating $\matr R(t)$ 
describe different propagation times, namely $t+\Delta t$ for the parts of the configurations associated with $\matr A$ and $t$ for the parts of the configurations associated with $\matr B$.
After back-propagation of $\matr R(t+\Delta t)$, $\matr R(t)$ is absorbed into $\matr B(t)$, which becomes the new root node of the TTNS; see   \autoref{fig:tddmrg}(h). 
The described procedure is repeated: $\matr B(t)$ is propagated and orthogonalized, the resulting $\matr R(t+\Delta t)$ from the QR-decomposition of $\matr B(t+\Delta t)$ is back-propagated and absorbed into the new root node $\matr C(t)$; see   \autoref{fig:tddmrg}(i). 
Doing this for each tensor in the TTNS finally leads to the fully propagated state $\ket{\Psi(t+\Delta t)}$.\footnote{As technical remark, care must be taken for tensors placed at tree branches to propagate them only once. These tensors need to be  skipped during parts of the sweep so that they are only propagated once in one sweep.
See Refs.~\cite{Tensor2019schroder,Time2020bauernfeind,Time2021lindoy,Time2021ceruti} for more details.
}

\begin{figure}
\centering
\includegraphics{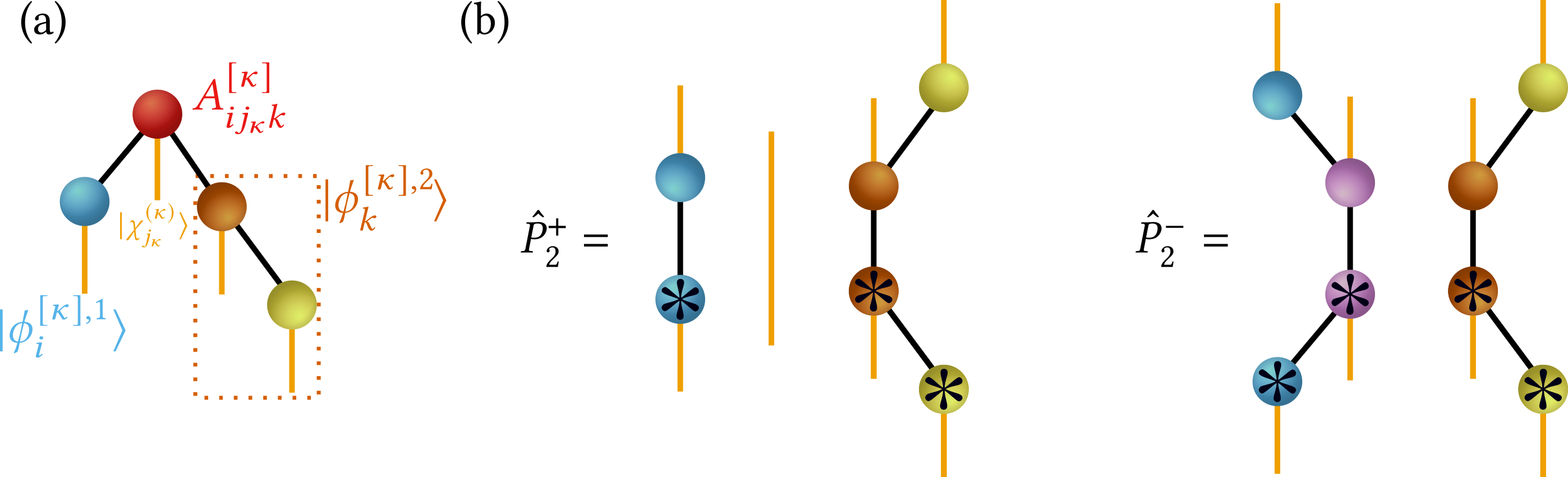}
\caption{Example MPS and parts of tangent space MPS projector.
(a) Example of the notation used in \autoref{eq:mps_ttns_form} for $\kappa=2$.
(b) Projectors $\hat P^+_2$ and $\hat P^-_2$ for the MPS shown in (a). These projectors are part of the tangent space projector, \autoref{eq:p_t}.
$\hat P^-_2$ makes use of \autoref{eq:mps_ttns_form_R}, where $A^{[\kappa]}_{ij_\kappa k}$ was replaced by $Q_{ij_\kappa x}^{[\kappa]}$ from \autoref{eq:phi_qr_mps}, thus the different color of that node.}
  \label{fig:mps_proj}
\end{figure}

We now sketch the derivation of TDVP-DMRG. To simplify expressions, we here only derive it for MPSs, but generalizing the TDVP-DMRG scheme for trees is straightforward, as we have demonstrated in Figs.~\ref{fig:ttns_qr} and \ref{fig:tddmrg}.
An MPS corresponds to a linear, unbalanced tree, and each node has two virtual bonds and one physical bond, save for the tensors at the end, which only have one virtual bond.
We can write an $F$-dimensional MPS whose canonical form (root node) is centered at dimension $\kappa$ as
\begin{equation}
  \ket{\Psi} = \sum_{i j_\kappa k} A^{[\kappa]}_{i j_\kappa k} \ket{\phi^{[\kappa],1}_i}\otimes \ket{\chi^{(\kappa)}_{j_\kappa}} \otimes \ket{\phi^{[\kappa],2}_k}.
  \label{eq:mps_ttns_form}
\end{equation}
See \autoref{fig:mps_proj}(a) for an example diagram. 
Here we use a slightly modified ML-MCTDH notation where we add the location of the root node/canonical form using $[\kappa]$ and where we have dropped the label for the first layer in the root tensor and the SPFs. 
To change the root node to dimension $\kappa +1$,
we QR-decompose the root tensor as
\begin{equation}
  A^{[\kappa]}_{i j_\kappa k} = \sum_x Q_{ij_\kappa x}^{[\kappa]} R_{xk}^{[\kappa]},
  \label{eq:phi_qr_mps_A}
\end{equation}
and use $\matr Q$ as expansion coefficients of the new SPFs for the root node at dimension $\kappa+1$ (compare with \autoref{fig:ttns_qr}):
\begin{equation}
\ket{\phi^{[\kappa+1],1}_x}  = \sum_{ij_\kappa}  Q_{ij_\kappa x}^{[\kappa]} \ket{\phi^{[\kappa],1}_i} \otimes \ket{\chi^{(\kappa)}_{j_\kappa}}.
  \label{eq:phi_qr_mps}
\end{equation}
Using the newly formed $\ket{\phi^{[\kappa+1],1}_x}$, we obtain an \emph{intermediate} state with $\matr R$ as root node:
\begin{equation}
  \ket{\Psi} = \sum_{x k} R^{[\kappa]}_{xk} \ket{\phi^{[\kappa+1],1}_i}\otimes \ket{\phi^{[\kappa],2}_k},
  \label{eq:mps_ttns_form_R}
\end{equation}
which combines $\ket{\phi^{[\kappa+1],1}_i}$ and $\ket{\phi^{[\kappa],2}_k}$ in one TTNS.
This representation is known as bond-canonical form in MPS language.
It is the one used for back-propagating $\matr R$ in \autoref{fig:tddmrg}.
Absorbing $\matr R$ into $\ket{\phi^{[\kappa],2}_k}$ as in \autoref{fig:ttns_qr}  then leads to the TTNS representation with orthogonalization center/root node at $\kappa+1$.

TDVP-DMRG makes use of the TDVP written 
in terms of the tangent space projector $\hat P_T$,
\autoref{eq:tdse_p},  using a special 
form of $\hat P_T$~\cite{Time2015lubicha,Unifying2016haegeman}
that utilizes different canonical forms of the MPS: 
\begin{equation}
  \hat P_T = \sum_{\kappa=1}^F \hat P^+_\kappa - \sum_{\kappa=1}^{F-1} \hat P^-_\kappa,\label{eq:p_t}
\end{equation}
with
\begin{align}
  \hat P_\kappa^+ &= \sum_i \ket{\phi^{[\kappa],1}_i} \bra{\phi^{[\kappa],1}_i} \otimes \hat 1_\kappa \otimes \sum_{k}  \ket{\phi^{[\kappa],2}_k}\bra{\phi^{[\kappa],2}_k},\label{eq:P+} \text{ and}\\
  \hat P_\kappa^- &= \sum_x \ket{\phi^{[\kappa+1],1}_x} \bra{\phi^{[\kappa+1],1}_x} \otimes \sum_{k}  \ket{\phi^{[\kappa],2}_k}\bra{\phi^{[\kappa],2}_k}.
\end{align}
See \autoref{fig:mps_proj}(b) for example diagrams.
Note that $\hat P_\kappa^-$ uses the SPFs from the bond-canonical form, \autoref{eq:mps_ttns_form_R}.
The first sum in \autoref{eq:p_t} projects onto all possible TTNSs that only differ at one site (compare with the first-order variations shown in  \autoref{fig:ml_deriv_derivs}). The second sum subtracts all components that are identical with the current state~\cite{Timeevolution2019paeckel}, save for the gauge. %

Using \autoref{eq:p_t}
and assuming a time-independent Hamiltonian,
we can then write the TDVP propagation, \autoref{eq:tdse_p}, from time $t$ to $t+\Delta t$  as~\cite{Time2015lubicha,Matrix2018kurashige}
\begin{align}
  \ket{\Psi(t+\Delta t)} =& \exp\left( -\ii\Delta t \hat P_T   \hat H\right) \ket{\Psi(t)} \\
  =& \exp\left[-\ii\Delta t \left(\hat P^+_1 - \hat P^-_1 + \cdots + \hat P^+_{F-1} - \hat P^-_{F-1} + \hat P^+_F\right) \hat H\right] \ket{\Psi(t)} \nonumber\\
  \approx& 
  \exp\left(-\ii  \Delta t\hat P^+_1 \hat H\right) 
  \exp\left(+\ii  \Delta t\hat P^-_1 \hat H\right) 
  \cdots
  \exp\left(-\ii \Delta t \hat P^+_{F-1}  \hat H\right) \nonumber\\
  &\exp\left(+\ii \Delta t \hat P^-_{F-1}  \hat H\right) 
  \exp\left(-\ii \Delta t \hat P^+_{F} \hat H\right)\ket{\Psi(t)} %
  + \mathcal O(\Delta t^2),\label{eq:tddmrg_expp}
\end{align}
where we approximated the exponential using Lie-Trotter splitting.
This leads to the scheme discussed above and shown in \autoref{fig:tddmrg} for a TTNS. 
$\exp\left(+\ii  \Delta t\hat P^-_\kappa \hat H\right)$ forward-propagates $F$ root nodes and $\exp\left(+\ii  \Delta t\hat P^-_\kappa \hat H\right)$ backward-propagates $F-1$  $\matr R$ matrices.
For more details on the derivation of different aspects we refer to Refs.~\cite{Projectorsplitting2014lubich,Time2015lubich,Time2015lubicha,Unifying2016haegeman,Tangent2018bonfanti,Timeevolution2019paeckel,Time2021lindoya,Time2021lindoy,Matrix2018kurashige,Projector2022gleisa}.
An expression similar to \autoref{eq:tddmrg_expp} may be obtained for TTNSs by ensuring that tensors with more than two virtual bonds in a TTNS will be propagated only once during one sweep (the time propagation can be omitted during instances of the sweep where these tensors become the new root node)~\cite{Time2020bauernfeind,Time2021lindoy,Time2021ceruti}; a mathematical proof of this is missing, however~\cite{Time2021ceruti}.
Using $\Delta t \to \Delta t/2$ as time step and combining \autoref{eq:tddmrg_expp} with a backward sweep 
leads to a Lie-Trotter splitting error of order $\mathcal O(\Delta t^3)$.
This is of the same order as that of the CMF integrator.
Compared to the CMF scheme for MCTDH, however, no adaptive time step so far has been implemented and typically different time steps needs to be tried out.
Using composition schemes, higher orders are possible as well, but this so far has not been explored. 

An interesting special case of TDVP-DMRG that resembles more ML-MCTDH than DMRG 
is the ``non-hierarchical''  propagation scheme introduced by Weike and Manthe~\cite{Symmetries2021weike}.
Therein, 
based on Löwdin-orthogonalized~\cite{Nonorthogonality1970lowdin} SHFs (by using $[{\matrgreek{ \rho}}^z]^{-1/2}$ as transformation matrix), 
a time-dependent and node-dependent gauge operator $\hat g^z(t)$ is defined that leads to symmetric equations of motions for the SHFs and for the SPFs.
This makes all SHFs orthonormal and the SPFs non-orthogonal. The equations of motions for \emph{all} tensors then take a form similar to that of the root tensor, \autoref{eq:roottensor_eom}, plus an additional term for the operator $\hat g^{1;\kappa}(t)$:
\begin{equation}
  \ii \dot{A}_{I}^{1} =\sum_J \matrixe{\Phi^{1}_I}{\hat H}{\Phi^1_J} \matr A_J^1 - \sum_{\kappa}^{d} \sum_J \matrixe{\Phi^{1}_I}{\hat g^{1;\kappa}(t)}{\Phi^1_J} A_J^1.
  \label{eq:roottensor_eom_gauge}
\end{equation}
Together with the Löwdin orthogonalization, 
these additional gauge-operator terms $\hat g^{1;\kappa}$ for each dimension of $\matr A^1$ roughly correspond to the decomposition and back-propagation of the $\matr R$ tensor in TDVP-DMRG, and here $\matr R$ is constrained to be $[{\matrgreek{ \rho}}^z]^{+1/2}$.
Thus, compared to TDVP-DMRG, there is no 
Lie-Trotter splitting
but instead, due to the specific orbital gauge, the effective Hamiltonian becomes time-dependent.
While making use of DMRG ideas (exploiting different orthogonality schemes),
this propagator  is straightforward to implement using existing MCTDH codes.
Further, it has the appealing property that all tensors can be propagated simultaneously using a CMF integrator with adaptive time step. This renders an implementation of the propagator embarrassingly parallel.
Another ``basis-update and Galerkin'' (BUG) integrator for Tucker decompositions  recently has been proposed by Ceruti and Lubich~\cite{Unconventional2022ceruti}.
It does not make use of backward propagation and enables parallel propagation, but the initial version does not conserve norm and energy.
A bond-dimension-adaptive and norm/energy-conserving extension of this for TTNSs is mentioned in \autoref{sec:subspace_enrichment}.

We end this section by noting that many other methods to solve the TDSE exist in the context of DMRG. Many of these are, however, targeted at certain model Hamiltonians. 
See Ref.~\cite{Timeevolution2019paeckel} for a recent review for TDVP-DMRG and other propagation methods for MPSs. 

\section{Time-independent DMRG}
\label{sec:tise_dmrg}
Since ML-MCTDH approximates the TDSE, eigenstates can be obtained through ML-MCTDH by propagation in imaginary time~\cite{Direct1986kosloff}. 
To overcome the same issues as in real-time propagation, a CMF variant for imaginary time evolution and standard MCTDH has been developed by Meyer et al.~and dubbed improved relaxation~\cite{Calculation2006meyer,Computation2008doriol}, where the propagation of the root tensor is replaced by   Davidson's  iterative diagonalization method~\cite{Iterative1975davidson}.
Improved relaxation can be viewed as a vibrational 
multiconfiguration self-consistent field approach~\cite{Multiconfigurational1994culot,Analytical1997drukker,Vibrational2010heislbetz},
where the optimization of the SPFs through direct minimization (\eg, via Jacobi rotations~\cite{Analytical1997drukker,Vibrational2010heislbetz}) 
is replaced by imaginary time propagation.
Alternatively, Manthe developed a modified Lanczos scheme that takes into  account the different SPF spaces of each MCTDH state during the Lanczos procedure~\cite{Iterative1996manthe,Multiconfigurational2012wodraszka}. 
State averaging can be used to target several states at once~\cite{Multiconfigurational1994culot,State2008manthe,Computation2008doriol,Iterative2012hammer}.
While improved relaxation can be straightforwardly generalized from MCTDH to ML-MCTDH~\cite{Iterative2014wang}, the same issues as discussed in \autoref{sec:tddmrg} remain. Further, it is very difficult to target many excited states and most (ML-)MCTDH eigenstate optimizations so far have targeted the few lowest eigenstates or excited states that are easily to identify, \eg, by a particular excitation in one mode. 

As alternative to the ML-MCTDH algorithm for targeting eigenstates, the TDVP-DMRG method can be used for eigenstates as well by imaginary time propagation.
Like improved relaxation, the imaginary time propagation of the root node can be replaced by directly diagonalizing the effective Hamiltonian (in its renormalized basis), $\matrixe{\Phi^{1}_I}{\hat H}{\Phi^1_J}$. 
Eigenstate optimization only requires following the energy gradient. Hence, the back-propagation in imaginary time of $\matr R$ is not necessary and $\matr R$ can directly be absorbed into the next tensor. 
This then leads to the (one-site) DMRG algorithm~\cite{Density1992white,Densitymatrix1993white},
which  also is known as alternating least squares (ALS) scheme~\cite{Alternating2012holtz,Tensor2015szalay,Tensor2019hackbusch}.
Thus, the DMRG algorithm actually is simpler than TDVP-DMRG and all that is required during the sweep is sequentially solving an effective eigenvalue problem. 

\begin{figure}
\centering
     \includegraphics{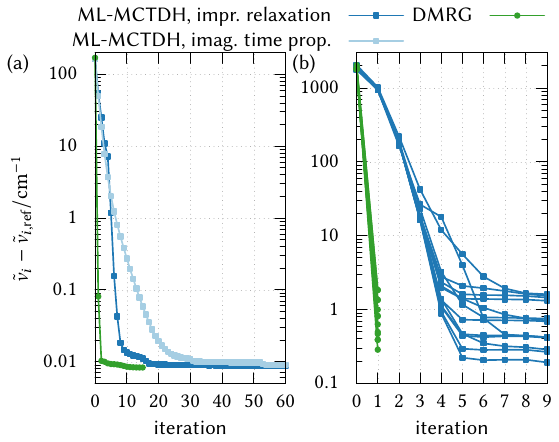}
     \caption{Performance of ML-MCTDH and TTNS/DMRG for a 12-dimensional model of acetonitrile, which employs a quartic Taylor expansion~\cite{Using2011avila}.
     Absolute errors of the ground state (a) and the lowest 13 states (b) using the state averaging approach.
     The green curve shows the performance of the DMRG algorithm and the blue curves show that of the ML-MCTDH method, using either improved relaxation (dark blue) or imaginary time propagation (pale blue; error after each time step of $1\, \mathrm{fs}$).
     All three methods use exactly the same initial TTNS for both optimizations without adapting the number of single particle functions (thus the ground state computation is more accurate). The iteration is proportional to the runtime of each method. All three methods have the same computational scaling but different prefactors.
     Reprinted from Ref.~\cite{Computing2019larsson}, with the permission of AIP Publishing.}%
     \label{fig:CH3CN_ML_vs_TTNS}
\end{figure}

The DMRG decouples the full nonlinear equations that would appear when following the energy gradient for all tensors simultaneously (which is done in ML-MCTH imaginary time propagation).
Actually, (one-site) DMRG resembles a fixed-point iteration where, for each tensor, an eigenvalue problem is solved iteratively, which is similar to the self-consistent field algorithm used in mean-field methods.
Indeed, the DMRG algorithm can be derived by introducing a  Lagrangian similar to the TDVP~\cite{Density2008chan}.
A numerical comparison of ML-MCTDH-based algorithms with the DMRG algorithm is shown in \autoref{fig:CH3CN_ML_vs_TTNS}. DMRG requires many fewer iterations to converge than ML-MCTDH.
Note, however, that state averaging can only be done approximately using the DMRG~\cite{TimeStep2017ronca,Computing2019larsson}, and for the same accuracy state-averaged DMRG requires a slightly larger bond dimension than state-averaged ML-MCTDH.

DMRG can also be extended to the computation of excited states. A simple way to do that is to use level shifting, that is, to compute the $N_S$ lowest states successively by shifting the $N_S-1$ previously computed states  $\ket{\Psi_i}$ in energy,
\begin{equation}
  \hat H \to \hat H + Z \sum_{i=1}^{N_S-1} \ketbra{\Psi_i}{\Psi_i},
\end{equation}
where $Z$ is some large number~\cite{CheMPS22014wouters}. 
While this approach may seem to break down when the states are very approximate, we have successively used it to accurately compute more than 1000 eigenstates of complicated problems such as the 15-dimensional vibrational problem of the fluxional Zundel ion~\cite{Stateresolved2022larsson}.
As long as many of the $N_S$ lowest states are actually of interest, this approach is fast, accurate, and straightforward to use~\cite{Computing2019larsson}.
Ultimately, however, the density of states becomes too large at large energies and 
more sophisticated approaches are required to directly target  highly-energetic eigenstates with desired properties. 
Some ways to achieve that are using the DMRG algorithm but targeting excited states in the effective Hamiltonian~\cite{Targeted2007dorando,Computing2019larsson}, solving the eigenvalue problem of $(\omega - \hat H)^2$ or $(\omega - \hat H)^{-1}$, where $\omega$ is the target energy~\cite{Optimization2019baiardi}, or using global approaches that describe the eigenstate as sum of TTNSs~\cite{Calculating2016rakhuba,ExcitedState2021baiardi}.

\section{Propagating unoccupied SPFs and subspace enrichment}
\label{sec:opt_unocc_spf}
Here, we will discuss how to deal with the ill-defined propagation of unoccupied SPFs and how to enlarge or ``enrich'' the SPF subspace. 
Both topics are related to each other and research on these topics recently has been very active both in the MCTDH and DMRG communities.

\subsection{Propagating unoccupied SPFs}
\label{sec:spf_regularization}
In most scenarios, the number of required SPFs (bond dimension of the TTNS) increases with time. 
Many simulations even start with a single Hartree product as initial state (only one SPF occupied in each dimension, $\nSPF=1$), and, after a few time steps, the state can rapidly become correlated. 
To take that into account,
one either must use an initial TTNS with a larger bond dimension (additional initially unoccupied SPFs), or one must dynamically add SPFs during the time evolution. 
The first one is the default case in MCTDH, and the second one a recent development~\cite{Dynamical2017larsson,Systematic2017mendive-tapia,Regularizing2020mendive-tapia}.
In either cases, the single-particle density matrix $\matrgreek \rho^z$ that needs to be inverted in the SPF equations of motions, \autoref{eq:spf_eom}, becomes singular (vanishing natural occupations).\footnote{One could avoid explicit inversion and solve a linear system instead. This is numerically more stable but does not alleviate the singularity problem, however.}
To avoid singular matrices, in practice they are being regularized during time propagation~\cite{Wave1992manthe}.
For many typical applications, regularization works well.
This is the case even when non-direct-product SPF spaces are being used and new SPFs are added and removed during the simulations~\cite{Dynamical2017larsson}.
Instead of inverting the density matrix, one can pseudo-invert the regularized coefficient tensor,\footnote{
Matricizing the tensor $\matr A^z$ and 
dropping the $z$ label, $\matrgreek\rho = \matr A^\dagger \matr A$. Then the SPF equations of motions contain $(\matr A^\dagger \matr A)^{-1}  \matr A^\dagger$, which is the Moore-Penrose pseudoinverse and which can be regularized directly, \eg, through SVD.}
which leads to improved stability~\cite{Multiconfigurational1990meyer,Regularizing2018meyer,Regularizing2018wang}.
Even with improved stability, however, the regularization can still become an issue both in terms of numerical stability and numerical efficiency~\cite{Regularity2007koch,Projectorsplitting2014lubich,Multiconfigurational2015manthe,Instabilities2016hinz}.

Formally, in TDVP-DMRG there is no inverse of the single-particle matrix involved, and no explicit regularization is required. 
While the TDVP-DMRG approach is robust with respect to small natural occupations~\cite{Discretized2016kieri},
orthogonalizing the SHFs is ill-defined for singular density matrices and then any orthogonalization procedure %
also becomes arbitrary.  %
However,  arguments that it is not possible to propagate unoccupied SPFs in TDVP-DMRG~\cite{Regularizing2018meyer}
do not hold,  as one can always modify the unoccupied SPFs appropriately, for example, either using the procedures described in \autoref{sec:subspace_enrichment} or even using the standard MCTDH regularization approach.

In general,  methods that use the TDVP can only describe SPFs to first order in time and thus there always is an arbitrariness regarding unoccupied SPFs~\cite{Multiconfigurational2015manthe}.
This holds both for (ML-)MCTDH and TDVP-DMRG, regardless of the way the regularization is performed (or whether it is  ``hidden'' by using matrix decompositions or gauges~\cite{Symmetries2021weike}).
Hence in certain scenarios care must be taken to converge simulations properly~\cite{Instabilities2016hinz,Timedependent2018kloss,Symmetries2021weike}, even though standard regularization works fine for a plethora of use cases. 

\subsection{Subspace enlargement: Optimizing unoccupied SPFs}
\label{sec:subspace_enrichment}
Instead of trying to propagate unoccupied SPFs, which is ill-defined, 
one can instead optimize them differently. 
There are different ways to optimize unoccupied SPFs, which in DMRG language also is known as subspace enrichment or subspace enlargement. 
Here, we will first discuss the two-site DMRG approach, then continue with MCTDH approaches, and finally discuss one-site DMRG approaches. Many of these are related to each other.

\subsubsection{Two-site DMRG approach}
\begin{figure}
\centering
\includegraphics{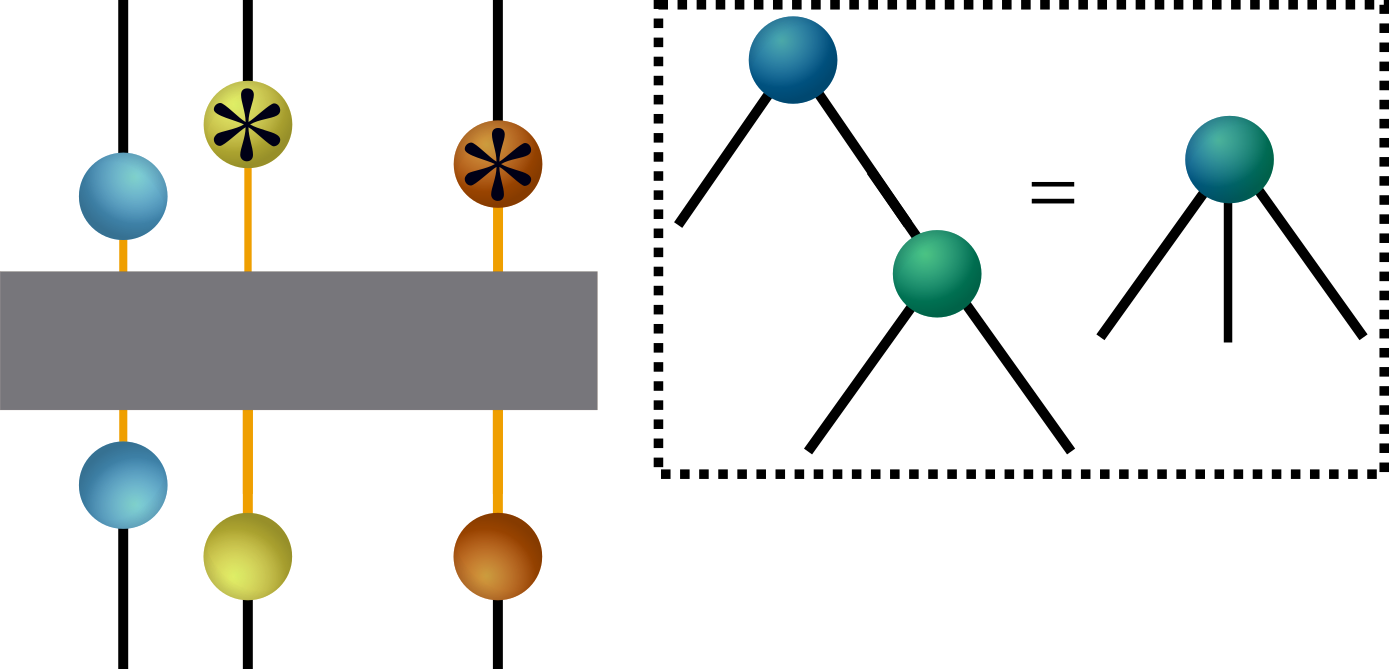}
\caption{Effective two-site Hamiltonian.
The TTNS from \autoref{fig:mltree} is used as example. 
Diagonalizing the effective Hamiltonian optimizes the tensor shown  in the dotted rectangle, which is a contraction of  $\matr A^1$ with $\matr A^{2;2}$.
The gray rectangle corresponds to the Hamiltonian $\hat H$.
Compare with \autoref{eq:two_site_Hamil}.
}
  \label{fig:ttns_2s}
\end{figure}

In practice, some but not all possible issues arising from the first-order description of the SPFs are alleviated in TDVP-DMRG by using the so-called two-site algorithm, where the local propagation problem is solved in the subspace of not one but \emph{two} tensors in the tree. This is done by first contracting the root tensor and one of its neighbors, then propagating the contracted tensor, and finally ``de-contracting'' the propagated tensor using, \eg,  SVD. 
In MCTDH language, 
we define the SHF configurations as 
\begin{equation}
\ket{\Xi_{J^{\kappa_l}}^{l; \kappa_1,\kappa_2,\dots,\kappa_{l}}} %
= \bigotimes_{\stackrel{\tilde\kappa_{l}=1}{\tilde\kappa_l \neq \kappa_l}}^{d_{\kappa_1,\kappa_2,\dots,\kappa_{l}}} \ket{\phi_{j_{\tilde \kappa_{l}}}^{l; \kappa_1,\kappa_2,\dots, \tilde\kappa_l}}, 
\label{eq:shf_configs}
\end{equation}
where $J^{\kappa_\lambda}$ contains all indices $j_{\tilde \kappa_l}$ but $j_{\kappa_l}$.
Then the two-site DMRG solves the Schrödinger equation in the  space of the SHF and SPF configurations, thus two-site DMRG effectively optimizes  an SHF-SPF coefficient tensor pair. This leads to an effective Hamiltonian with matrix elements
\begin{equation}
   \matrixe{\Xi_{I^\kappa}^{1;\kappa}\Phi^{2;\kappa}_{I'}}{\hat H}{\Xi_{J^\kappa}^{1;\kappa}\Phi^{2; \kappa}_{J'}}, 
   \label{eq:two_site_Hamil}
\end{equation}
compare with \autoref{fig:ttns_2s}.

The two-site variant can avoid convergence issues and is  particularly important for symmetry-adapted DMRG~\cite{Densitymatrix2011schollwock}.
Before one can orthogonalize to the next node in the sweep,
the two-site tensor needs to be decontracted using, \eg, an  SVD.
This allows for a straightforward way to dynamically adjust the bond dimension (number of SPFs). 
By observing the sum of discarded singular values,
one obtains an error estimate, which can be used for energy extrapolations~\cite{Highly2002chan,Controlling2003legeza,Abinitio2015olivares-amaya}. 
In fact, standard DMRG initially was formulated as two-site algorithm~\cite{Density1992white} and the one-site version is a more recent development~\cite{Density2005white}. 
Note, however, that two-site TDVP-DMRG still is based on the TDVP and thus possible issues due to  unoccupied SPFs may not be fully resolved.
Two-site DMRG also is known as modified ALS~\cite{Alternating2012holtz}.

\subsubsection{Lee/Fischer/Manthe approach}
Developing algorithms to systematically extend the SPF space/bond dimension is straightforward and both for MCTDH and for one-site TDVP-DMRG many schemes have been developed to extend the number of SPFs/bond dimension dynamically during time~\cite{Systematic2017mendive-tapia,Dynamical2017larsson,Regularizing2020mendive-tapia,Efficient2021dunnett,TensorTrain2022lyu,RankAdaptive2021dektor,Stochastic2022xu}. 
It is more difficult to actually optimize the newly added and thus unoccupied SPFs. 
To optimize unoccupied SPFs/to enlarge the subspace, in the DMRG and MCTDH communities, different approaches have been developed.
We first discuss solutions developed in the context of MCTDH.
There, the general idea is that instead of propagating SPFs that are unoccupied at time $t$, which is ill-defined, they are optimized by maximizing a quantity related to the overlap of the then-occupied SPFs at time $t+\Delta t$. This allows for a natural expansion of the SPF space during the time evolution.
Such an expression can be derived in various ways, but the general form of the expression derived by different scientists is similar.

\begin{figure}
\centering
\includegraphics{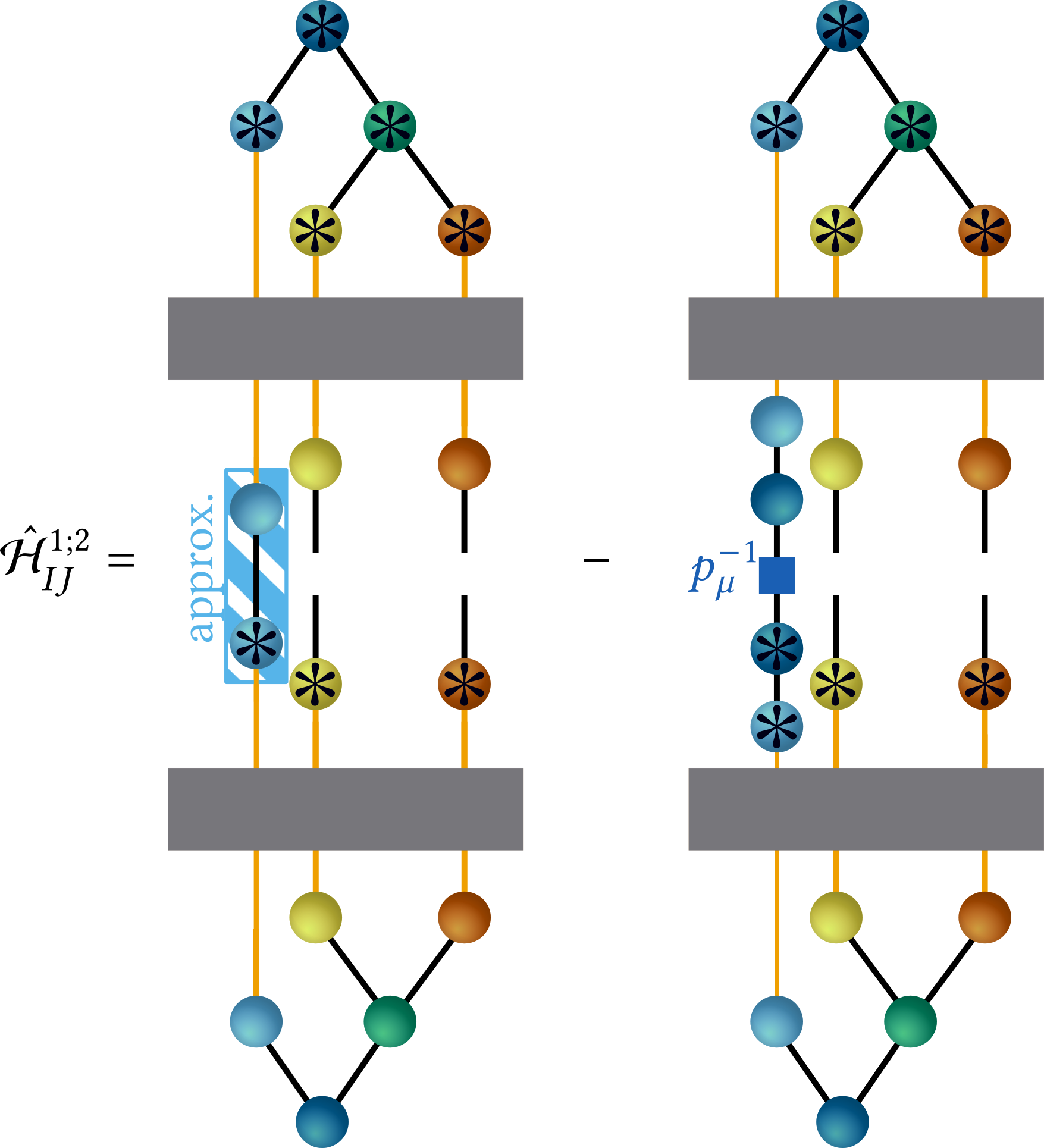}
\caption{Operator related to optimal unoccupied SPFs. 
Compare with \autoref{eq:h2_op}.
The TTNS from \autoref{fig:mltree} is used as example.
The gray rectangle corresponds to the Hamiltonian $\hat H$ and the blue square corresponds to the inverse of the natural occupations (overlaps of the natural SHFs).
The blue-striped  region corresponds to the approximation introduced in \autoref{eq:h2_op_dieter}.
This avoids computing $\hat H^2$ in the space of the primitive basis.
}
  \label{fig:enlarge}
\end{figure}

Lee and Fischer derived an expression for optimal SPFs for bosonic  MCTDH~\cite{Role2007streltsov,Multiconfigurational2008alon,Colloquium2020lode}
by minimizing the error
$\matrixe{\Psi}{[\ii \partd{}{t} - \hat H]^\dagger [\ii \partd{}{t} - \hat H]}{\Psi}$~\cite{Truncated2014lee}.
Manthe derived the same expression in the language of ordinary (ML-)MCTDH by analyzing the  expansion of the single-particle density matrix to second order in time~\cite{Multiconfigurational2015manthe,Optimized2018manthe}.
To analyze the expression Manthe arrived at, we consider here only natural SHFs, denoted by $\ket{\widetilde \Psi^z_\mu}$.\footnote{Extending this to non-natural SHFs is straightforward.}
By inspecting the natural occupations,
we divide the SHFs and corresponding SPFs in occupied ($\mu \in \text{occ}$) and unoccupied ones. Note that this introduces a numerical parameter. %
The expression used for finding optimal unoccupied SPFs then contains the following operator represented by the configurations $\{\ket{\Phi_J^{z'+1}}\}_J$:
\begin{align}
  \mathcal{\hat H}_{IJ}^z 
   &=  \matrixe{\Psi}{\hat H}{\Phi_J^{z'+1}}\left( \hat 1 - \mathcal{\hat P_\text{occ}}^z\right)\matrixe{\Phi_I^{z'+1}}{\hat H}{\Psi},
  \label{eq:h2_op}
\end{align}
where  $\mathcal{\hat P_\text{occ}}^z$ projects onto the space of the occupied natural SHFs:
\begin{equation}
  \mathcal{\hat P_\text{occ}}^z = \sum_{\mu\in \text{occ}}\ket{\widetilde \Psi^z_\mu}  \frac{1}{p_{\mu}^z} \bra{\widetilde\Psi^z_\mu}.
\end{equation}
The optimal unoccupied SPFs are then identified by the eigenstates with the largest eigenvalue of the operator 
\begin{equation}
\hat \Delta_\text{occ}^{z} = (\hat 1-\hat P_\text{occ}^z) \mathcal{\hat H}^z (\hat 1-\hat P_\text{occ}^z),\label{eq:delta_op}
\end{equation}
where $\hat P_\text{occ}^z$ projects onto the occupied natural SPFs. Since $\hat \Delta_\text{occ}^{z}$ contains both  $(\hat 1-\hat P_\text{occ}^z)$ and $(\hat 1 - \mathcal{\hat P_\text{occ}}^z)$, it  covers the space that is not included in the tangent space.
The eigenstates can be computed efficiently using iterative eigenvalue solvers such as the Lanczos solver.
See \autoref{fig:enlarge} for an example diagram.

While \autoref{eq:delta_op} contains the unoccupied SPFs that are optimal to second-order in time,
the term  $\matrixe{\Psi}{\hat H}{\Phi_J^{z'+1}}\matrixe{\Phi_I^{z'+1}}{\hat H}{\Psi}$ %
in  \autoref{eq:h2_op} is costly to compute as it contains $\hat H^2$ partially evaluated in the primitive basis. 
For SoP operators (and similarly for MPOs), $\hat H^2$ 
can be expressed using the partition from \autoref{eq:H_partition} as 
\begin{align}
\matrixe{\Psi}{\hat H}{\Phi_J^{z'+1}}\matrixe{\Phi_I^{z'+1}}{\hat H}{\Psi} &=
\sum_{s,s'}^S \matrixe{\Psi}{ \mathcal{\hat H}^z_s \hat h^{z}_s}{\Phi_J^{z'+1}}\matrixe{\Phi_I^{z'+1}}{ \mathcal{\hat H}^z_{s'} \hat h^{z}_{s'}}{\Psi}.
\end{align}
Inserting the SHF-SPF expansion \autoref{eq:shf_expansion} leads to 
\begin{equation}
\matrixe{\Psi}{\hat H}{\Phi_J^{z'+1}}\matrixe{\Phi_I^{z'+1}}{\hat H}{\Psi} =
\sum_{s,s'}^S \sum_{ij} \matrixe{\Psi^z_i}{ \mathcal{\hat H}^z_s
 \mathcal{\hat H}^z_{s'}}{\Psi^z_j} 
 \matrixe{\phi^z_i}{\hat h^{z}_s}{\Phi_J^{z'+1}}
\matrixe{\Phi_I^{z'+1}}{\hat h^{z}_{s'}}{\phi^z_j},
\label{eq:uwe_op_doublesum}
\end{equation}
which contains a double sum over all $S$ terms in the SoP Hamiltonian.
To avoid a scaling of $\mathcal O(S^2)$,  \autoref{eq:uwe_op_doublesum} can be approximated
by including only the ``diagonal'' terms with $s=s'$, which then leads to a favorably-scaling implementation. Manthe showed that this approximation leads to sufficiently optimal unoccupied SPFs~\cite{Optimized2018manthe}.

\subsubsection{Mendive-Tapia/Meyer approach}
Using a different derivation, Mendive-Tapia and Meyer arrived 
at an expression for the \emph{coefficients}  of the optimal unoccupied SPFs~\cite{Regularizing2020mendive-tapia}. 
They further derived an expression for the optimal unoccupied SPFs that is identical to \autoref{eq:h2_op} except for the costly $\matrixe{\Psi}{\hat H}{\Phi_J^{z'+1}}\matrixe{\Phi_I^{z'+1}}{\hat H}{\Psi}$, which is approximated by projecting this expression  onto the configurations of the SHFs.
This leads to an efficient implementation for an approximation of $\hat \Delta$.
Using another derivation that resembles more that of Lee and Fischer~\cite{Truncated2014lee}, Martinazzo and Burghardt arrived at essentially the same expression~\cite{LocalinTime2020martinazzo}.

\begin{figure}
\centering
\includegraphics{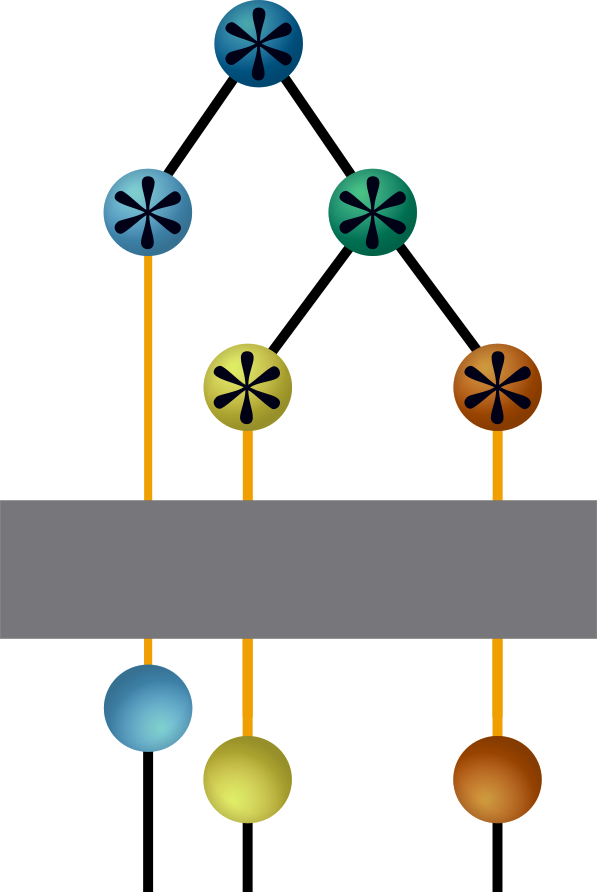}
\caption{State used for finding optimal unoccupied SPFs based on  Eqns.~\ref{eq:fop} and \ref{eq:h2_op_dieter}.
Compare with \autoref{fig:ttns_2s} and \ref{fig:enlarge}.}
  \label{fig:enlarge_2s}
\end{figure}

We now show 
that the two-site TDVP-DMRG algorithm
relates to the expression from Mendive-Tapia and Meyer. 
Their derivation was for ordinary MCTDH, but this can be straightforwardly extended to ML-MCTDH.
Using the SHF configurations, \autoref{eq:shf_configs},
we can define the following state
\begin{align}
\ket{\widetilde F^z_{J^{\kappa_l}}} =\left( \hat 1 - \mathcal{\hat P}^z\right) \matrixe{\Xi^z_{J^{\kappa_l}}\Phi_I^{z'+1}}{\hat H}{\Psi}.
\label{eq:fop}
\end{align}
Then $\mathcal{\hat{H}}_{IJ}^z$ from \autoref{eq:h2_op}  projected onto the space of the SHF configurations, dubbed here  $\mathcal{\hat{\widetilde H}}_{IJ}^z$, can be written as\footnote{This applies the SHF projector twice. As projectors are idempotent, this is identical to only applying the projector once, as is done in \autoref{eq:h2_op}.}
\begin{equation}
  \mathcal{\hat{\widetilde H}}_{IJ}^z = \sum_{J^{\kappa_l}} \ketbra{\widetilde F^z_{J^{\kappa_l}}}{\widetilde F^z_{J^{\kappa_l}}}. \label{eq:h2_op_dieter}
\end{equation}
In a similar fashion, by projecting $\ket{\widetilde F^z_{J^{\kappa_l}}}$ onto $(1-\hat P_\text{occ}^z)$ we can obtain an expression for an approximated $\Delta_\text{occ}^{z}$.
How does this relate to the two-site algorithm? 
$\ket{\Xi^z_{J^{\kappa_l}}\Phi_I^{z'+1}}$ used in \autoref{eq:fop} contains both the configurations of the SHF \emph{and} the configurations of the corresponding SPF. 
The SHFs and SPFs themselves are obtained by contracting with a coefficient tensor. So \emph{two} coefficient tensors are required to retrieve the total TTNS from $\ket{\Xi^z_{J^{\kappa_l}}\Phi_I^{z'+1}}$, which accordingly describe the configurations used in the two-site DMRG algorithm!
This is easily seen by comparing \autoref{fig:enlarge_2s}, where $ \matrixe{\Xi^z_{J^{\kappa_l}}\Phi_I^{z'+1}}{\hat H}{\Psi}$ is depicted,
with the corresponding two-site effective Hamiltonian shown in \autoref{fig:ttns_2s}.
Note that the computational cost of using \autoref{eq:h2_op_dieter} is smaller than that of using the two-site DMRG algorithm as the configurations in one node are contracted in \autoref{eq:h2_op_dieter}.\footnote{The SHF and SPF projectors do not increase the computational cost of \autoref{eq:h2_op_dieter}.} %
Hence, the expression found by Mendive-Tapia and Meyer can be interpreted as an approximation of the two-site DMRG algorithm,\footnote{This is also discussed in Ref.~\cite{klossComm}.}
and either this or Manthe's expressions  can be used not only in ML-MCTDH but also in one-site DMRG to increase the SPF space (thus increasing the bond dimension).
Note that these expressions can be used both for imaginary and for real-time propagation, and thus they can be combined with both DMRG and TDVP-DMRG.

\subsubsection{One-site DMRG approaches}
We now discuss solutions  to avoid dealing with the issue of propagating unoccupied SPFs developed in the context of the DMRG.
Yang and White suggested enlarging the TNS by using additional, approximate global Lanczos propagations~\cite{Timedependent2020yang}, where $\hat H$ explicitly (but approximately) acts onto $\ket{\Psi}$.
Another, more common approach initiated by White is based on perturbing the state during the DMRG sweep~\cite{Density2005white}.
There the starting point is slightly different to that of MCTDH. 
While the SPFs are extended directly using eigenvectors of $\hat \Delta^z_\text{occ}$ from \autoref{eq:delta_op}, 
in DMRG context this happens in SHF representation by adding some new states $\ket{F^z_x}$ to the SHF space and 
modifying the SHF-SPF orthogonalization procedure.
This can be done by either 
adding rows to the matricized coefficient tensor $\matr A^{z'}$  with entries $A^{z'}_{iJ}$ representing the SHF~\cite{Strictly2015hubig}, 
or by creating an SHF density matrix $\matrgreek \gamma = \matr A^{z' \dagger} \matr A^{z'}$ and perturbing it using~\cite{Density2005white}
\begin{equation}
\matrgreek \gamma \to \matrgreek \gamma + \sum_x \ketbra{F^z_x}{F^z_x}.\label{eq:subspace_enrichment_rdm}
\end{equation}
Note that $\matrgreek \gamma$ represents the SHF space and is not identical to the single-particle density matrix. 
Both procedures, adding rows to $\matr A^{z'}$ or perturbing $\gamma$, are identical but have different computational scalings~\cite{Strictly2015hubig,Computing2019larsson}.
Next to the difference that the SHF space and not the SPF space is enlarged, 
another major difference to MCTDH subspace enlargement schemes is that time-independent DMRG schemes may include an additional compression of the new state with larger $\nSPF$
to the original number of SPFs. This compression alters $\ket{\Psi}$, hence the name subspace enrichment. It is used for avoiding local minima with unwanted symmetries in DMRG.

\subsubsection{White approach}
The most commonly used enrichment scheme in DMRG is the one developed by White~\cite{Density2005white}. It is based on \autoref{eq:subspace_enrichment_rdm} but has also been extended to the SVD-based enrichment~\cite{Strictly2015hubig}. 
To make the connection to the MCTDH expansion schemes more clear, we will exchange the SPFs and SHFs in the following (or equivalently, system and environment in DMRG language). 
Thus we will not
perturb the SHF density matrix,  
\autoref{eq:subspace_enrichment_rdm} but rather
diagonalize the SPF operator \autoref{eq:delta_op}.\footnote{
Note that in an actual implementation,
due to gauge invariance,
exchanging SPFs and SHFs  
leads to different numerical values, depending on how the SHFs are orthogonalized.}
In this representation, using first-order perturbation theory and considering the SHF-SPF decomposition of the Hamiltonian, \autoref{eq:H_partition},
White derived the following term, which we show here in MCTDH notation:
\begin{align}
  \mathcal{\hat{\widetilde H}}_{IJ}^z 
= \sum_{s,s'}^S \sum_{ij}\frac{1}{\epsilon_s \epsilon_{s'}}
M_{sis'j}
 \matrixe{\phi^z_i}{\hat h^{z}_s}{\Phi_J^{z'+1}}
\matrixe{\Phi_I^{z'+1}}{\hat h^{z}_{s'}}{\phi^z_j},\label{eq:h2_op_white}
\end{align}
where $\epsilon_s$  corresponds to energies for each sum term evaluated in the SPF basis, and where
\begin{equation}
  M_{sis'j} =\matrixe{\Psi^z_i}{ \mathcal{\hat H}^z_s(\hat 1 - \mathcal{\hat P}^z)
 \mathcal{\hat H}^z_{s'}}{\Psi^z_j}. \label{eq:white_m}
\end{equation}
Note that aside from $\epsilon_s$,  and an exchange of SPFs and SHFs in White's original expression (which we do not show here),
\autoref{eq:h2_op_white}
is identical to the expression derived by Manthe, compare with Eqns.~\eqref{eq:h2_op} and \eqref{eq:uwe_op_doublesum} 
(there $\mathcal{\hat P}^z$ is not included).
For SoP operators or, equivalently, MPOs
with $S$ terms,
Manthe avoided the $\mathcal O(S^2)$ scaling by
evaluating the $\mathcal{\hat P}^z$-part of \autoref{eq:white_m} exactly but only considering the terms with $s=s'$ for the $\hat 1$-part.
White made a more severe approximation and assumed $M_{sis'j}\approx \delta_{ss'}\delta_{ij}$. Then \autoref{eq:h2_op_white} evaluates to 
\begin{equation}
 \mathcal{\hat{\widetilde H}}_{IJ}^z \approx
\sum_{s}^S \sum_{i} a_s
 \matrixe{\phi^z_i}{\hat h^{z}_s}{\Phi_J^{z'+1}}
\matrixe{\Phi_I^{z'+1}}{\hat h^{z}_{s'}}{\phi^z_i},
\end{equation}
with $a_s = 1/\epsilon_s$. Instead of evaluating $a_s$ explicitly, White set it to an empirical parameter that does not depend on $s$.
The more severe approximations done by White are useful as White considered time-independent DMRG and was seeking for a way to overcome local minima and to improve convergence. 
\subsubsection{Energy variance approaches}
Related subspace expansion methods have been developed that are based on
minimizing the two-site contribution to the energy variance~\cite{Error2018hubig,Variational2018zauner-stauber},
\begin{equation}
  \matrixe{\Psi}{(\hat H - E) {\hat Q}_{T2} (\hat H-E)}{\Psi},%
  \label{eq:energy_variance}
\end{equation}
where ${\hat Q}_{T2}$ project onto the space that is orthogonal to  the two-site tangent space (the zero- and one-site contributions are already included in one-site DMRG)~\cite{Error2018hubig,Projector2022gleisa}.\footnote{${\hat Q}_{T2}$ can be formed either using SHF and SPF projectors or explicitly by creating a null space~\cite{Error2018hubig}.}
While \autoref{eq:energy_variance} leads to  a very different derivation for finding optimal unoccupied SPFs,
this, essentially, is the same approach as that of Mendive-Tapia and Meyer, \autoref{eq:h2_op_dieter}.
A recent scheme to efficiently approximate \autoref{eq:energy_variance} uses sequences of SVDs~\cite{Timedependent2022lia,Controlled2023gleis}.\footnote{%
Note that these SVD-based approximations~\cite{Strictly2015hubig,Controlled2023gleis} 
make use of temporary matrices that scale with the bond dimension and the number of sum terms in an SoP Hamiltonian (or, equivalently, the MPO bond dimension). As there can be more than $\mathcal O(100-1000)$ terms in typical applications, this can lead to prohibitively large matrices and additional approximations need to be introduced, \eg, by only considering parts of the total Hamiltonian for the subspace enrichment.
}
\subsubsection{BUG integrator approach}
An extension of the non-energy-conserving BUG integrator~\cite{Unconventional2022ceruti},
which we briefly mentioned in \autoref{sec:tddmrg},
further allows for subspace extension and is norm and energy-conserving~\cite{Rankadaptive2022ceruti,Parallel2023ceruti}.
Therein, the SPF space at time $t$ is augmented by that at time $t+\Delta t$ and the bond dimension can be increased by up to a factor of two at each time step. It has recently been extended from the Tucker format to TTNSs~\cite{RankAdaptive2023ceruti}.

\section{Tree structures}
\label{sec:tree_opt}

Here, we will discuss why TTNSs are commonly used for molecular (ro-)vibrational quantum dynamics simulations whereas MPSs and not TTNSs are mostly used in other fields. As this discussion depends on the quantum problem, we restrict our comparison  to  molecular systems and compare molecular electronic structure with molecular vibrational dynamics. 
In addition, and related to this, we will discuss some strategies to find optimal tree structures for TTNSs.

\subsection{TTNSs or MPSs?}

Compared to MPSs, TTNSs  can more efficiently capture different groups of correlated degrees of freedom and efficiently separate them from less-correlated degrees of freedom. 
These degrees of freedom correspond to either orbitals in  electronic structure theory or vibrational (and rotational) modes/coordinates in vibrational quantum dynamics.  For example, a reaction can typically be described by some reaction modes that are highly correlated, and some spectator or ''bath`` modes that do not take direct part in the reaction. TTNSs allow for an efficient separation of these groups of correlated or uncorrelated modes.
Compared to MPSs, TTNSs thus often converge faster with respect to the number of required SPFs, $\nSPF$,
both in vibrational dynamics~\cite{Computing2019larsson} and in molecular electronic structure~\cite{Simulating2010murg,Efficient2013nakatani,Tree2015murg,T3NS2018gunst,Tensor2015szalay}.
If that is the case, why are TTNSs rarely used for electronic structure simulations? The reason is a different computational scaling.
MPSs have tensors of size $\nSPF \times \nSPF \times N$. MPSs describing electronic structures are most efficiently represented in Fock space~\cite{Tensor2015szalay}.
The physical dimension is then $N\le 4$ and thus negligible.
Thus, every electronic structure MPS tensor operation scales at least as $\nSPF^2$, where $\nSPF$ ranges from $500$ to even $30,000$~\cite{Abinitio2015olivares-amaya,Chromium2022larsson,Massively2023menczer}.
In contrast, TTNS tensors scale at least as 
$\nSPF^3$, since the minimal ''useful`` tensor dimension in a tree is $3$.
Accordingly, 
the scaling of TTNS operations in electronic structure is a factor of $\sim \nSPF$ \emph{larger} than that of MPSs~\cite{Efficient2013nakatani,T3NS2018gunst}. 
Even if TTNSs require a smaller $\nSPF$ for convergence, 
overall MPSs are computationally more advantageous in many electronic structure situations. 
This may change if the tree structure can also be exploited for the Hamiltonian (using a TTNO and not an MPO), together with a numerical compression of the operator~\cite{Matrix2016chan,Block22023zhai}.
A similar scaling argument applies to the use of mode combination for electronic structure simulations, which can be advantageous only in special cases~\cite{Minimal2020larsson,Matrix2022larsson}.

In contrast, vibrational dynamics simulations are most often performed in Hilbert space, and there the physical dimension is of similar magnitude than $\nSPF$: $N\sim \nSPF\sim \mathcal O(10)$.\footnote{With mode combination, the effective physical basis size can even be much larger than $\nSPF$; see \autoref{fig:eigen_ttns_opt} for an example.}
Then the argument above does not  apply any more and TTNSs frequently offer a clear computational advantage over MPSs. 
This has been numerically verified in Ref.~\cite{Computing2019larsson} and a reduction of the tensor sizes and a decrease of the computational runtime was observed when using TTNSs instead of MPSs for the 12-dimensional vibrational acetonitrile system. 
Note that vibrational dynamics can also be performed in Fock space~\cite{Second2004christiansen,Systematic2020madsen,MRMCTDH2020madsen} and MPSs can efficiently describe vibrational dynamics as well~\cite{Timedependent2019xie,LargeScale2019baiardi,Optimization2019baiardi,Matrix2018kurashige,Timedependent2021ren,Stochastic2022xu,Flexible2023glaser}.
Note further that $\nSPF$ is much smaller than in typical molecular electronic structure simulations. For example,  our simulations on the Zundel system used  $\nSPF\le 150$ to obtain $\sim 1000$ vibrational excited states with an energy error of  $\ll\unit[1]{cm^{-1}}$~\cite{Stateresolved2022larsson}.
Despite the ''small`` $\nSPF$, 
compared to electronic structure, 
in terms of $\nSPF$ 
this is one of the largest  vibrational TTNS/ML-MCTDH simulations done on realistic non-model systems. %
Many applications do not require such a tight energy convergence (\eg, as other observables are of interest) and thus require an even smaller $\nSPF$. 
The computational challenge is then different and lies in the efficient simulation of time evolution or computing \emph{many} excited states, sophisticated post-processing to obtain the required observables, and setting up realistic Hamiltonians~\cite{Quantum2007evenhuis,Fulldimensional2007vendrell,Multidimensional2009meyer,Transforming2020schroder,Coupling2022schroder}.

\subsection{Optimal tree structure}

The efficiency of MPSs depends on the ordering of the degrees of freedom in the MPS. Strongly coupled degrees of freedom should be placed closely in an MPS.
Given the establishments of many sophisticated algorithms using global optimization approaches and insights from quantum information theory~\cite{Highly2002chan,Convergence2005moritz,Quantuminformation2011barcza,Abinitio2015olivares-amaya,Fly2022li},
this nontrivial optimization problem can be considered as solved.

Less attention has been devoted to optimize tree structures, which is more complicated. 
One main guiding principle to set up an efficient tree is that the \emph{optimal} tensor dimension is three~\cite{Classical2006shi,Simulating2010murg,Efficient2013nakatani,T3NS2018gunst,Computing2019larsson}.
Any tree structure can be expressed using three-dimensional tensors and
restricting the tensor dimensions to three allows for an easier development of programs~\cite{T3NS2018gunst}.
Why using tree-dimensional tensors and not higher-dimensional tensors? The size of each tensor scales with $\nSPF^d$, where $d$ is the dimension. 
Applications of SoP operators onto a tensor scales as $\nSPF^{d+1}$~\cite{New1990manthe,General1993bramley}.
Thus, the higher the tensor dimension, the worse the scaling! The lowest possible tensor dimension that still allows for generating trees should be used, which is three.
Two-dimensional tensors are sometimes used in ML-MCTDH applications (including our example in \autoref{fig:mltree}), but they cannot express tree branching,  and they either can be restricted to be  diagonal~\cite{Multiconfiguration1993jansen,Comment1994manthe},
or they can be absorbed in a neighboring tensor, as is the case for  the $\matr R$ matrix in DMRG sweeps (see \autoref{fig:tddmrg}).

If three-dimensional tensors are the most optimal ones and two-dimensional tensors should be avoided to get TTNSs with most optimal scaling with respect to the number of parameters, why are seemingly non-optimal tensor dimensions being used in actual ML-MCTDH applications?
We are not aware that this has been discussed in ML-MCTDH literature
and we here provide three possible answers. %
(1) One practical reason is that mode combination immediately leads to high-dimensional tensors. For model systems mode combination can be avoided, but for non-model systems, mode combination is crucial for optimally fitting the potential into SoP form~\cite{Studying2012meyer,Transforming2017schroder,Transforming2020schroder}. 
(2) Another, more speculative reason is that  higher-dimensional tensors allow for a ''closer`` connection of highly correlated groups. Solely using three-dimensional tensors leads to more layers and more tensors in the tree. 
While we have not observed numerical difficulties using DMRG algorithms, 
many tensors and layers in a TTNS may lead to difficulties when integrating the ML-MCTDH equations of motion.  
(3) Trees with many layers lead to computational disadvantages when using quadrature to evaluate the potential operator in ML-MCTDH~\cite{Multilayer2008manthe}, although recent developments have improved this~\cite{Nonhierarchical2022ellerbrock}.
Thus, while three-dimensional tensors may give the most optimal scaling  with respect to number of parameters and operator applications, there are other reasons that do not allow for solely using three-dimensional tensors. Further studies are needed to clarify this.

Optimizing the tree structure is difficult for non-model systems (and sometimes even for model systems)
and in typical ML-MCTDH application, the setup of the tree structure is  mostly guided by chemical intuition and trial and error~\cite{Multilayer2011vendrell,Reaction2012welsch,Fulldimensional2014meng}.
To improve this, 
we introduced a systematic and automated way to find optimal tree based on some starting guess~\cite{Computing2019larsson}, that is, an initial, unoptimized tree, is ''disentangled`` to a more optimal tree where correlated and uncorrelated degrees of freedom are better separated from each other.
As the ordering in an MPS, this is an NP-hard problem and heuristic optimization methods need to be used~\cite{Global2011hartke}.
Our ''disentangling`` method was initially based on a greedy approach and contains four steps:
(1) A tensor and its neighbor are randomly chosen,  contracted, and set as new root node.
(2) The dimensions of the resulting two-site tensor are then randomly perturbed.
(3) At a randomly chosen bipartition, the two-site tensor is ''decontracted`` using an SVD. 
(4) The resulting two tensors are kept whenever the size of the so-obtained pair of tensors is smaller than that of the initial pair of tensors. 
These four steps
are repeated as long as there is a considerable reduction in the total number of parameters. 
This simple but fast procedure leads to very good results, and we could find a more optimal tree structure for the 12-dimensional vibrational acetonitrile  system~\cite{Computing2019larsson}.
We recently extended our greedy approach to simulated annealing, which allows us to overcome local minima found by the greedy approach. Further, we added the option to remove and to add nodes, which makes this tree structure optimization completely general.

\begin{figure}
\centering
\includegraphics[]{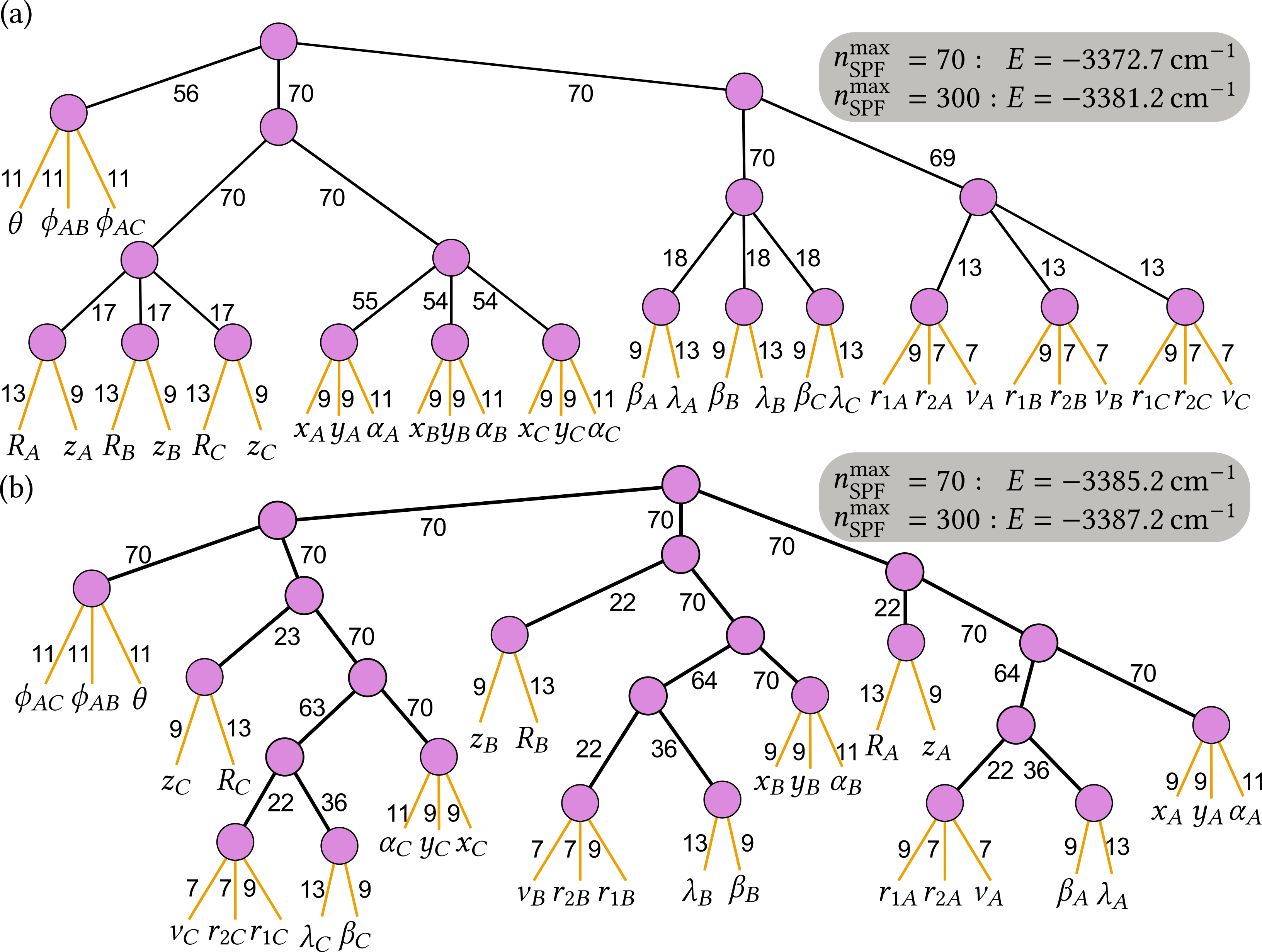}
\caption{TTNSs with (un-)optimized structure of the 33-dimensional vibrational Eigen ion.
(a) Tree structure used in Ref.~\cite{Coupling2022schroder,markus_schroder_2022_7064870}. 
(b) Optimized structure using the algorithm from Ref.~\cite{Computing2019larsson}. 
The symbols denote the specific coordinates~\cite{Coupling2022schroder}.
The numbers denote the bond dimension used during a DMRG ground state optimization with an adaptive number of SPFs with maximum value set to $\nSPF^\text{max}=70$.
The DMRG-optimized energies for  two values of $\nSPF^\text{max}$  are shown in the gray boxes.
The mode combinations (groups of coordinates) have not been optimized during the tree structure optimization, but this is possible.
}
  \label{fig:eigen_ttns_opt}
\end{figure}

An example of our tree structure optimization is shown in \autoref{fig:eigen_ttns_opt} for the problem of finding an optimal tree for the 33-dimensional Eigen ion~\cite{Coupling2022schroder,markus_schroder_2022_7064870}.
Our tree optimization procedure leads to a TTNS (\autoref{fig:eigen_ttns_opt}(b)) that for a bond dimension of $70$ gives an energy  that is $\sim\unit[13]{cm^{-1}}$ lower than the unoptimized one shown in  \autoref{fig:eigen_ttns_opt}(a), which served as initial guess and was taken from Ref.~\cite{Coupling2022schroder,markus_schroder_2022_7064870}.
Even with $\nSPF=300$ the energy of the unoptimized tree still is lower than that of the optimized tree with $\nSPF=70$!
Note that, save for the mode combination, the unoptimized tree contains four-dimensional tensors whereas the optimized tree only contains three-dimensional tensors, which is consistent with our discussion above. 
Our optimization procedure also allows us to get insights into the molecular system as it identifies strongly coupled modes.
Our tree structure optimization can be used for each eigenstate, and it can be used during a time propagation by optimizing the structure after every few time steps.

In condensed matter physics, related tree structure optimization were recently proposed that are either based on entanglement bipartitions~\cite{Entanglement2023okunishi}
or on structure optimization similar to ours but done \emph{during} the DMRG optimization~\cite{Automatic2023hikihara}. 
These and our procedures are based on some initial guess for the tree  and an initial Hamiltonian. The former is straightforward to obtain but for the latter  for non-model systems one often 
needs to fit the potential~\cite{Studying2012meyer,Transforming2017schroder,Transforming2020schroder}, which sometimes requires finding optimal mode combinations.
A recently suggested approach to do this is to 
analyze correlations from classical molecular dynamics simulations
in ways that are very similar to quantum information approaches used in DMRG~\cite{Optimal2023mendive-tapia}.
Alternatively, one can first use a non-optimal mode combination to get a non-optimal potential fit and then use the tree structure optimization.
By inspecting which modes are grouped together, this automatically also gives optimal mode combinations.

\section{Conclusions}
\label{sec:conclusions}
The ML-MCTDH method and the DMRG are powerful algorithms with a common mathematical background, but they were developed in different communities and  very different languages are used to describe the same mathematical expressions.
Here, we gave a direct and thorough comparison of the ML-MCTDH and DMRG theories by translating MCTDH expressions to tensor network diagrams and by comparing ML-MCTDH and DMRG algorithms to solve the Schrödinger equation.

Many independent developments in the  ML-MCTDH and in DMRG communities are very similar to each other, but different languages are used and often similar expressions are derived based on different approaches and contexts.
We highlighted this by showing the similarities of theories used in both fields for the advanced topic of subspace enrichment or, equivalently, finding optimal unoccupied single-particle functions. 
Finally, we discussed why trees are ubiquitously used in molecular quantum dynamics but why MPSs currently are preferred in molecular electronic structure, and how to optimize tree structures.

Different mathematical ``languages'' have different expressive powers.
In our opinion, diagrams are particularly useful for highlighting tensor contractions, and they offer convenient visual representations of the mathematical equations that can also be used for derivations, as we have shown here by deriving the ML-MCTDH equations.
In contrast, in our opinion, 
the MCTDH language is very convenient for highlighting different subspaces in a TTNS and hierarchical bases. We hope that this contribution will help to foster more adaption of using diagrams in MCTDH literature as well as the use of MCTDH concepts in DMRG literature.

Many topics have not been addressed here, and we did omit some developments from applied mathematics, as well as alternatives to the ML-MCTDH method and the DMRG.
For example, other topics that have been addressed by both communities but that we did not discuss here
relate to the tensor network representation of the Hamiltonian and how to account for distinguishable particles, density matrices and finite temperature.
As the ML-MCTDH method and the DMRG are typically used in different areas of research, some topics are more actively developed by one of the respective communities.
For example, topics %
that are more actively used in MCTDH methods are pruning (configuration selection) and mixed basis-grid representations. 
Vice versa,  topics %
that are more actively used in the context of the DMRG are symmetries, spectra calculations in frequency domain, and
global approaches that directly apply the Hamiltonian onto the tensor network.
Given these examples, there is still much room for mutual cross-fertilization of ideas. %
We hope that this article will help to foster more exchange of knowledge that will advance the overarching topic of tensor network states.

\section*{Acknowledgements}
H.R.L.~thanks A.~M.~Nierenberg for helpful discussions,  and 
Uwe Manthe, Hans-Dieter Meyer, and 
Imam Wahyutama for comments on a preliminary version of this article.
\section*{Funding}

This work was supported by the US
National Science Foundation (NSF) via grant no.~CHE-2312005.
Additional support came from University of California Merced start-up funding, %
and through 
computational time on the Pinnacles and Merced clusters at University of California Merced (supported by NSF OAC-2019144 and ACI-1429783).

\if\USEMOLPHYS1
\bibliographystyle{tfo}
\fi

\end{document}